\documentclass[%
 reprint,
 amsmath,amssymb,
 aps,
]{revtex4-2}

\usepackage{graphicx}% Include figure files
\usepackage{dcolumn}% Align table columns on decimal point
\usepackage{bm}% bold math
\usepackage{amsmath, amsthm, amssymb }
\usepackage{float,epsfig}
\usepackage{color}
\usepackage{multirow}
\usepackage{cleveref}
\usepackage{braket}
\usepackage{subfigure}

\begin{document}

\newtheorem{theorem}{\bf Theorem}[section]
\newtheorem{proposition}[theorem]{\bf Proposition}
\newtheorem{definition}[theorem]{\bf Definition}
\newtheorem{corollary}[theorem]{\bf Corollary}
\newtheorem{example}[theorem]{\bf Example}
\newtheorem{exam}[theorem]{\bf Example}
\newtheorem{remark}[theorem]{\bf Remark}
\newtheorem{lemma}[theorem]{\bf Lemma}
\newtheorem{statement}[theorem]{\bf Statement}
\newcommand{\nrm}[1]{|\!|\!| {#1} |\!|\!|}

\newcommand{\calL}{{\mathcal L}}
\newcommand{\calX}{{\mathcal X}}
\newcommand{\calA}{{\mathcal A}}
\newcommand{\calB}{{\mathcal B}}
\newcommand{\calC}{{\mathcal C}}
\newcommand{\calK}{{\mathcal K}}
\newcommand{\C}{{\mathbb C}}
\newcommand{\R}{{\mathbb R}}
\newcommand{\U}{{\mathrm U}}
\renewcommand{\SS}{{\mathbb S}}
\newcommand{\LL}{{\mathbb L}}
\newcommand{\st}{{\star}}
\def\kernel{\mathop{\rm kernel}\nolimits}
\def\sigan{\mathop{\rm span}\nolimits}

\newcommand{\klasse}{{\boldsymbol \Delta}}

\newcommand{\ba}{\begin{array}}
\newcommand{\ea}{\end{array}}
\newcommand{\von}{\vskip 1ex}
\newcommand{\vone}{\vskip 2ex}
\newcommand{\vtwo}{\vskip 4ex}
\newcommand{\dm}[1]{ {\displaystyle{#1} } }

\newcommand{\be}{\begin{equation}}
\newcommand{\ee}{\end{equation}}
\newcommand{\beano}{\begin{eqnarray*}}
\newcommand{\eeano}{\end{eqnarray*}}
\newcommand{\inp}[2]{\langle {#1} ,\,{#2} \rangle}
\def\bmatrix#1{\left[ \begin{matrix} #1 \end{matrix} \right]}
\def \noin{\noindent}
\newcommand{\evenindex}{\Pi_e}

\newcommand{\tb}[1]{\textcolor{blue}{ #1}}
\newcommand{\tm}[1]{\textcolor{magenta}{ #1}}
\newcommand{\tre}[1]{\textcolor{red}{ #1}}
\newcommand{\snote}[1]{\textcolor{blue}{Shantanav: #1}}

%\newcommand {\proof} {\par{\it Proof}. \ignorespaces}
%\newcommand {\eproof}
%      {\sigace
%        {\ \vbox{\hrule\hbox{\vrule height1.3ex\hskip0.8ex\vrule}\hrule}}
%        \par}

%%%%%%%%%%%%%%%%%%%%%%%%%%%%%%%%%%%%%%%%%%%%%%%%%%%%%%%%%%%%%%%%%%%%%%%%%%

%\def \Gr{{\mathsf{Grov}}}
\def \K{{\mathbf k}}
\def \N{{\mathbb N}}
\def \R{{\mathbb R}}
\def \F{{\mathbb F}}
\def \C{{\mathbb C}}
\def \Q{{\mathbb Q}}
\def \Z{{\mathbb Z}}
\def \I{{\mathbb I}}
\def \D{{\mathcal D}}
\def \H{{\mathcal H}}
\def \P{{\mathcal P}}
\def \M{{\mathcal M}}
\def \B{{\mathcal B}}
\def \O{{\mathcal O}}
\def \calG{{\mathcal G}}
\def \PO{{\mathcal {PO}}}
\def \X{{\mathcal X}}
\def \Y{{\mathcal Y}}
\def \calW{{\mathcal W}}
\def \pf{{\bf Proof: }}
\def \lam{{\lambda}}
\def\lc{\left\lceil}   
\def\rc{\right\rceil}
\def \N{{\mathbb N}}
\def \Ls{{\Lambda}_{m-1}}
\def \Gb{\mathrm{G}}
\def \Hb{\mathrm{H}}
\def \Delta{\triangle}
\def \Rar{\Rightarrow}
\def \p{{\mathsf{p}(\lam; v)}}

\def \D{{\mathbb D}}

\def \tr{\mathrm{Tr}}
\def \cond{\mathrm{cond}}
\def \lam{\lambda}
\def \sig{\sigma}
\def \sign{\mathrm{sign}}

\def \ep{\epsilon}
\def \diag{\mathrm{diag}}
\def \rev{\mathrm{rev}}
\def \vec{\mathrm{vec}}

\def \ham{\mathsf{Ham}}
\def \herm{\mathsf{Herm}}
\def \sym{\mathsf{sym}}
\def \odd{\mathsf{sym}}
\def \en{\mathrm{even}}
\def \rank{\mathrm{rank}}
\def \pf{{\bf Proof: }}
\def \dist{\mathrm{dist}}
\def \rar{\rightarrow}

\def \rank{\mathrm{rank}}
\def \pf{{\bf Proof: }}
\def \dist{\mathrm{dist}}
\def \Re{\mathsf{Re}}
\def \Im{\mathsf{Im}}
\def \re{\mathsf{re}}
\def \im{\mathsf{im}}

\def \sym{\mathsf{sym}}
\def \sksym{\mathsf{skew\mbox{-}sym}}
\def \odd{\mathrm{odd}}
\def \even{\mathrm{even}}
\def \herm{\mathsf{Herm}}
\def \skherm{\mathsf{skew\mbox{-}Herm}}
\def \str{\mathrm{ Struct}}
\def \eproof{$\blacksquare$}

\def \bS{{\bf S}}
\def \cA{{\cal A}}
\def \E{{\mathcal E}}
\def \X{{\mathcal X}}
\def \F{{\mathcal F}}
\def \cH{\mathcal{H}}
\def \cJ{\mathcal{J}}
\def \tr{\mathrm{Tr}}
\def \range{\mathrm{Range}}
\def \adj{\star}

\preprint{APS/123-QED}

\title{Limit theorems and localization of three state quantum walks on a line \\ defined by generalized Grover coins}% Force line breaks with \\
%\thanks{A footnote to the article title}%

\author{Amrita Mandal$^1$}
\email{mandalamrita55@gmail.com}
\author{Rohit Sarma Sarkar$^{1}$}
\email{rohit15sarkar@yahoo.com}
\author{Shantanav Chakraborty$^{2,3}$}
\email{shchakra@iiit.ac.in}
\author{Bibhas Adhikari$^1$}
\email{bibhas@maths.iitkgp.ac.in}
%\author{Sonjoy Majumder$^1$}
%\email{sonjoym@phy.iitkgp.ac.in}

\affiliation{$^1$Department of Mathematics, Indian Institute of Technology Kharagpur,\\ Kharagpur 721302, India}
%\affiliation{Department of Mathematics and Center for Theoretical Studies, Indian Institute of Technology Kharagpur, Kharagpur 721302, India}
\affiliation{$^2$Centre for Quantum Science and Technology, International Institute of Information Technology,\\ Hyderabad, 500032, India}
\affiliation{$^3$ Centre for Security, Theory and Algorithmic Research, International Institute of Information Technology,\\ Hyderabad, 500032, India}

\date{\today}% It is always \today, today,
             %  but any date may be explicitly specified

\begin{abstract}
%Localization is a feature of discrete-time quantum walks wherein there is a non-zero probability of the walker to remain in its initial position even after the time-averaged limiting distribution is reached.
In this article, we undertake a detailed study of the limiting behaviour of a three-state discrete-time quantum walk on one dimensional lattice with generalized Grover coins. %\tb{two families} of parametrized coins, called generalized Grover coins. \tb{Interestingly, one family contains the widely used Grover coin as a special case and the negative of Grover coin belongs to the other family.} 
Two limit theorems are proved 
%that includes computable expression of the limiting value of the probability measure for finding the walker at any position on the line. 
and consequently we show that the quantum walk exhibits localization at its initial position, for a wide range of coin parameters. Finally, we discuss the effect of the coin parameters on the peak velocities of probability distributions of the underlying quantum walks. 

%This answers a question posed by Stefanak et al., [The European Physical Journal D 66 (5) 2012] by showing the localization phenomena inherited by a class of a parametric family of coin operators that includes the Grover coin. 

%\begin{description}
%\item[Usage]
%Secondary publications and information retrieval purposes.
%\item[Structure]
%You may use the \texttt{description} environment to structure your abstract;
%use the optional argument of the \verb+\item+ command to give the category of each item. 
%\end{description}
\end{abstract}

%Use showkeys class option if keyword
                              %display desired
\maketitle

%\tableofcontents

%\section{\label{sec:level1}First-level heading:\protect\\ The linebreak was forced \lowercase{via} \textbackslash\textbackslash}

\section{Introduction}\label{sec:intro}

Quantum walks, the quantum analogue of classical random walks \cite{aharonov2001quantum} represent an universal model for quantum computation \cite{childs2009universal,lovett2010universal}. Besides being a useful primitive to design quantum algorithms \cite{childs2003exponential, ambainis2007quantum, magniez2011search, apers2021quadratic}, quantum walks also provide a useful framework to model transport in quantum systems \cite{mohseni2008environment, rebentrost2009environment, chin2010noise}. Just as in the classical scenario, quantum walks can be defined in both discrete and in continuous-time \cite{meyer1996kt, farhi1998quantum, venegas2012quantum}. Often for several of the aforementioned applications, studying the long-time behaviour of both discrete-time and continuous-time quantum walks is extremely crucial. Quantum evolutions are unitary and hence quantum walks do not naturally converge to a limiting distribution, unlike classical random walks. However, for a quantum walk, one can define the limiting distribution as the long-time average probability distribution of finding the walker in each node of the graph. This long-time behaviour of quantum walks has been fundamental to demonstrate the speedup of several quantum algorithms \cite{childs2003exponential, chakraborty2020finding, atia2021improved}, to prove the equivalence between the circuit and Hamiltonian models of quantum computing \cite{aharonov2008adiabatic, caha2018clocks} and also to understand the phenomenon of mixing of quantum walks \cite{aharonov2001quantum, richter2007quantum, chakraborty2020fast, chakraborty2020analog}. Consequently, it is important to develop a more comprehensive understanding of the limiting distributions of the various quantum walks and to highlight the features that distinguish them from their classical counterparts.

Localization has been widely studied in the context of discrete-time quantum walks and its precise definition varies across articles (For example see \cite{inui2004localization, inui2005localization, inui2005one, segawa2013localization, kollar2015strongly, tate2019eigenvalues} and the references therein). The probability of finding the walker at a fixed lattice point converges to zero after infinite long time in the case of Hadamard walk on a line with two inner states \cite{konno2009localization}. In this paper our goal is to study the limiting distribution and the localization phenomena for the coined three state discrete-time quantum walk on a line. We consider two distinct families of parameterized coin operators which we have introduced in \cite{sarkar2020periodicity}: one that includes the widely used Grover coin as a special case, while the other does not. In particular, for this graph, we demonstrate, for the first time, localization on a general family of coins that does not include the Grover coin.

Inui et al. derived the first long-time limit theorem for the three state quantum walk on a line with a Grover coin (generally referred to as the Grover walk) and show that the probability of finding the particle at the origin does not converge to zero after infinite time steps, demonstrating localization \cite{inui2005one}. On the other hand, the three-state quantum walk with an asymmetrical jump and three-state quantum walk on a triangular lattice, both using the Grover coin, do not exhibit localization \cite{inui2005localization,kollar2010recurrences}. 
%While the three-state quantum walk on a line using the Grover coin exhibits localization at the origin  \cite{inui2005localization}, on a triangular lattice, localization is not reported \cite{kollar2010recurrences}. 
The probability of staying at the initial position also vanishes for some special set of escaping initial states in the case of a two dimensional Grover walk on the Cartesian lattice, leading to partial trapping \cite{inui2004localization}. Furthermore, there exists quantum coins for such escaping states (corresponding to strong trapping) \cite{kollar2015strongly}. In \cite{kollar2020complete}, the authors have classified explicitly the coins leading to the trapping effect for walk on the two-dimensional square lattice based on the existence or non-existence of an escaping state.

Attempts have been made to generalize the Grover walk in the literature by considering parametric coin operators that contain the Grover matrix for some specific values of the parameters \cite{inui2004localization,vstefavnak2012continuous,sarkar2020periodicity,MANDAL2022}. Discrete-time quantum walks with a generalized Grover coin are generally referred to as generalized Grover walks (GGWs). One of the endeavours of the community has been to understand the dynamics of GGWs in order to distinguish them from their classical counterparts \cite{watabe2008limit,sarkar2020periodicity,mandal2021localization}. The dependence of the underlying quantum walk dynamics on the values of the coin parameters has been explored in Ref.~\cite{di2011alternate}. The limit distribution of a four state quantum walk with a parametric coin has been investigated in  \cite{watabe2008limit}. The authors of Ref.~\cite{ide2011return} have provided analytical results for the asymptotic value of the return probability and discuss the localization of a two-state one-dimensional DTQWs as $t\to \infty$. In \cite{vstefavnak2012continuous} {\v{S}}tefa{\v{n}}{\'a}k et al. have introduced two families of parametric coin operators: one by deforming the eigenvalues and the other by deforming the eigenvectors  of the Grover coin. Further, in \cite{vstefavnak2014limit} {\v{S}}tefa{\v{n}}{\'a}k et al. have presented a rigorous analysis of the limiting behavior of these coined walks and the role of eigenstates for the localization phenomena. In \cite{mandal2021localization} we have explored the localization property of four state quantum walks on two dimensional lattices with generalized Grover coins.

%While both the walks preserve localization, for quantum walks using coins of the first kind, the features of the three-state Grover walk gradually diminish as the coin parameter increases \cite{vstefavnak2014limit}.

In terms of GGWs, it is undoubtedly challenging to find the parametric coin matrices that preserve the property of Grover walks. With an appropriate choice of coin operator from the generalized coin collections, the result regarding localization at a position may help us to design effective quantum algorithms based on the corresponding quantum walk \cite{chen2016defect}. Furthermore, it is interesting to ask whether it is indeed possible to study the long-time dynamics of discrete-time quantum walks using a class of parametric coin operators that do not include the Grover coin. Can such quantum walks exhibit localization? In this article, we make novel inroads towards answering these questions by considering two families of parametric  orthogonal coin operators of dimension $3 \times 3$. %Interestingly, one of these is a generalization of the Grover matrix, the other does not include in them the Grover coin operator. 
Mathematically, these coin operators are linear sum of permutation matrices and hence they can have convenient quantum circuit representation \cite{bataille2022quantum}. These coins  have been used recently to study properties of quantum walks dynamics such as periodicity of evolution \cite{sarkar2020periodicity}. In this article, we derive the long-time limit probability distribution and study the localization of three-state DTQWs with these families of parametric coin operators. Using this limiting value of the probability measure we are able to justify localization phenomena at a vertex for a given initial state.

Furthermore, the spread of the quantum walk wavefunction on a line can be modelled using the theory of wavefront propagation \cite{knight2003quantum, kempf2009group, orthey2019connecting}. In this framework, Ref.~ \cite{vstefavnak2012continuous} analyzed the peak velocities of the wavefunctions of two GGWs and explore the dependence of the coin parameter on the rate of propagation of the walk through the line. Furthermore, the localization of the wavefunction is dependent on the eigenvalues of the evolution operator that are independent of the wavenumber (constant). These correspond to the peak of the probability distribution which does not propagate, ensuring localization. Based on the constant eigenvalues of the evolution operator, a necessary and sufficient condition is obtained in \cite{tate2019eigenvalues} for quantum walks on infinite lattices corresponding to periodic transition operator that does not localize at a vertex.

%For the Grover walk on a line, this was explored in Ref.~\cite{inui2004}. For lattices with periodic boundary conditions, the existence of constant eigenvalues of the time-evolution operator and the existence of localization has been shown to be equivalent \cite{segawa2016generator}.

In this article, following \cite{vstefavnak2014limit}, we also find the dependence of the peak velocities of the wavefunction as a function of the parameters corresponding to general families of coin operators we have considered. We revisit the localization phenomena from the aspect of zero travelling velocity of one of the peaks. Consequently, we obtain new families of $3\times 3$ generalized Grover coin operators, preserving  localization, which is different from the two classes mentioned in \cite{vstefavnak2012continuous}. This answers a question posed by {\v{S}}tefa{\v{n}}{\'a}k et al. in \cite{vstefavnak2012continuous}. In this work we note that the permutation coins play a key role in the behavioral changes of probability distribution at a time and sign changes of group velocities. These observations make the proposed walk intriguingly different from the existing GGWs in literature. 

The organization of this paper is as follows. In Section \ref{sec:three state walk} we formally define the quantum walk we consider, namely three state DTQWs on a line with parametric coin operators. We present two limit theorems on probability distributions at infinite time steps and establish the localization property in Section \ref{sec:limit law}. In Section \ref{sec:peak velocity} we determine the right and left travelling peak velocities of the probability distribution as a function of coin parameter and discuss the impact of the peak velocities on the localization phenomena of the walks. Finally, we conclude this article with some problems that can be considered in near future.

\section{Three state walk on a line and generalized Grover coins} \label{sec:three state walk}
In this section we review three-state quantum walks on a line \cite{inui2005one, vstefavnak2012continuous} and we briefly discuss generalized Grover coins of dimension $3\times 3$ that are recently introduced in literature \cite{sarkar2020periodicity}. 

Discrete time coined quantum walks (DTQWs) are defined on the Hilbert space $\H=\H_p \otimes \H_c,$ where $\H_p$ is the position space and $\H_c$ is the coin space. The Hilbert space $\H_p$ is considered as separable and is isomorphic to $l^2(\Z),$ the Hilbert space of all square summable functions on $\Z$.  Hence the superposition of the canonical basis vectors subject to the unit norm condition form the position state vector $\ket{ m}\in \H_p$ for the position $m \in \Z$ \cite{venegas2012quantum}.  Otherwise, in a more conventional way we can say that $\H_p=Span\{\ket{m}| m \in \Z\}$ \cite{vstefavnak2012continuous}, \cite{vstefavnak2008recurrence}.

At every step the intrinsic behaviour of the quantum particle allows it to move to the right or to the left or stay at its current position. Hence the dimension of $\H_c$ is the number of internal degrees of freedom (called chirality) of the particle at each step is three.  The vector of the standard basis assigned to each of these three displacements spans the space $\H_c$ i.e. $\H_c=Span\{\ket{1},\ket{2},\ket{3}\},$ where $\{\ket{l}|~l=1,2,3\}$ is the canonical ordered basis of $\mathbb{C}^3$. 
  Thus the total state space $\H=Span\{\ket{m}\otimes\ket{{l}}|~m\in \mathbb{Z},l\in\{1,2,3\}\}$ and is isomorphic to $ l^2(\mathbb{Z})\otimes\C^3$ \cite{ko2016one}.
  
One single step of a DTQW is described by a unitay operator $U=\mathsf{S}(I\otimes C),$ where $I$ is the identity operator acting on $\H_p$, $C$ is the coin operator acting on the space $\H_c$ while $\mathsf{S}$ is the conditional shift operator, allows the walker to transition into the next step, controlled by the state of the coin register. The operator $\mathsf{S}$, acting at a particle in position $m\in\mathbb{Z}$ results in  
\beano 
\mathsf{S} = \sum_{m\in \mathbb{Z}}  && \left(\ket{m-1}\bra{m} \otimes \ket{1}\bra{1} + \ket{m}\bra{m}\otimes \ket{2}\bra{2} \right.\\ && \left.+  \ket{m+1}\bra{m}\otimes \ket{3}\bra{3}\right). 
\eeano
 The state vector $\ket{\psi(t)}\in \H$ of the walker after $t$ time steps is given by \begin{eqnarray} 
\ket{\psi(t)} &=& U\ket{\psi(t-1)}=U^t\ket{\psi(0)} \nonumber \\ &=& \sum_{ m\in \mathbb{Z}}\sum_{l\in \{1,2,3\}}\psi_{l}( m,t)\ket{ m}\otimes \ket{l},\label{eqn:psi t}
\end{eqnarray} where the product state of initial position state and initial coin state is the initial state vector $\ket{\psi(0)}\in \H$  and $\psi_l(m,t)$ is the probability amplitude at time $t$ and at the position $m,$ with coin state $\ket{l},~l\in\{1,2,3\}.$ 
Then at time $t$ and at $m \in \mathbb{Z}$ the probability amplitude vector for chirality state being left, center or right is \be \label{eqn:prob am} \ket{\psi( m,t)}=\left[\psi_1( m,t),\psi_2( m,t),\psi_3( m,t)\right]^T.\ee 
Therefore we write $$\ket{\psi(t)}=[\ldots,\ket{\psi(-1,t)},\ket{\psi(0,t)},\ket{\psi(1,t)},\ldots]^T.$$ 
  
Now for a complex valued function  $f\in l^2(\Z),$ its discrete Fourier transform  ${\bf f}:\I \rightarrow \C$ is defined as  ${\bf f}(k)=\sum_{m\in \Z}{f(m)e^{i m k}}\in L^2(\I),$ where $\I=[-\pi,\pi]$ and $L^2(\I)$ is the Hilbert space consisting of all square integrable functions on $\I.$ Thus the Fourier transform from $l^2(\Z)$ to
$L^2(\I)$ extends to the Fourier transformation of $\ket{\psi(m,t)}$ as \be\label{eq:Fourier tr}\ket{\Psi(k,t)}=\sum_{m\in \Z}{\ket{\psi(m,t)}e^{i k m}},\ee
where $\ket{\Psi(k,t)}=\left[\Psi_1(k,t),\Psi_2(k,t),\Psi_3(k,t)\right]^T,\Psi_l(k,t)=\sum_{m\in \mathbb Z}{\psi_l(m,t)e^{i k m}}, l=1,2,3$ for $k\in \I.$ The inverse Fourier transformation is given as 
$$
\ket{\psi(m,t)}=\frac{1}{(2\pi)}\int_{[-\pi,\pi]}e^{-i k m}\ket{\Psi(k,t)} \, dk.
$$
Now we obtain the time evolution of $\ket{\psi(m,t)}$ from Eq.~\eqref{eqn:psi t} and Eq.~\eqref{eqn:prob am} as
 \beano \ket{\psi(m,t+1)} &=& \ket{1}\bra{1}C\ket{\psi(m-1,t)}+\ket{2}\bra{2}C\ket{\psi(m,t)} \\ && +\ket{3}\bra{3}C\ket{\psi(m+1,t)},
 \eeano 
where $C$ is the coin operator. Then the following holds from \eqref{eq:Fourier tr}:
 \beano
 \ket{\Psi(k,t+1)} &=& \sum_{m\in \mathbb Z}(\ket{1}\bra{1}C\ket{\psi(m-1,t)}+\ket{2}\bra{2}C\ket{\psi(m,t)} \\ && +\ket{3}\bra{3}C\ket{\psi(m+1,t)})e^{i m k} \\ 
 &=& \sum_{m\in \mathbb Z}e^{ik}\ket{1}\bra{1}C\ket{\psi(m-1,t)}e^{i (m-1) k}  \\ && +\ket{2}\bra{2}C\ket{\psi(m,t)}e^{i m k} \\ && +e^{-ik}\ket{3}\bra{3}C\ket{\psi(m+1,t)}e^{i (m+1) k} \\ 
 &=& \left(e^{ik}\ket{1}\bra{1}C+\ket{2}\bra{2}C  +e^{-ik}\ket{3}\bra{3}C\right)\ket{\Psi(k,t)}\\
 &=& \mbox{diag}(e^{ik},1,e^{-ik})C\ket{\Psi(k,t)}.
 \eeano
 Let $\tilde U(k)=D(k)C$ for $D(k)=\mbox{diag}(e^{ik},1,e^{-i k}).$ Then from above the time evolution in momentum space is given as \cite{vstefavnak2008polya,vstefavnak2008recurrence}.

	\begin{equation} \label{eq:psi}
    \ket{{\Psi}(k,t)}=\tilde{U}(k)\ket{{\Psi}(k,t-1)}=\tilde{U}^t(k)\ket{\Psi(k,0)}.
    \end{equation}
    
Let the eigenvalues of the unitary operator $\tilde U(k)$ be of the form $e^{iw_j(k)}$ with the corresponding eigenvector $\ket{v_j(k)}$, for $j\in \left\{1,2,3\right\}$. Then the vector of probability amplitudes can be determined by Eq.~\eqref{eq:Fourier tr} and Eq.~\eqref{eq:psi} as follows:
\begin{eqnarray} \label{eqn:spectral de}
&& \ket{\psi(m,t)} \nonumber\\ &=&\frac{1}{(2\pi)}\int_{k\in (-\pi,\pi]} e^{-i k m}\ket{\Psi(k,t)} dk \nonumber\\
%&=&\int_{k} e^{-i k.m} \tilde U^t(k) \Psi(\K,0) dk \nonumber\\
%&=&\int_{k}  e^{-i k.m} \sum_{j=1}^r {e^{i w_j(k) t} \ket{v_j(k)}\bra{v_j(k)}} \Psi(\K,0) dk\nonumber\\
&=&\sum_{j=1}^{3}\frac{1}{(2\pi)}\int_{k\in (-\pi,\pi]}  e^{i (-k m+ w_j(k)t)}\braket{v_j(k)|\Psi(k,0)}   \ket{v_j(k)} dk.\nonumber \\ &&
\end{eqnarray}
  
From (\ref{eqn:psi t}), the probability of finding the walker at time $t$ and at position $ m$ is \begin{eqnarray} \label{eqn:prob} P( m,t) &=& \langle \psi( m,t)|\psi( m,t)\rangle \nonumber \\ &=& |\psi_1( m,t)|^2+|\psi_2( m,t)|^2+|\psi_3( m,t)|^2,\end{eqnarray}
and it can be derived from Eq.~\eqref{eqn:spectral de}. Whereas, the probability $P(0,t)$ of finding the walker at position $m=0$ after $t$ time steps, which is also called the return probability of the walker to the initial position ${0},$ can be evaluated as  $P(0,t)=\langle \psi(0,t)|\psi(0,t)\rangle.$
%as \be P( m,t)=\langle \psi( m,t)|\psi( m,t)\rangle.\ee
Note that if the walk starts at $m=0$ and is localized there, then from Eq.~\eqref{eqn:psi t}, we get that $$
\ket{\psi(0)}=\sum_{l=1}^3\psi_l(0,0)\ket{0}\otimes \ket{l},
$$ 
whereas the other probability amplitudes $\psi_l(m,0)=0$ for $ m\neq 0$. Hence from Eq.~\eqref{eq:Fourier tr} we have that $\ket{\Psi(k,0)}=\ket{\psi(0,0)}$. We assume that the initial state of the walker is at the origin. Thus, the initial quantum states are determined by $\ket{\psi(0,0)}=[\alpha,\beta,\gamma]^T,$ where $\alpha,\beta,\gamma \in \mathbb{C}$ and $|\alpha|^2+|\beta|^2+|\gamma|^2=1.$

We emphasize that the localization phenomena is extensively studied in literature for DTQWs on different topologies of the underlying system with a variety of coin operators, as mentioned in Section \ref{sec:intro}. However, the mathematical definition of localization associated with the probability measure of finding the walker at a vertex differs across articles. For example, in \cite{inui2004localization}, the positive value of the total time averaged probability $\lim_{T\rightarrow \infty} \frac{1}{T}\sum_{t=0}^{T-1}  P(m,t), T\geq 1,$ is considered as the condition for localization; whereas in Ref. \cite{machida2014limit} positive value of $\sum_{m\in \Z} \lim_{t \to \infty} P(m,t)$ is taken to be the criterion for localization. Indeed, it has been shown that the localization phenomena at the initial position depends on the initial state of the walker \cite{inui2005one}. In this paper, we follow  the localization criterion outlined in Ref.~\cite{segawa2016generator}. Thus we consider the localization of a DTQW at a location $m$ corresponding to an initial state if $\lim_{t \to \infty} P(m,t)>0.$ A computable formula of this limiting value is obtained in terms of the coin parameters and a generic initial state. Therefore, the limiting value can be obtained with a given initial state and the coin parameters and consequently, the localization phenomena can be tested. For completeness, we also classify all the initial states for which the proposed walks exhibit localization. Finally we provide a condition based on the coin parameters satisfying which the proposed walks show localization in the sense of Ref. \cite{machida2014limit}. 
%It is easy to see that this definition of localization implies that the conditions of localization discussed in \cite{machida15,inui2004,inui2005one} also hold. 
 
%\tre{We say that the walk localizes at the vertex $m\in \mathbb{Z}$ if there exists an initial coin state $\ket{\Psi(k,0)}$ such that $\lim_{t \to \infty} P(m,t)>0$  \tre{\cite{machida15}, \cite{ko2016one}}.}
  
\subsection{Generalized Grover coins}
In this work we consider $3\times 3$ parametric coin operators to describe the three-state DTQWs on the line. In particular, we choose $C\in \X$ and $C \in \Y$ \cite{sarkar2020periodicity}, where \begin{widetext}
\beano
\X&=& \left\{ \bmatrix{
x & y &1-x-y\\
1-x-y & x & y\\
y &1-x-y & x
} : x^2+y^2-x-y+xy=0,-\frac{1}{3}\leq x \leq 1 \right\},\label{eqn:x}\\
\mathcal{Y}&=&\left\{ \bmatrix{
x & y &-1-x-y\\
-1-x-y & x & y\\
y &-1-x-y & x
} : x^2+y^2+x+y+xy=0, -1\leq x\leq \frac{1}{3} \right\}. \label{eqn:y}
\eeano
\end{widetext}

It is easy to check that the Grover coin is an element of $\X$ by setting $x=-1/3$ and $y=2/3;$ whereas the negative of the Grover coin is a member of $\Y$ setting $x=1/3$ and $y=-2/3.$ It is also shown in \cite{sarkar2020periodicity} that $\X$ and $\X\cup \Y$ both form groups under matrix multiplication. Further, we consider one-parameter trigonometric parametrization of $\X$ and $\Y$ as $\X_\theta$ and $\Y_\theta$ respectively (see Section \ref{sec:limit law}). Then it is obvious that if $A(\theta)\in \X_\theta\cup \Y_\theta$ then $[A(\theta)]^{-1}=[A(\theta)]^T=A(-\theta)$ where $\theta\in [-\pi, \, \pi].$ Another salient feature of these parametric orthogonal matrices is that these are permutative matrices, that is, any row of any such matrix is a permutation of any other row, a combinatorial structure of the Grover matrix. Besides, matrices in $\X \cup \Y$ can be expressed as linear sum of permutation matrices. Consequently, these coins can conveniently be implemented by parametrized quantum circuits \cite{klappenecker2003quantum}. Finally, we establish from the numerical computation that $\lim_{t\rightarrow \infty} P(m,t)$ at the initial position $m$ is equal for the coin operators $A(\theta)$ and its inverse $A(-\theta).$ 

As mentioned in the introduction, attempts have been made in literature to study coined quantum walks by generalizing the Grover diffusion matrix with parametric unitary matrices. However, the existing classes of parametric coins do not have any algebraic structure such as a group structure. This makes the proposed classes of coins significantly different from the existing parametric coins in the literature. Moreover, note that $\mbox{Det}(A)=1$ if $A\in\X$ and $\mbox{Det}(A)=-1$ if $A\in Y$ (see \cite{sarkar2020periodicity}). Hence the matrices in $\X$ and $\Y$ belong to different connected components in the Lie group formed by orthogonal matrices of order $3$. 

%\tre{Besides, the matrix class $\X$ includes the Grover coin and hence we call the walks with coins from $\X$ as generalized Grover walks. It is needless to say that the matrix collection $\Y$ contains the matrix which is negative of the Grover coin. }

In the next two sections we determine the limit laws of the probability distribution for infinite time steps. Besides, we discuss the localization phenomena of the walks from two different aspects: existence of the non-zero return probability $\lim_{ t \to \infty} P(m,t)$ and the zero velocity of probability distribution peak of the particle to stay at the position $m=0$ on the line.

\section{Limit theorems for walks with generalized Grover coins} \label{sec:limit law}
The asymptotic probability distribution and the localization of three state walks on one and two dimensional lattices with parametric coin operators have been studied \cite{machida2014limit,machida2015localization}.
In this section, we consider the three-state walks on a line with parametric coin operators $C\in \X\cup \Y$. We present the limit value of the probability that a walker can be found at a vertex on $\mathbb{Z}$ for long time. Besides, we show the localization phenomena of these walks.  

\subsection{Coins from $\X$}

The evolution operator of three state walk with generalized Grover coin in Fourier domain is given by $\tilde U_{\mathcal X}(k)=D(k)C,$ where $C\in \mathcal{X}$ and $k\in (-\pi,\pi].$
Indeed we write explicitly
\begin{equation}\label{eq:evolu 1d line}
\tilde{U}_{\mathcal X}(k)= \bmatrix{xe^{ik} & ye^{ik} & (1-x-y)e^{ik}\\
1-x-y & x & y\\
ye^{-ik} & (1-x-y) e^{-ik} & x e^{-ik}},
\end{equation}
where $x^2+y^2+xy-x-y=0$. 

First we derive the eigenvalues and eigenvectors of $\tilde{U}_{\mathcal X}(k)$  and $\tilde{U}_{\mathcal Y}(k)$ via two theorems. They are defined as follows:

\begin{theorem} \label{thm:eig 1d line}
Consider $\tilde{U}_{\mathcal X}(k)$ as defined in Eq.~\eqref{eq:evolu 1d line}. Then the set of eigenpairs $(\lambda_j(k),v_j(k))$ of $\tilde{U}_{\mathcal X}(k)$ are 
\beano \lambda_j &=& e^{i\omega_j(k)},\\ \ket{v_j(k)} &=& \frac{1}{\|f_j(\omega(k))\|}\left[\frac{y+\lambda_j(1-x-y)}{(1-x-y)} \right. \\ && \left. +\lambda_j ye^{-ik},1,\frac{(1-x-y)+\lambda_j y}{y+\lambda_j (1-x-y)e^{ik}}\right]^T,\eeano
 where $\omega_1(k)=0,\omega_2(k)=-\omega_3({ k})=\omega({k})$ with $\cos{\omega(k)}=x\cos{k}-\frac{(1-x)}{2}$ and 
 \beano \|f_j(\omega(k))\|^2 &=& \frac{{1+x}-{2x}\cos{\omega(k)}}{{1+x}-{2x}\cos{({\omega(k)}-k)}}  + 1 \\ &&+ \frac{{1+x}-{2x}\cos{\omega(k)}}{{1+x}-{2x}\cos{({\omega(k)}+k)}},\eeano
for $j=\{1,2,3\}$.
\end{theorem}
\begin{proof} Note that the characteristic polynomial of $\tilde U_{\mathcal X}(k)$ is $\kappa_{\mathcal X}(\lambda)=\lambda^3-(2 \cos k+1) x \lambda^2+(2 \cos k+1) x \lambda-1.$ Thus the eigenvalues and the corresponding eigenvectors can be obtained by simple algebraic computations.
\end{proof}

Let us take $\ket{\Psi(k,0)}=[\alpha,\beta,\gamma]^T\in \C^3,$ where $|\alpha|^2+|\beta|^2+|\gamma|^2=1.$ 
From Eq.~\eqref{eqn:spectral de}, we have
\begin{eqnarray} \label{wave function:prob ampli} && \ket{\psi(m,t)} \nonumber \\ &=& \sum_{j=1}^{3}\frac{1}{2\pi}\int_{-\pi}^{\pi} e^{i (- k m+ w_j( k)t)}\braket{v_j( k)|\Psi(k,0)} \ket{v_j( k)} dk \nonumber \\
&=& \sum_{j=1}^3[\psi_j(m,t,1,\alpha,\beta,\gamma),\psi_j(m,t,2,\alpha,\beta,\gamma), \nonumber \\ && \psi_j(m,t,3,\alpha,\beta,\gamma)]^T, 
\end{eqnarray}
where 
\beano && \psi_j(m,t,l,\alpha,\beta,\gamma) \\ &=& \frac{1}{2\pi}\int_{-\pi}^{\pi} e^{i (- k m+ w_j( k)t)} \braket{v_j(k)|\Psi(k,0)}\braket{l|v_j(k)},\eeano
and $\{\ket{l}| l=1,2,3\}$ is the canonical basis of $\mathbb{C}^3$. 

Clearly $\psi_l(m,t)=\sum_{j=1}^{3}\psi_j(m,t,l,\alpha,\beta,\gamma)$, for $l=1,2,3$ and hence 
$$P(m,t)=\sum_{l\in \{L,S,R\}}|\psi_l(m,t)|^2=\sum_{l=1}^3\sum_{j=1}^3|\psi_j(m,t,l,\alpha,\beta,\gamma)|^2.
$$ 
Let us denote the probability of finding a particle at node $m$ after time steps $t$ with chirality $l$ by $P(m,t,l),$ then $P(m,t,l)=|\psi_\mathsf{l}(m,t)|^2$, where $l=1,2,3.$

 Now we state the long-time limit theorem of the DTQWs with coin operators from $\X$ at vertex position $m$ on the line as follows. In order to derive these theorems, we make use of the well-known Riemman-Lebesgue Lemma~\cite{stein2011fourier} which is stated as follows:
 
  \begin{theorem}(Riemman-Lebesgue Lemma)
  Let $f$
  %$f:\mathbb{R}\rightarrow \mathbb{R}$ 
  be an integrable real valued function on $[0,2\pi]$. Then $\lim_{n \to \infty}\int_{0}^{2\pi}f(x) \cos(nx)\,dx=0$ and $\lim_{n \to \infty}\int_{0}^{2\pi}f(x) \sin(nx)\,dx=0.$
\end{theorem} 

We prove the following theorem which finds the asymptotic value of the probability measure for finding the particle as $t\to \infty$ for discrete-time quantum walks using coin operators $C\in \mathcal{X}$.

\begin{theorem} \label{thm: limit law 1d}
 Let $C\in \X$ be not a permutation matrix i.e. $x \neq 0,1$ and $\Psi(k,0)=[\alpha,\beta,\gamma]^T$ be the initial state of the walker. Then for $m\in \mathbb{Z}$  
 \begin{widetext}
 \beano
 \lim_{t\to \infty}{P}(m,t) &\sim&  \frac{1}{3(1-x)(x+3)}\{|A\nu^{|m|} + B \nu^{|-m+1|}|^2 + \left|\alpha(1-x-y)\nu^{|m|}+\alpha y \nu^{|m+1|}+\beta(1+x)\nu^{|m|}\right. \\ && \left. -\beta x \nu^{|-m+1|} -\beta x \nu^{|m+1|}+ \gamma(1-x-y)\nu^{|-m+1|}+\gamma y \nu^{|m|}\right|^2 + |A\nu^{|m+1|}+B\nu^{|m|}|^2\}
\eeano
\end{widetext}
where
\begin{eqnarray*}
\nu&=& - \frac{(x-3)+\sqrt{3(1-x)(x+3)}}{2x},\\
A&=&{\alpha(1-x)+\beta(1-x-y)} , \\
B&=&\gamma(1-x)+\beta y.
\end{eqnarray*}
 If $C\in \X$ are permutation matrices and  $\ket{\Psi(k,0)}=[\alpha,\beta,\gamma]^T$ is the initial state of the walker, then for $m\in \mathbb{Z}$,
  \beano  \lim_{t\to \infty}{P}(m,t)\sim  
  \left\{
  \begin{array}{l}
   |\beta|^2 \,\, \mbox{if} \,\, m=0, C=I \\ \\
   \frac{1}{3}(|\alpha|^2 +2|\beta+\gamma|^2)  \, \mbox{if} \,\, m=0, C=P_{(123)} \\ \\
   \frac{1}{3}|\beta+\gamma|^2  \,\, \mbox{if} \,\, m=-1, C=P_{(123)} \\ \\
    \frac{2}{3}|\alpha|^2 \,\, \mbox{if} \,\, m=1, C=P_{(123)} \\ \\
     \frac{1}{3}(|\gamma|^2 +2|\alpha+\beta|^2) \,\, \mbox{if} \,\, m=0, C=P_{(132)} \\ \\
      \frac{2}{3}|\gamma|^2 \,\, \mbox{if} \,\, m=-1, C=P_{(132)} \\ \\
      \frac{1}{3}|\alpha+\beta|^2 \,\, \mbox{if} \,\, m=1, C=P_{(132)} \\ \\
      0, \,\, \mbox{otherwise.}
  \end{array}
\right. \eeano
\end{theorem}
\begin{proof}
 The asymptotic nature of the amplitude vector can be derived by Riemann-Lebesgue Lemma for infinite time steps (see Appendix \ref{appdx:A} for the detail) so that
 
\beano
&& \ket{\psi(m,t)} \\ &=&\sum_{j=1}^{3}\frac{1}{2\pi}\int_{-\pi}^{\pi} e^{i (- k m+ w_j( k)t)}\braket{v_j( k)| \Psi(k,0)} \ket{v_j( k)} dk\\
&\sim& \frac{1}{2\pi}\int_{-\pi}^{\pi} e^{- i k m}\braket{v_1(k)|\Psi(k,0)} \ket{v_1( k)} dk\,\,\, (\mathrm{for~}t\to \infty).
\eeano

Thus, 
\beano
&& \ket{\psi(m,t)} \sim \frac{1}{2\pi}\int_{-\pi}^{\pi}\frac{1}{\|f_1(\omega(k))\|^2}\left[\frac{1-x}{(1-x-y)+ye^{-ik}},1,\right. \\ && \left. \frac{1-x}{y+(1-x-y)e^{ik}}\right]^T
\left[\alpha \frac{1-x}{(1-x-y)+ye^{ik}}+\beta+ \right. \\ && \left.\gamma \frac{1-x}{y+(1-x-y)e^{-ik}}\right]e^{-ikm} dk    ~~~~(\mathrm{for~}t\to \infty)\\
&=&[\psi_1(m,t,1,\alpha,\beta,\gamma),\psi_1(m,t,2,\alpha,\beta,\gamma),\psi_1(m,t,3,\alpha,\beta,\gamma)]^T
\eeano

By Theorem \ref{thm:eig 1d line} we have 
$$\frac{1}{\|f_1(\omega(k))\|^2}=\frac{1+x-2x\cos{k}}{3-x-2x \cos{k}},$$ 
for $\lambda_1=e^{i \omega_1(k)}=1$ and  
hence the coefficient of $\alpha$ in the expression of $\psi_1(m,t,1,\alpha,\beta,\gamma)$ is equal to
\beano
\frac{1}{2\pi}\int_{-\pi}^{\pi} && \frac{(1-x)^2}{(1-x-y+ye^{-ik})(1-x-y+ye^{ik})} \\ && \left(\frac{1+x-2x\cos{k}}{3-x-2x \cos{k}}\right)e^{-ikm} dk.
%\frac{1}{2\pi}\int_{-\pi}^{\pi}\frac{(1-x)^2}{(1-x-y)^2+y^2+2y(1-x-y)\cos k}\left(\frac{1+x-2x\cos{k}}{3-x-2x \cos{k}}\right)e^{-ikm} dk\\&=\frac{1}{2\pi}\int_{-\pi}^{\pi}\frac{(1-x)^2}{(1-x^2)+2(x^2-x)\cos k}\left(\frac{1+x-2x\cos{k}}{3-x-2x \cos{k}}\right)e^{-ikm} dk\\&=\frac{1}{2\pi}\int_{-\pi}^{\pi}\frac{1-x}{1+x-2x\cos{k}}\left(\frac{1+x-2x\cos{k}}{3-x-2x \cos{k}}\right)e^{-ikm} dk.
\eeano

 Whenever $x\not \in \{0,1\}$ i.e. $C\in \X$ is not a permutation matrix, the above expression simplifies to 
\beano
&& \frac{1}{2\pi}\int_{-\pi}^{\pi}\frac{(1-x)}{3-x-2x \cos{k}}e^{-ikm}dk \\ && =
\frac{1}{2\pi}\int_{-\pi}^{\pi}\frac{(1-x)\cos{km}}{3-x-2x \cos{k}}dk\\ &&=\frac{1}{\sqrt{3(1-x)(x+3)}}(1-x)\nu^{|m|}, 
\eeano
where $\nu=- \frac{(x-3)+\sqrt{3(1-x)(x+3)}}{2x}.$ 
Similarly we calculate for the others and come up with the following.
\begin{widetext}
\begin{align*}
\ket{\psi(m,t)}\sim &\frac{1}{\sqrt{3(1-x)(x+3)}}\left[\alpha(1-x)\nu^{|m|}+\beta(1-x-y)\nu^{|m|}+\beta y \nu^{|-m+1|}+ \gamma(1-x)\nu^{|-m+1|},\right.\\ 
&\left. \alpha(1-x-y)\nu^{|m|}+\alpha y \nu^{|-m-1|}+\beta(1+x)\nu^{|m|}-x\beta\nu^{|-m+1|} -x\beta\nu^{|-m-1|}+\right.\\ &\left.\gamma(1-x-y)\nu^{|-m+1|}+y\gamma\nu^{|m|}, \alpha(1-x)\nu^{|-m-1|}+\beta y \nu^{|m|}+\beta(1-x-y)\nu^{|-m-1|}+\gamma(1-x)\nu^{|m|}\right].
\end{align*}
\end{widetext}

Thus the statement of the theorem follows after computing the norm of $\psi(m,t)$. Now let us take $C$ to be the permutation coin. This can be obtained from $\mathcal{X}$ by substituting either $x=0$ or $x=1$.

First let $x=1,$ then $C$ is the identity matrix. Clearly the eigenvalues of corresponding $\tilde{U}(k)$ are $e^{ i k},1,e^{-i k}$. Thus,
\begin{align*}
\ket{\psi(m,t)}\sim&\frac{1}{2\pi}\int_{-\pi}^{\pi} \braket{2|\Psi(k,0)}e^{-ikm}\ket{2} dk~~~(\mathrm{for~}t \to \infty),
\end{align*} 
where $\ket{2}=[0 ,1, 0]^T$. This results $\lim_{t\to \infty}P(m,t)=|\beta|^2$ for $m=0$ and $\lim_{t\to \infty}P(m,t)=0$ for $m\neq 0.$

Let $x=0$ then either $y=0$ or $y=1.$ If $y=1$ then $C=P_{(123)}.$ 
%and $$\tilde{U}(k)= \bmatrix{0 & e^{ik} & 0\\0 & 0 & 1\\e^{-ik} & 0 & 0}.$$
%$\bmatrix{0 & 1 & 0\\0 & 0 & 1\\1 & 0 &0}=P_{(213)}$ 
Correspondingly, the eigenvalues of $\tilde{U}(k)$ are $\omega,\omega^2$ and $1$ where 
$\omega=e^{\frac{2\pi i}{3}}$. In this case,
\beano \ket{\psi(m,t)} \sim \frac{1}{2\pi}\int_{-\pi}^{\pi} && \left[\alpha+ (\beta+\gamma)e^{i k}, \alpha e^{-ik}+ (\beta+\gamma), \right. \\ && \left. \alpha e^{-ik}+ (\beta+\gamma)\right]^Te^{-i k m} dk ~~(t \to \infty). 
\eeano

If $x=0$ and $y=0,$ then $C=P_{(132)}.$ 
%and $$\tilde{U}(k)=\bmatrix{0 & 0 & e^{ik}\\1 & 0 & 0\\0 & e^{-i k} &0}.$$ 
The eigenvalues of $\tilde{U}(k)$ are $\omega,\omega^2$ and $1$ where $\omega=e^{\frac{2\pi i}{3}}$. Hence we get 
%\begin{widetext}
\beano \ket{\psi(m,t)} \sim \frac{1}{2\pi}\int_{-\pi}^{\pi} && \left[(\alpha+ \beta)+\gamma e^{ik},    (\alpha+ \beta)+\gamma e^{ik}, \right.\\ && \left. (\alpha+ \beta)e^{-i k}+\gamma\right]^T e^{-i k m} dk,
\eeano setting $t \to \infty.$ This completes the proof. \end{proof}
%\end{widetext}

%\textcolor{red}{Better to write these two equations using 'ket' notation.}
 
%Thus the statement of the theorem follows. 

\begin{figure}[H]
    \centering
    \subfigure[$x=-\frac{1}{3}$]{\includegraphics[height=4 cm,width=4 cm]{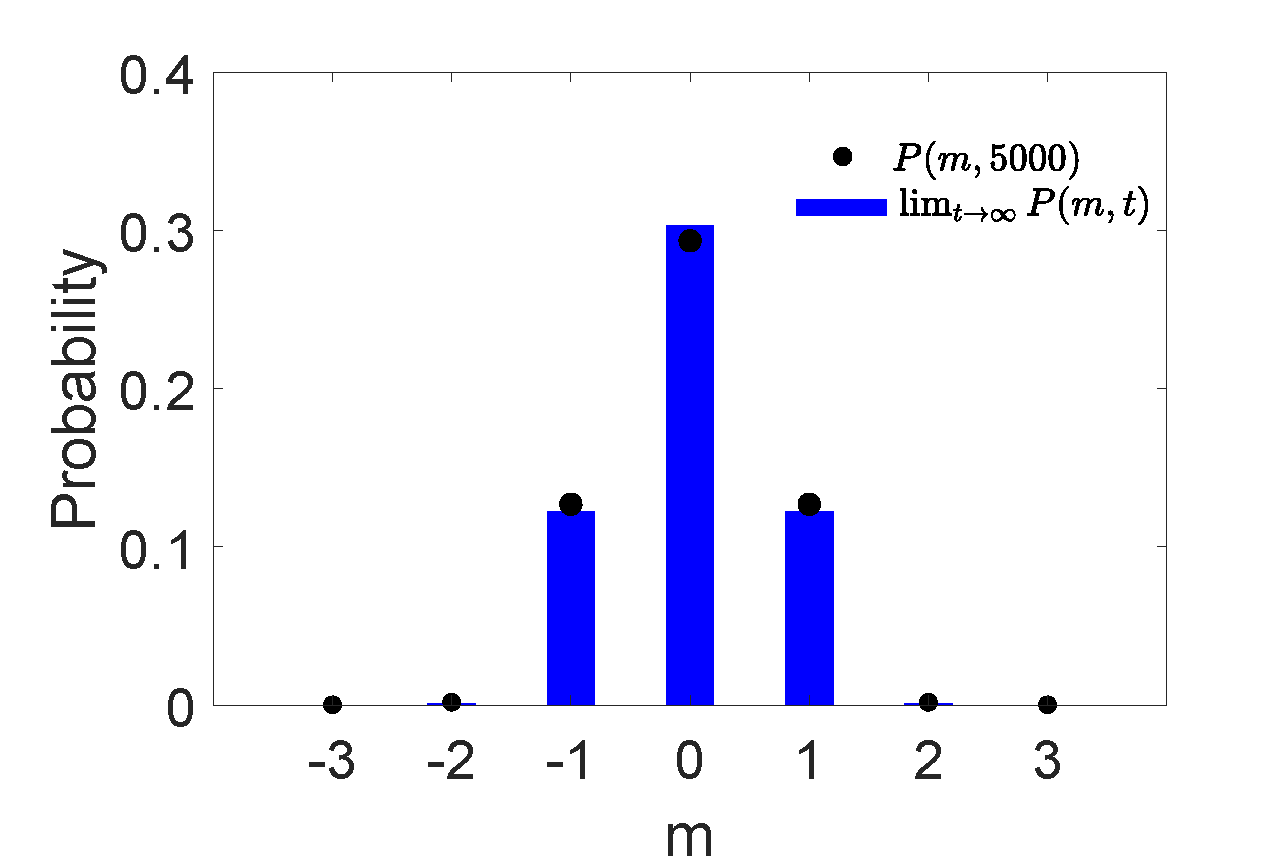}}
\hspace{0.25cm}
\subfigure[$x=\frac{1+\sqrt2}{3}$]{\includegraphics[height=4 cm,width=4 cm]{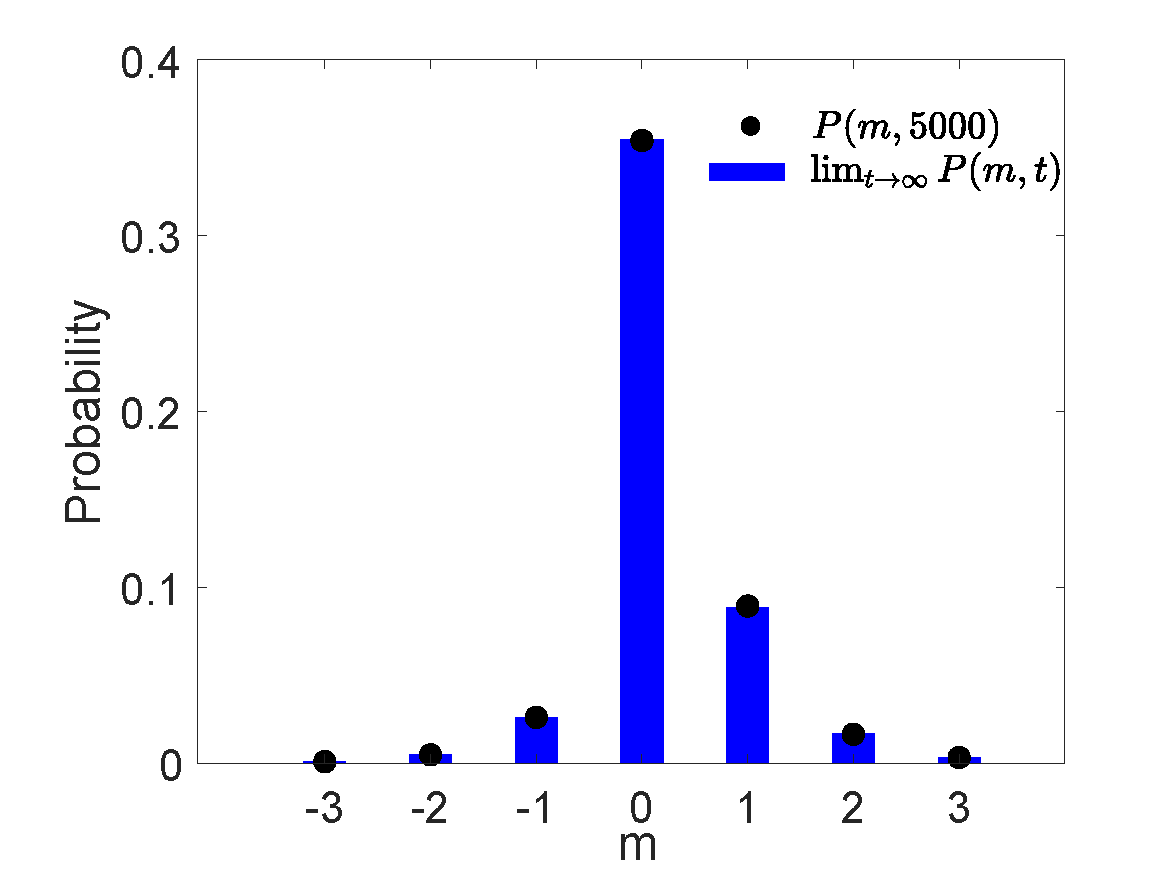}} 
\caption{Figure $(a)$ and Figure $(b)$ compare the probability distribution $P(m,t)$ at time $t = 5000$ (black points) and the limit measure $\lim_{t\to \infty}{P}(m,t)$  (blue bars) when the underlying coin operators are from the collection $\X$ correspond to $x=-\frac{1}{3}$ and $x=\frac{1+\sqrt2}{3}$,
%$\X_{\pi}$ and $\X_{\frac{\pi}{4}},$ 
respectively; with initial coin state $[\frac{1}{\sqrt{3}},\frac{1}{\sqrt{3}},\frac{1}{\sqrt{3}}]^T.$ 
We get $P(0, 5000)=0.2934,\lim_{t\to \infty}{P}(0,t)=0.3031$ for $(a)$ and $P(0, 5000)=0.3542,\lim_{t\to \infty}{P}(0,t)=0.3548$  for $(b).$ 
%In both the cases we see that the probabilities at the position $m=0$ is higher than the probability values at other positions on the line. Also, except the positions at $m=0,1,-1$ the probabilities are very small. 
}\label{fig:limit prob x}
\end{figure}

In order to corroborate the theoretical findings of Theorem \ref{thm: limit law 1d}, we plot Figure \ref{fig:limit prob x}. We compare the limiting probability of the walker to be at some vertex $m$ of the line, i.e.\ $\lim_{t\to\infty} P(m,t)$, obtained in the aforementioned theorems with the actual probability of the walker to be at $m$ after a large but finite number of walk steps. For Figure \ref{fig:limit prob x}, we consider a DTQW starting from origin and using a coin operator from the family $\mathcal{C}\in\mathcal{X}.$ In all the figures we observe that the probability of finding the particle is maximum around the origin.

Using the computable expression of $\lim_{t\to\infty} P(m,t)$ as derived in Theorem \ref{thm: limit law 1d}, the localization condition for the DTQWs according to the definition given in \cite{machida2014limit} is  \beano && \left[\frac{1+\nu^2}{1-\nu^2}\left(2|A|^2+2|B|^2+|T_1|^2+|T_2|^2+|T_3|^2\right) \right. \\ && \left. +\frac{4 \nu}{1-\nu^2}\left(\mbox{Re}(A\overline{B})+\mbox{Re}(T_1\overline{T_2})+\mbox{Re}(T_1\overline{T_3})\right) \right. \\ && \left. +2\nu^2\frac{3-\nu^2}{1-\nu^2}\mbox{Re}\left(T_2\overline T_3\right)\right]>0\eeano where $T_1=\alpha(1-x-y)+\beta(1+x)+\gamma y,T_2=\alpha y -\beta x,T_3=-\beta x+\gamma(1-x-y)$. Note that $-5+2\sqrt{6}\leq \nu< 1$ when $x\neq 0,1$ so that $1-\nu^2\neq 0$. This follows from the derivation in Appendix~\ref{appdx:B}.

%If $x\neq 1,0,$ the walks localize at $m\in \mathbb{Z}$ with an initial state $[\alpha,\beta,\gamma]^T$ such that $\lim_{t \to \infty}P(m,t)>0$ and consequently $\sum_{m\in \mathbb{Z}}\lim_{t \to \infty}P(m,t)>0.$

%Using the above derivation we can also justify the localization phenomena of the walks under consideration if
%Thus the condition that the walks localize yields from above as 
% \beano && \left[\frac{1+\nu^2}{1-\nu^2}\left(2|A|^2+2|B|^2+|T_1|^2+|T_2|^2+|T_3|^2\right) \right. \\ && \left. +\frac{4 \nu}{1-\nu^2}\left(\mbox{Re}(A\overline{B})+\mbox{Re}(T_1\overline{T_2})+\mbox{Re}(T_1\overline{T_3})\right) \right. \\ && \left. +2\nu^2\frac{3-\nu^2}{1-\nu^2}\mbox{Re}\left(T_2\overline T_3\right)\right]>0\eeano adapting the definition of localization in \cite{machida15}.

\begin{figure}[H]
    \centering
    \subfigure[$m=0$]{\includegraphics[height=4 cm,width=4 cm]{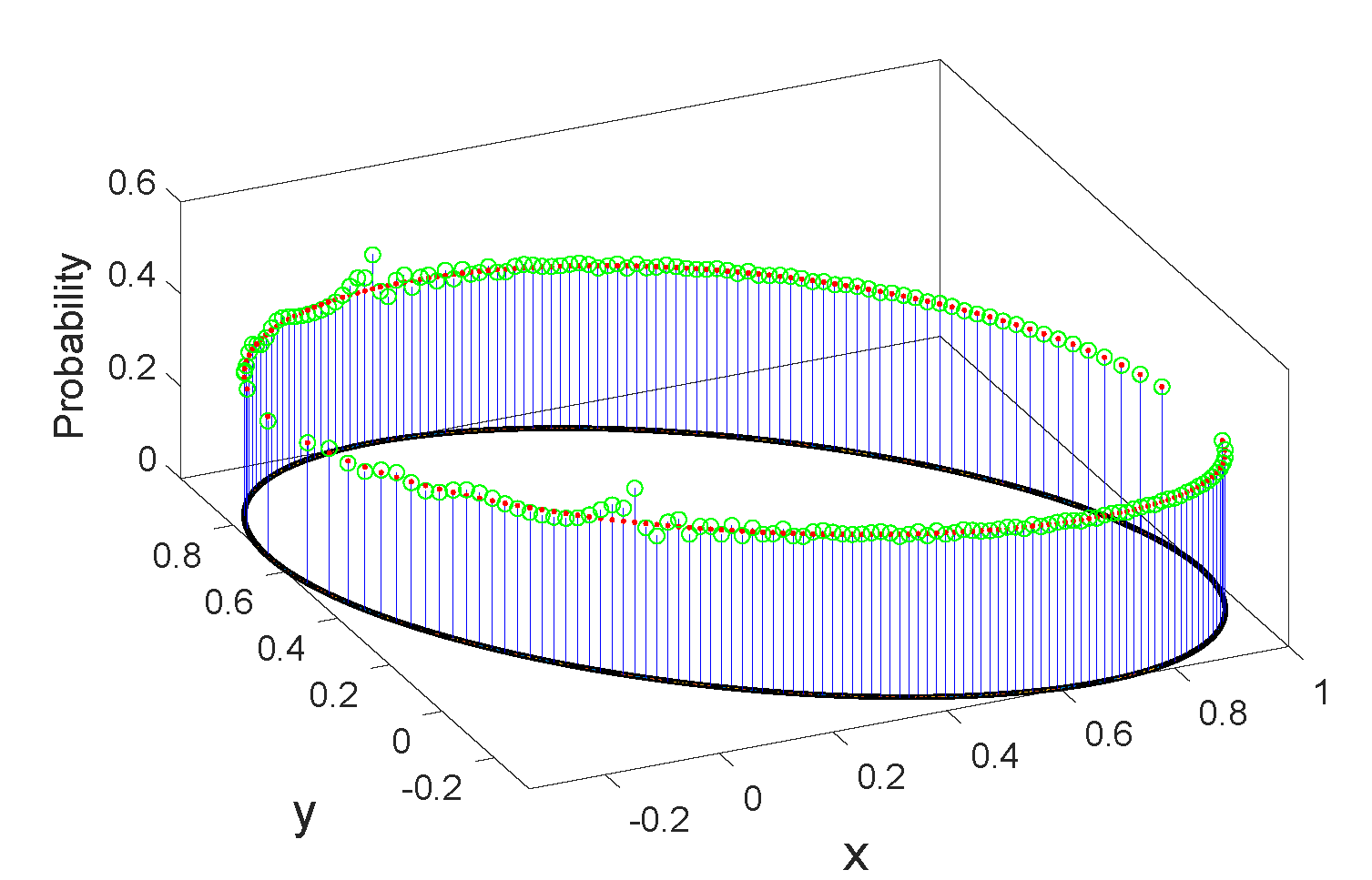}}
     \subfigure[$m=0$]{\includegraphics[height=4 cm,width=4 cm]{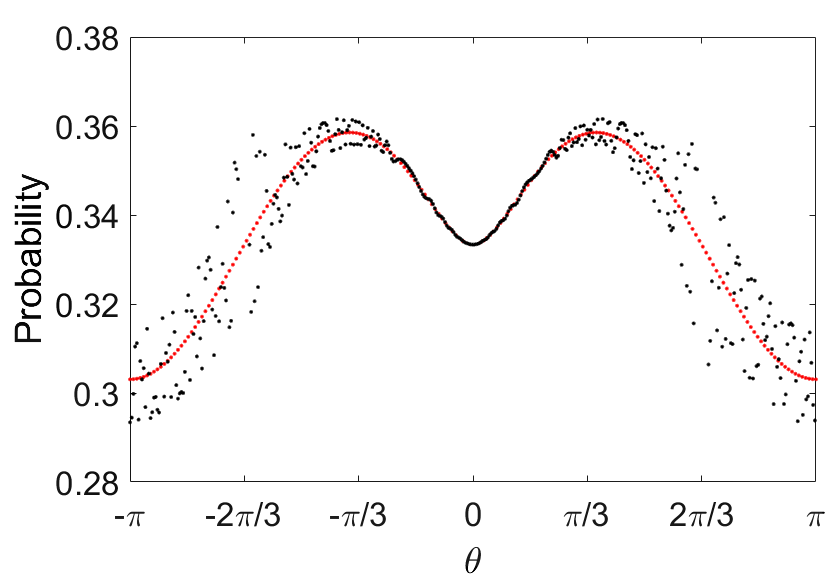}}
     \subfigure[$m=1$]{\includegraphics[height=4 cm,width=4 cm]{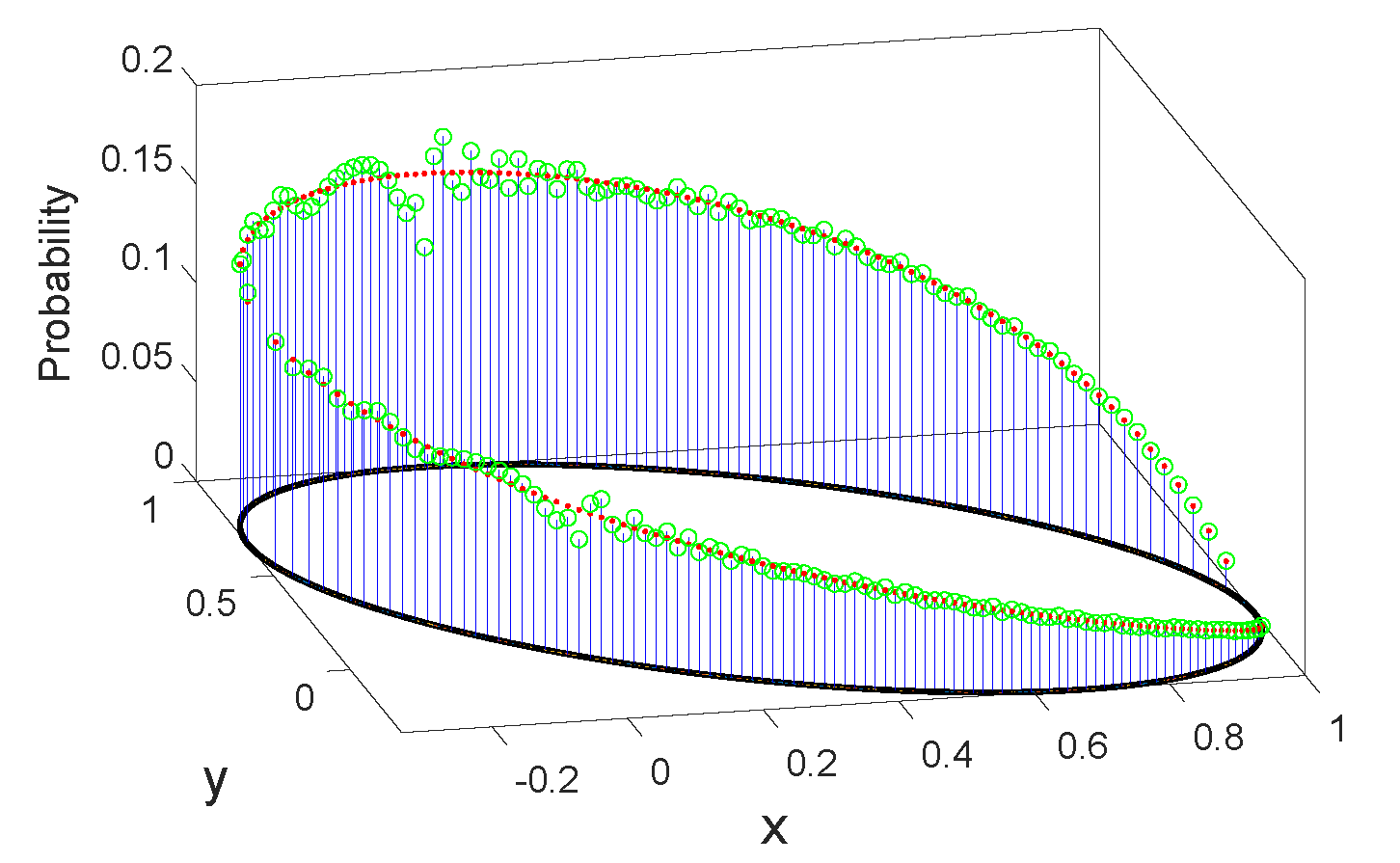}}
    \hspace{0.1 cm}
    \subfigure[$m=1$]{\includegraphics[height=4 cm,width=4 cm]{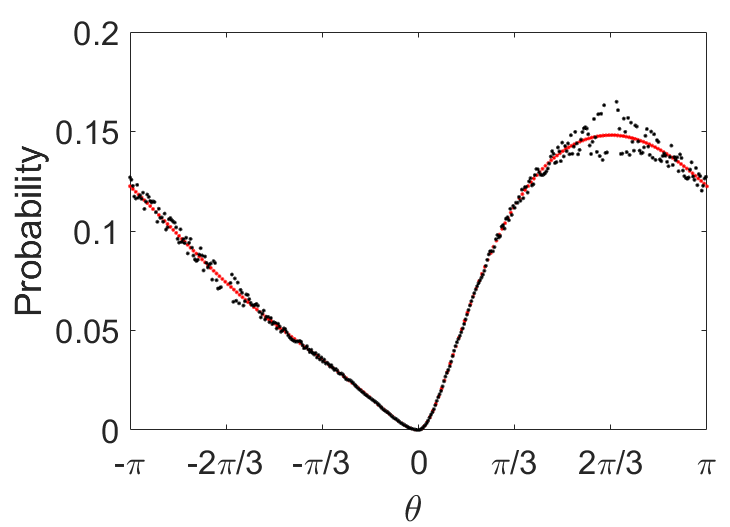}}
    \caption{The initial coin state is taken as $[\frac{1}{\sqrt{3}},\frac{1}{\sqrt{3}},\frac{1}{\sqrt{3}}]^T$ for all the walks. 
    In $(a)$ and $(c)$ we compare the values $P(m,5000)$ (green points) and $\lim_{t\to \infty}{P}(m,t)$ (red point) for the walker's positions at the origin i.e. $m=0$ and $m=1,$ respectively; with coin operators correspond to different values of $(x,y)$ changes along $x^2+y^2-x-y+xy=0,-\frac{1}{3}\leq x,y \leq 1.$ Besides, we draw the probabilities $P(m,5000)$ (black points) and  $\lim_{t\to \infty}{P}(m,t)$ (red points)  due to the change in $\theta \in [-\pi,\pi],\theta \neq0, \frac{2\pi}{3},-\frac{2\pi}{3},$ where $x=\frac{1+2\cos{\theta}}{3},$ for positions $m=0$ and $m=1,$ given in $(b)$ and $(d),$ respectively.  }\label{fig:limit prob theta x}
\end{figure}

 Let us now choose the initial coin state $\alpha=\frac{1}{\sqrt3}=\beta=\gamma.$ 
Then from Theorem \ref{thm: limit law 1d},
\beano && \lim_{t \to \infty}P(0,t)=\frac{1}{3(1-x)(x+3)} \\ && \left[|A + B \nu|^2 +\frac{1}{3}\left|2+\nu (1-3x)\right|^2 + |A\nu+B|^2\right].\eeano Since $2+\nu (1-3x)\neq 0$ for $x\neq 0,1,$ we get $\lim_{t \to \infty}P(0,t)>0.$ Correspondingly,
the walks with coins from $\X,x\neq 0,1$ show localization at $m=0.$ 
%whenever the initial coin state is $[\frac{1}{\sqrt3}, \frac{1}{\sqrt3}, \frac{1}{\sqrt3}]^T,$ 
 Figure \ref{fig:limit prob theta x} supports this fact. 
 
  In fact we establish the following corollary from Theorem \ref{thm: limit law 1d} regarding the initial states that lead zero value to the limiting probabilities.

\begin{corollary}
Let $C\in \X$ and $x\neq 0,1.$ Then $\lim_{t \to \infty}P(0,t)=0$ if and only if $\ket{\Psi(k,0)}=\frac{1}{N}[(1-x-y),-(1-x),y]^T,N=\sqrt{(1-x-y)^2+(1-x)^2+y^2}.$
\end{corollary}
\begin{proof}
The limiting probability takes zero value for the walks under consideration, whenever $A=B=0$ and $x\neq 0,1.$ Hence the corresponding initial state reads $\ket{\Psi(k,0)}:=\ket{\Psi_{\X}}=\frac{1}{N}[(1-x-y),-(1-x),y]^T,$ for which $\lim_{t \to \infty}P(0,t)$ of the walks with coins from $\X,x\neq 0,1$ vanishes.
\end{proof}

Indeed, it can be checked that $\ket{\Psi_{\X}}$ is orthogonal to the eigenvector $\ket{v_1(k)}$ given in Theorem \ref{thm:eig 1d line} corresponds to the eigenvalue $1.$

Now we redefine the set of matrices in $\X$ can be represented by a single parameter $\theta$. To this end, for $-\pi\leq \theta \leq \pi$, setting 

$$
(x,y)=\left(\frac{1+2\cos{\theta}}{3},\frac{1-\cos{\theta}}{3}+ \frac{1}{\sqrt{3}}\sin{\theta}\right)
$$ yields one-parameter representation for the matrices in $\mathcal{X}$ in terms of the parameter $\theta$ \cite{sarkar2020periodicity}. We call this matrix class as $\mathcal{X}_{\theta},$ where 
\begin{widetext}
$$ \mathcal{X}_\theta = \left\{\bmatrix{\frac{2\cos{\theta}+1}{3} & \frac{(1-\cos{\theta})}{3}+ \frac{1}{\sqrt{3}}\sin{\theta} & \frac{(1-\cos{\theta})}{3}- \frac{1}{\sqrt{3}}\sin{\theta}\\\frac{(1-\cos{\theta})}{3} - \frac{1}{\sqrt{3}}\sin{\theta} & \frac{2\cos{\theta}+1}{3} & \frac{(1-\cos{\theta})}{3} + \frac{1}{\sqrt{3}} \sin{\theta}\\\frac{(1-\cos{\theta})}{3} + \frac{1}{\sqrt{3}} \sin{\theta} & \frac{(1-\cos{\theta})}{3} - \frac{1}{\sqrt{3}} \sin{\theta} & \frac{2\cos{\theta}+1}{3}} : -\pi\leq \theta \leq \pi \right\}.$$
\end{widetext}

%appear in right side where $(x,y)$ is mapped as $(\frac{1+2\cos{\theta}}{3},\frac{1-\cos{\theta}}{3}+ \frac{1}{\sqrt{3}}\sin{\theta})$ in terms of the parameter $\theta$.

It is clear from the parametrization that if $A(\theta)\in\X_{\theta}$ then $[A(\theta)]^{-1}=A(-\theta).$ In Figure \ref{fig:limit prob theta x}, we compare the long time limit probabilities with the probability at a large but finite time, at two different positions $m=0,1$, across the range of coin parameters from $\X.$
In each case we set $\alpha=\beta=\gamma=\frac{1}{\sqrt{3}}.$  The diagrams $(a), (c)$ in Figure \ref{fig:limit prob theta x} show the change in probability values along the ellipse $x^2+y^2-x-y+xy=0,-\frac{1}{3}\leq x,y \leq 1;$ the diagrams $(b), (d)$ are plotted recording the changes in parameter $\theta$ for $[-\pi,\pi].$ We see that the probabilities take the relative minimum value around $(x,y)=(1,0)$ or consequently around $\theta=0.$ A relative maximum values for both the probability at $t=5000$ and limit measure appear for $(x,y)=(0,1)$ and $\theta= \frac{2\pi}{3}$ in $(c)$ and $(d),$ respectively. Besides, it is easy to see from Figure \ref{fig:limit prob theta x} that $\lim_{t\rightarrow \infty} P(m,t)$ at the initial position $m=0$ is equal for the coin operators $A(\theta)$ and its inverse $[A(-\theta)].$ 

\subsection{Coins from $\Y$}

In what follows, we provide a limit theorem for the proposed walks with coin operators from $\Y.$ We perform a similar analysis as above. For  $C\in \mathcal{Y},k\in (-\pi,\pi],$ $\tilde U_{\mathcal Y}(k)=D(k)C$ is given by 
\begin{equation}\label{eq:evolu 1d line y}
\tilde{U}_{\mathcal Y}(k)= \bmatrix{xe^{ik} & ye^{ik} & (-1-x-y)e^{ik}\\
-1-x-y & x & y\\
ye^{-ik} & (-1-x-y) e^{-ik} & x e^{-ik}},
\end{equation}
where $x^2+y^2+xy+x+y=0$.

The following theorem provides computable expressions for eigenvalues and eigenvectors of $\tilde{U}_{\mathcal Y}(k)$.

\begin{theorem} \label{thm:eig 1d line y}
Consider $\tilde{U}_{\mathcal Y}(k)$ in Eq.~\eqref{eq:evolu 1d line y}. Then the set of eigenpairs $(\lambda_j(k),v_j(k))$ of $\tilde{U}_{\mathcal Y}(k)$ are  \beano \lambda_j &=& e^{i\omega_j(k)},\\
 \ket{v_j(k)} &=& \frac{1}{\|f_j(\omega(k))\|}\left[\frac{-y+\lambda_j(1+x+y)}{(1+x+y)-\lambda_j ye^{-i k}}, 1,\right. \\ && \left. \frac{(1+x+y)-\lambda_j y}{-y+\lambda_j(1+x+y)e^{i k}}\right]^T,\eeano
 where $\omega_1(k)=\pi,\omega_2(k)=-\omega_3({ k})=\omega({k})$ with $\cos{\omega(k)}=-x\cos{k}-\frac{(1+x)}{2}$ and 
 \beano 
 \|f_j(\omega(k))\|^2 &=& \frac{{1-x}-{2x}\cos{\omega(k)}}{{1-x}-{2x}\cos{(\omega(k)-k)}}+ 1 +\\ && \frac{{1-x}-{2x}\cos{\omega(k)}}{{1-x}-{2x}\cos{(\omega(k)+k)}},\eeano
for $\{j=1,2,3\}$.
\end{theorem}
\begin{proof} Note that the characteristic polynomial of $\tilde U_{\mathcal Y}(k)$ is $\kappa_{\mathcal Y}(\lambda)=\lambda^3-(2 \cos k+1) x \lambda^2-(2 \cos k+1) x \lambda+1.$ Hence the desired results follow after simple algebraic calculations.
\end{proof}

The next theorem concerns the asymptotic value of the probability measure for finding the particle to be at any vertex $m$, as $t\to \infty$, for a DTQW using coin operators $C\in \mathcal{Y}$.

\begin{theorem} \label{thm:limit law 1d y}
  Let $C\in \Y$ be not a permutation matrix i.e. $x \neq 0,-1$ and $\Psi(k,0)=[\alpha,\beta,\gamma]^T$ be the initial state of the walker, then for $m\in \mathbb{Z}$ we get 
 \begin{widetext}
  \beano
\lim_{t\to \infty}{P}(m,t) &\sim& \frac{1}{3(1+x)(3-x)}\{|D\mu^{|m|} + E \mu^{|-m+1|}|^2 +
\left|\alpha(1+x+y)\mu^{|m|}-\alpha y \mu^{|m+1|}+\beta(1-x)\mu^{|m|}+\beta x(\mu^{|-m+1|} \right.\\ && \left. +\mu^{|m+1|})+ \gamma(1+x+y)\mu^{|-m+1|}-\gamma y\mu^{|m|}\right|^2 + |D\mu^{|m+1|}+E\mu^{|m|}|^2\}\eeano
\end{widetext}
where
\begin{eqnarray*}
\mu&=& \frac{-(x+3)+\sqrt{3(1+x)(3-x)}}{2x}\\
D&=&{\alpha(1+x)+\beta(1+x+y)}\\
E&=&\gamma(1+x)-\beta y.
\end{eqnarray*}
If $C\in \Y$ are permutation matrices and  $\Psi(k,0)=[\alpha,\beta,\gamma]^T$ is the initial state of the walker, then for $m\in \mathbb{Z}$ we have the following.
 \beano  \lim_{t\to \infty}{P}(m,t)\sim  
  \left\{
  \begin{array}{l}
   |\beta|^2 \,\, \mbox{if} \,\, m=0, C=I \\ \\
   \frac{1}{3}(|\alpha|^2 +2|\beta+\gamma|^2)  \, \mbox{if} \,\, m=0, C=-P_{(213)} \\ \\
   \frac{1}{3}|\beta+\gamma|^2  \,\, \mbox{if} \,\, m=-1, C=P_{(123)} \\ \\
    \frac{2}{3}|\alpha|^2 \,\, \mbox{if} \,\, m=1, C=P_{(123)} \\ \\
     \frac{1}{3}(|\gamma|^2 +2|\alpha+\beta|^2) \,\, \mbox{if} \,\, m=0, C=P_{(132)} \\ \\
      \frac{2}{3}|\gamma|^2 \,\, \mbox{if} \,\, m=-1, C=P_{(132)} \\ \\
      \frac{1}{3}|\alpha+\beta|^2 \,\, \mbox{if} \,\, m=1, C=P_{(132)} \\ \\
      0, \,\, \mbox{otherwise.}
  \end{array}
\right. \eeano
\end{theorem}
\begin{proof}
The proof follows adapting a similar approach as in Theorem \ref{thm: limit law 1d}.
\end{proof}

%\textcolor{red}{Maybe mention that the proof is similar to that of Theorem 3.4?}

\begin{figure}[H]
    \centering
    \subfigure[ $x=\frac{1}{3}$ ]{\includegraphics[height=4 cm,width=4 cm]{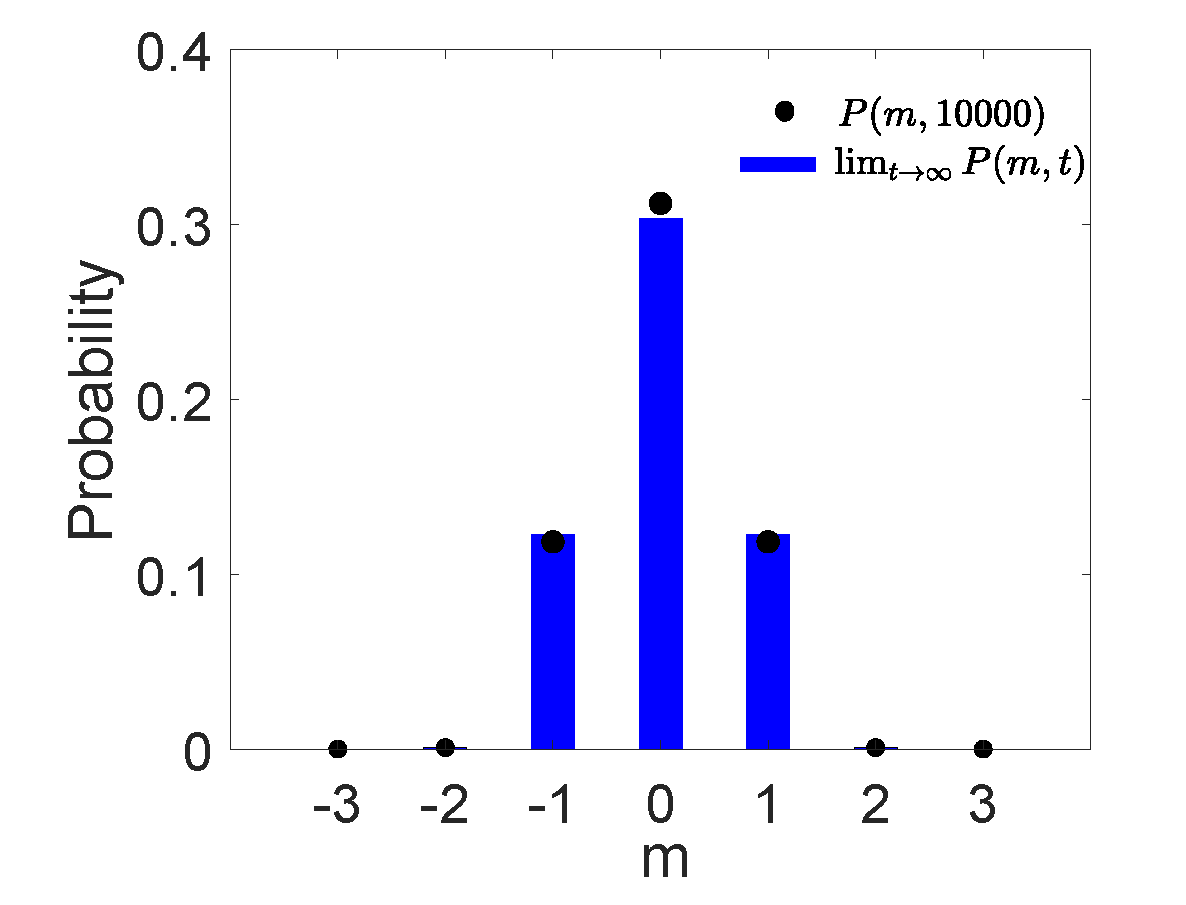}} 
\hspace{0.2cm}
\subfigure[ $x=\frac{\sqrt3-1}{3}$]{\includegraphics[height=4 cm,width=4 cm]{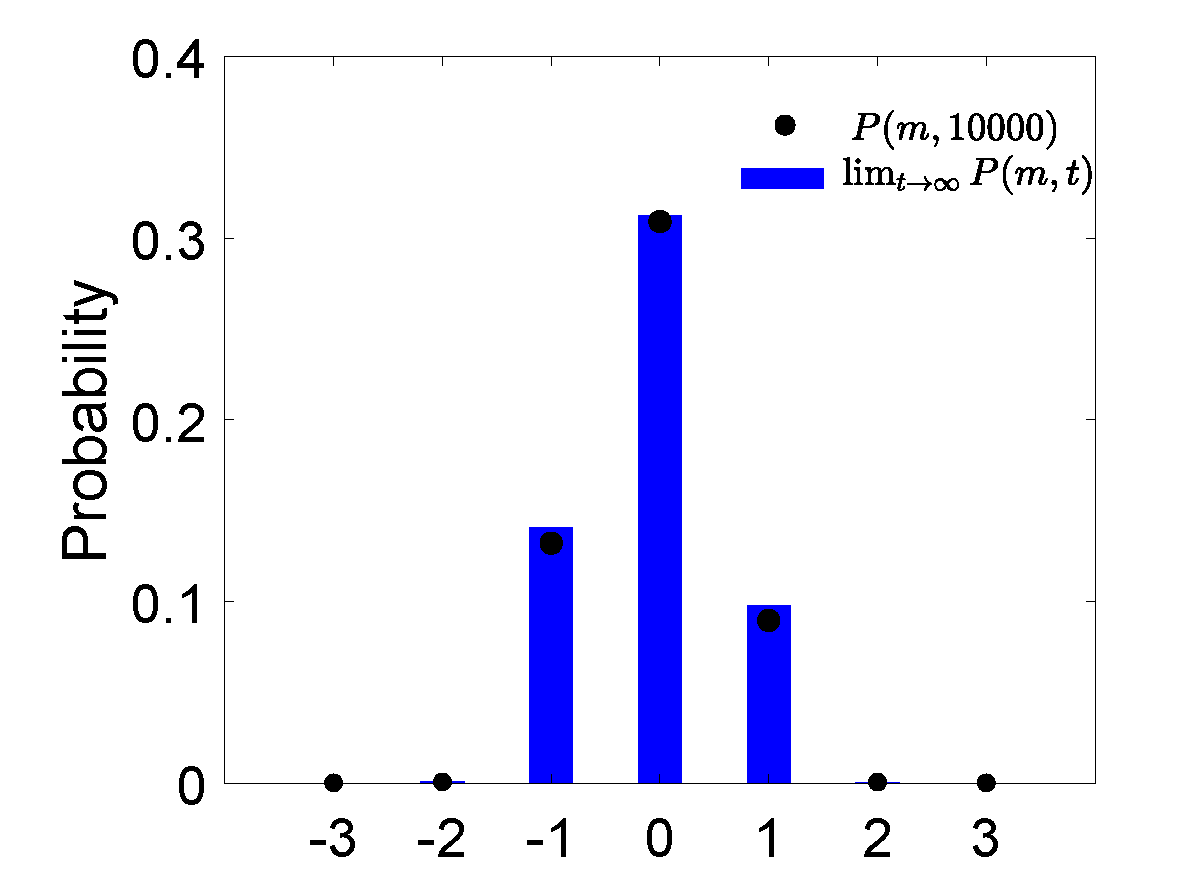}} 
\caption{Figure $(a)$ and Figure $(b)$ compare $P(m,t)$ for $t = 10000$ (black points) and $\lim_{t\to \infty}{P}(m,t),$ (blue bars) when the underlying coin operators are from $\Y$ correspond to $x=\frac{1}{3}$ and $x=\frac{\sqrt3-1}{3}$,
%$\Y_{0}$ and $\Y_{\frac{\pi}{6}},$ 
respectively; with initial coin state $[\frac{1}{\sqrt{3}},\frac{1}{\sqrt{3}},\frac{1}{\sqrt{3}}]^T.$ We get $P(0, 10000)=0.3120,\lim_{t\to \infty}{P}(0,t)=0.3031$ for $(a)$ and $P(0, 10000)=0.3092,\lim_{t\to \infty}{P}(0,t)=0.3123$ for $(b).$  
%In both the cases we see that the values of the probabilities for the walker's position on the line except the positions at $m=0,1,-1$ are very small, whereas at $m=0$ it is maximum. 
}\label{fig:limit prob y}
\end{figure}

In Figure \ref{fig:limit prob y}, we compare the numerical values of $P(m,t),t=10000$ and the limiting value of the probability measure fixing the same initial state, when the initial position of the walker is origin, that is, $m=0$ for two different coins from $\mathcal{Y}$. The figures support the localization phenomena.

The walks with coin operators from $\Y$ localize according to \cite{machida2014limit} if 
\beano && \left[\frac{1+\mu^2}{1-\mu^2}\left(2|D|^2+2|E|^2+|S_1|^2+|S_2|^2+|S_3|^2\right)+ \right.\\ && \left. \frac{4 \mu}{1-\mu^2}\left(\mbox{Re}(D\overline{E})+\mbox{Re}(S_1\overline{S_2})+\mbox{Re}(S_1\overline{S_3})\right) \right.\\ && \left. +2\mu^2\frac{3-\mu^2}{1-\mu^2}\mbox{Re}\left(S_2\overline S_3\right)\right]>0,\eeano where $S_1=\alpha(1+x+y)+\beta(1-x)-\gamma y,
S_2=-\alpha y +\beta x, S_3=\beta x+\gamma(1+x+y)$ and $-5+2\sqrt{6}\leq \mu< 1,\mu^2<1.$ See Appendix \ref{appdx:B} for the derivation.

\begin{figure}[H]
    \centering
    \subfigure[$m=0.$]{\includegraphics[height=4 cm,width=4 cm]{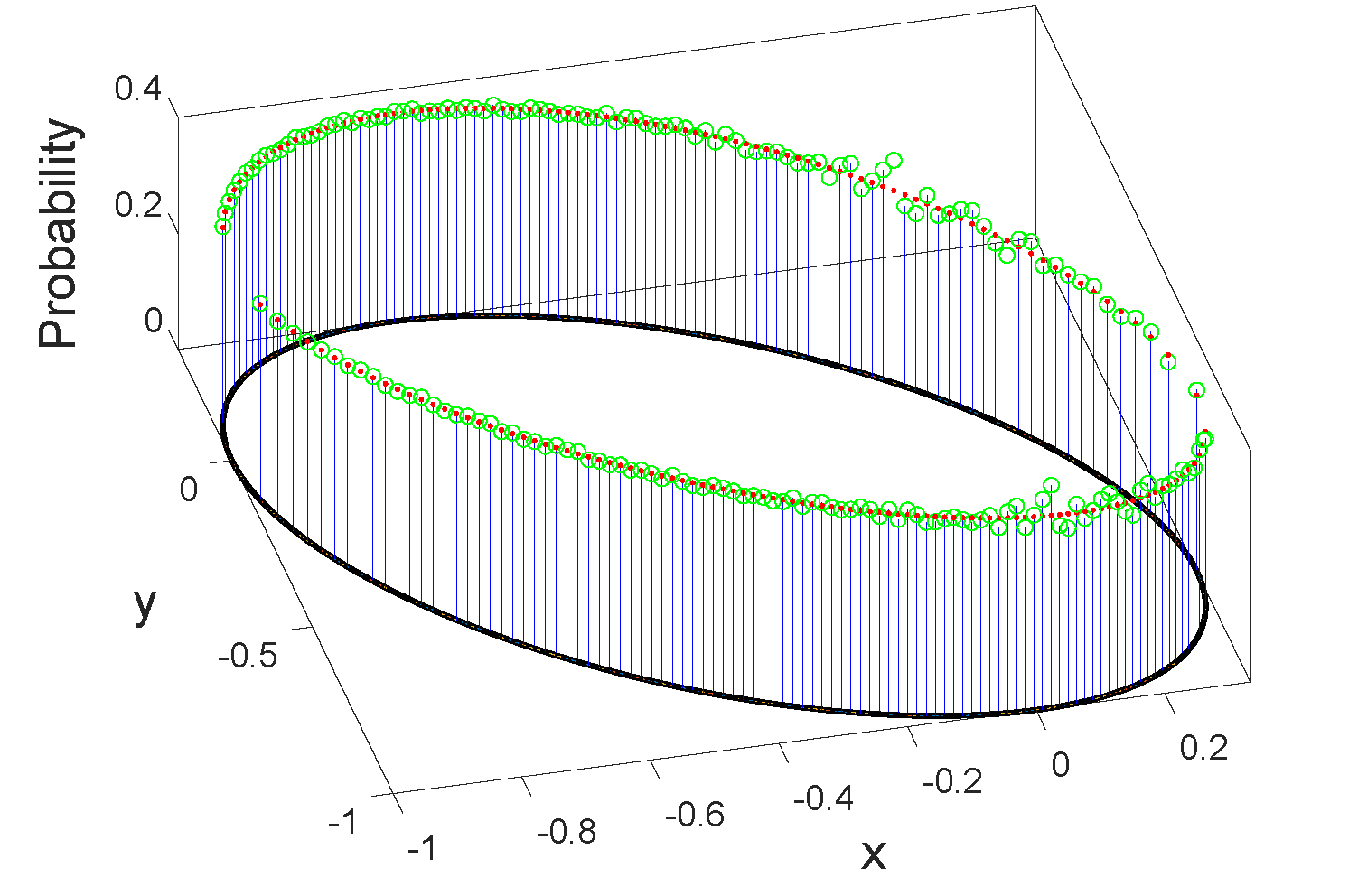}}
     \hspace{0.1 cm}
    \subfigure[$m=0.$]{\includegraphics[height=4 cm,width=4 cm]{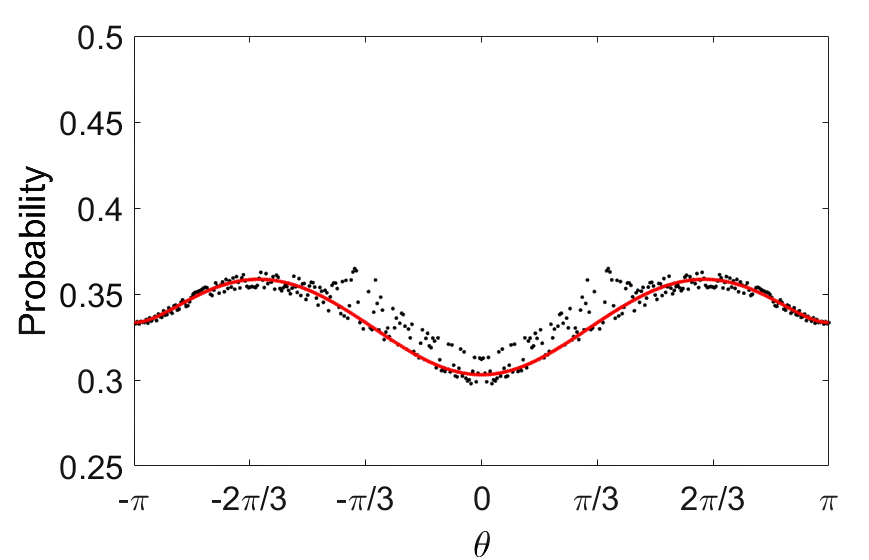}}\\
    \subfigure[$m=1.$]{\includegraphics[height=4 cm,width=4 cm]{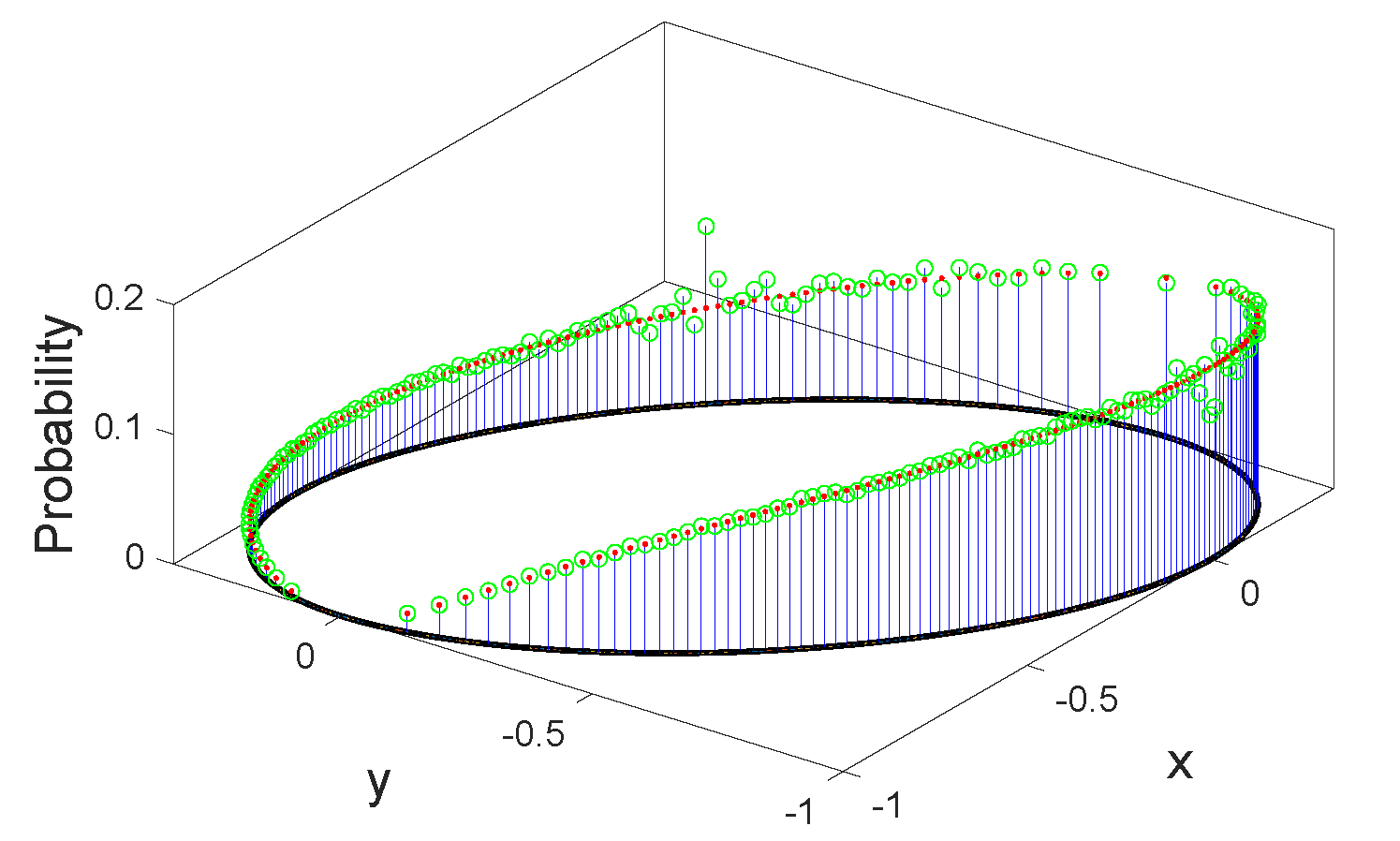}}
     \hspace{0.1 cm}
    \subfigure[$m=1.$]{\includegraphics[height=4 cm,width=4 cm]{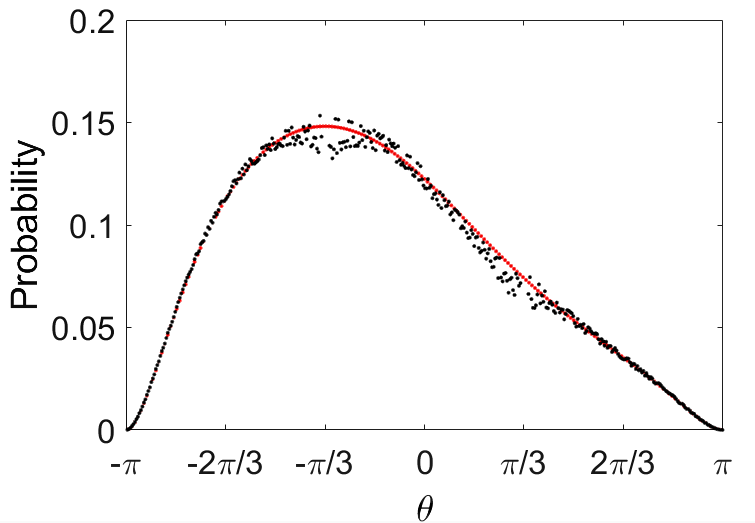}}
     \caption{Here the coin operators under consideration are chosen from $\Y,x\neq -1,0$ and the initial coin state is $[\frac{1}{\sqrt{3}},\frac{1}{\sqrt{3}},\frac{1}{\sqrt{3}}]^T$ in all the cases. 
    In $(a)$ and $(c)$ we compare the values for probability at $t=10000$ (green points) and limit probability measure (red point) for the walker's positions at $m=0$ and $m=1,$ respectively, with different coin operators. 
    %correspond to values of $(x,y)$ changes along $x^2+y^2+x+y+xy=0,-1\leq x,y \leq \frac{1}{3}.$ 
    Also we draw the probabilities $P(m,10000)$ (black points) and  $\lim_{t\to \infty}{P}(m,t)$ (red points) for different values of $\theta \in [-\pi,\pi],\theta \neq \pi,-\pi, \frac{\pi}{3},-\frac{\pi}{3},$ %where $x=\frac{2\cos{\theta}-1}{3},$ for positions $m=0$ and $m=1,$ given in $(b)$ and $(d),$ respectively. 
       } \label{fig:limit prob theta y}
\end{figure}

Also, from Theorem \ref{thm:eig 1d line y} we write,
\beano && \lim_{t \to \infty}P(0,t)=\frac{1}{3(1+x)(3-x)} \\ && \left[|D + E \mu|^2 +\frac{1}{3}\left|2+\mu (1+3x)\right|^2 + |D\mu+E|^2\right].\eeano Since $2+\mu (1+3x)\neq 0$ for $x\neq 0,-1,$ we get $\lim_{t \to \infty}P(0,t)>0$ and we say the walks with coins from $\Y,x\neq 0,-1$ show localization at $m=0.$  In account to this fact we have   Figure \ref{fig:limit prob theta y}. 
 In fact we establish the following corollary from Theorem \ref{thm:limit law 1d y} regarding the initial states that lead zero value to the limiting probabilities.

\begin{corollary}
Let $C\in \Y$ and $x\neq 0,-1.$ Then $\lim_{t \to \infty}P(0,t)=0$ if and only if $\ket{\Psi(k,0)}=\frac{1}{N}[-(1+x+y),(1+x),y]^T,N=\sqrt{(1+x+y)^2+(1+x)^2+y^2}.$
\end{corollary}
\begin{proof}
We get $\lim_{t \to \infty}P(0,t)=0$ whenever $D=E=0$ and $x\neq 0,-1.$ Therefore the corresponding initial state is $\ket{\Psi(k,0)}:=\ket{\Psi_{\Y}}=\frac{1}{N}[-(1+x+y),(1+x),y]^T,$ for which the limiting probability at $m=0$ of the walks with coins from $\Y,x\neq 0,-1$ vanishes. \end{proof}
Indeed, it can be checked that $\ket{\Psi_{\Y}}$ is orthogonal to the eigenvector $\ket{v_1(k)}$ given in Theorem \ref{thm:eig 1d line y} corresponds to the eigenvalue $-1.$

Now we redefine the set of matrices in $\Y$ can be represented by a single parameter $\theta$. To this end, for $-\pi\leq \theta \leq \pi$, for
the set of matrices in $\Y$, setting
$$
(x,y)=\left(\frac{2\cos{\theta}-1}{3},-\frac{1+\cos{\theta}}{3}+ \frac{1}{\sqrt{3}}\sin{\theta}\right)
$$
yields one-parameter representation for the matrices in $\mathcal{X}$ and $\Y$ in terms of the parameter $\theta$ \cite{sarkar2020periodicity}. We call this matrix class as  $\mathcal{Y}_{\theta},$ where 
\begin{widetext}
$$\mathcal{Y}_\theta = \left\{\bmatrix{
\frac{2\cos{\theta}-1}{3} & -\frac{(1+\cos{\theta})}{3}+ \frac{1}{\sqrt{3}}\sin{\theta} & -\frac{(1+\cos{\theta})}{3}- \frac{1}{\sqrt{3}}\sin{\theta}\\
-\frac{(1+\cos{\theta})}{3} - \frac{1}{\sqrt{3}}\sin{\theta} & \frac{2\cos{\theta}-1}{3} & -\frac{(1+\cos{\theta})}{3} + \frac{1}{\sqrt{3}} \sin{\theta}\\
-\frac{(1+\cos{\theta})}{3} + \frac{1}{\sqrt{3}} \sin{\theta} & -\frac{(1+\cos{\theta})}{3} - \frac{1}{\sqrt{3}} \sin{\theta} & \frac{2\cos{\theta}-1}{3}
} : -\pi\leq \theta \leq \pi \right\}.$$
\end{widetext}

%We see a some discrepancy between the probability for $t=10000$ and the limit measure for $m=0$ as compared to $m=1.$ \textcolor{red}{Not immediately clear to me why this discrepancy should be there. Does this happen for any large value of $t$?}

Note that if $\theta \in \{ \pi,-\pi, \frac{\pi}{3},-\frac{\pi}{3} \}$, we obtain the negative of the permutation matrices in $\mathcal{Y}_{\theta}.$ Also, if $A(\theta)\in\Y_{\theta}$ then $[A(\theta)]^{-1}=A(-\theta).$ In Figure \ref{fig:limit prob theta y}, we see the limiting values of the probability measure for the walks where the coin parameter $(x,y)$ changes along the ellipse $x^2+y^2+x+y+xy=0,-1\leq x,y\leq \frac{1}{3};$ or the coin parameter $\theta \in [-\pi,\pi].$  In $(b)$ and $(d)$ the probabilities take the relative minimum and maximum values are around $\theta=0$ and $\theta=-\frac{\pi}{3}$ respectively. 
%In contrast to that relative maximum for the probability at $t=10000$ and limit measure appear for $(x,y)$ equal to $(0,1)$ in $(c)$ or consequently for $\theta=\frac{2\pi}{3}$ in $(d).$ In contrast to that for the walker position at $m=0,$ there is inconsistency between two probabilities as shown in figure $(a)$ and $(b),$ where the relative maximum  appear at different positions.
In $(c)$ and $(d)$ the probability values are getting  near to zero in the neighbours of $(x,y)=(-1,0)$ and $\theta=\pm \pi$ respectively. Finally, observe from Figure \ref{fig:limit prob theta y} that $\lim_{t\rightarrow \infty} P(m,t)$ at the initial position $m=0$ is equal for the coin operators $A(\theta)$ and its inverse $[A(-\theta)].$

\section{Peak velocities of the walks} \label{sec:peak velocity}
In section \ref{sec:limit law} we established that the discrete-time quantum walks with coin operators from the family $\X_{\theta}$  and $\Y_{\theta}$ localize. This was done by demonstrating that there exists a non-vanishing probability for the quantum walk to remain at a vertex even as the time steps of the walk approach infinity. In this section, we discuss the localization phenomena of the walks from a  different perspective. 

In Ref.~\cite{vstefavnak2012continuous}, the authors establish a relationship between the choice of the coin parameters and the peak velocity with which the underlying quantum walk wavefunction spreads. The authors demonstrate this by considering a generalized Grover operator as coin, obtained by modifying its eigenvalues and eigenvectors. Peak velocities control the limit distribution and in turn influences localization.

 Here at first, we determine the peak velocities of the probability distribution of the walker's position when the coin operators are from $\X_{\theta}$ and $\Y_{\theta}$. This allows us to provide a new family of parametrized coin operators $\X_{\theta}$, outside those in Ref.~\cite{vstefavnak2012continuous}, which preserves the localization of the Grover walk. We describe how the coin parameter $\theta$ relates the spreading of the walks through the line.
 
 Recall from Eq.~\eqref{wave function:prob ampli}, we have that
\begin{eqnarray}\label{eq:prob amp vec}
\ket{\psi(m,t)} = \sum_{j=1}^{3}\frac{1}{2\pi}\int_{-\pi}^{\pi} && e^{i (- k m/t+ w_j( k))t}\braket{v_j(k)|\Psi(k,0)} \nonumber\\ && \ket{v_j( k)} dk.
\end{eqnarray}

Now to determine the behavior of $\ket{\psi(m,t)}$ as $t\to \infty$, we employ the idea given in Ref.~\cite{vstefavnak2012continuous} by using the method of stationary phase approximation (SPA) \cite{wong2001asymptotic}.

In general, the SPA states that for a propagating wavepacket, the contributions of the fast oscillating phases average out and only the amplitudes corresponding to the stationary phases contribute significantly to the overall wavefunction of the wavepacket. The propagation of the wavefunction corresponding to a quantum walk on an infinite line with periodic boundary conditions can be modelled using the theory of SPA. 

Consider the phase $\bar{\omega_j}(k)=\omega_j(k)-mk/t$ from Eq.~\eqref{eq:prob amp vec}. Now, the rate of the decay is given by the order of the stationary points of the phase $\bar{\omega_j}(k).$ By solving the equation $\frac{d^2}{dk^2}{\bar{\omega}_{j}(k)}=0$, we find the stationary points corresponding to the maximum value of $\frac{d}{dk}{\bar{\omega}_{j}(k)}.$ If $k_0$ is the solution of the second order derivative then by equating the first order equation $\frac{d}{dk}{\bar{\omega}_{j}(k)}=\frac{d}{dk}{\omega}_{j}(k)-m/t$ with zero we get the position of the peak after time $t$ as $m=\left.\frac{d}{dk}{\omega_{j}(k)}\right\vert_{k_0}t,$  whereas the peaks in the probability distribution curve propagates with the maximal group velocity $\left.\frac{d}{dk}{\omega_{j}(k)}\right\vert_{k_0}$. Here we recall that the group velocity of the propagating wavefunction corresponding to the quantum walk is simply $\frac{d}{dk}{\omega_{j}(k)}$, where $k$ and $\omega_j(k)$ are the wave number and angular frequency respectively \cite{kempf2009group}.

%\tm{The group velocity of the propagating wavefunction corresponding to the quantum walk is simply $\frac{d}{dk}{\omega_{j}(k)}$, where $k$ and $\omega_j(k)$ is the wave number and frequency respectively \cite{kempf2009group}.}

%\tm{By solving the equation $\frac{d^2}{dk^2}{\bar{\omega}_{j}(k)}=0$, we find the stationary points corresponding to the maximum value of $\frac{d}{dk}{\bar{\omega}_{j}(k)}.$ If $k_0$ is the solution of the second order derivative then by equating the first order equation with zero we get the position of the peak after time $t$ as $m=\left.\frac{d}{dk}{\omega_{j}(k)}\right\vert_{k_0}t,$  whereas the peaks in the probability distribution curve propagates with maximal group velocity $\left.\frac{d}{dk}{\omega_{j}(k)}\right\vert_{k_0}$.}

Next we obtain peak velocities for the wavefunction corresponding to a discrete-time quantum walk on a line with generalized coin operators chosen from the set $\X_\theta$ and $\Y_\theta$.

\subsection{With coin operators $C\in \X_{\theta}$ }

First we consider the walk with coins from $\X.$ Now for $x=\frac{1+ 2\cos\theta}{3},$ by Theorem \ref{thm:eig 1d line} the eigenvalues of $\tilde{U}_{\mathcal X}(k)$ are $\lambda_j=e^{i\omega_j(k)},j=1,2,3$ where
 $\omega_1(k)=0,\omega_2(k)=-\omega_3({ k})=\omega({k})$ with %$\cos{\omega(k)}=x\cos{k}-\frac{(1-x)}{2}.$  
 $\cos{\omega(k)}=-\frac{1}{3}(1-\cos{k})+\frac{\cos{\theta}}{3}(1+2 \cos{k}).$ 

   Since $\omega_{1}(k)$ is independent of the wave-number $k$, the localization effect of the quantum walks using $\mathcal{X}_{\theta}$ as the coin operator is preserved as that of the Grover walk \cite{inui2004localization}, \cite{inui2005one}. Indeed, the existence of constant eigenvalues of the time evolution operator and the localization are equivalent for DTQWs on infinite lattice under cyclic boundary conditions \cite{segawa2016generator}.

We shall now determine the peak velocities using the coin operators from $\X_{\theta}$ for $\theta=0,\frac{2 \pi}{3}, -\frac{2 \pi}{3},$ which are permutation matrices. For ${\theta=0},$ $\X_{\theta}$ becomes the identity matrix. Therefore, $\omega_{2,3}(k)=\pm k,$ consequently the right and left going peak velocities are 
$$
v_R=\frac{d}{dk}{{\omega}_{3}(k)}=-1,v_L=\frac{d}{dk}{{\omega}_{2}(k)}=1,
$$ 
whereas $\frac{d^2}{dk^2}{{\omega}_{2,3}(k)}=0$.  
~\\~\\
Similarly, if ${\theta=\frac{2 \pi}{3}}$ or ${\theta=-\frac{2 \pi}{3}}$ we get permutation matrices $P_{(123)}\in\X_{\theta}$ or $P_{(132)}\in\X_{\theta}$ respectively. For both the cases the eigenvalues $\omega_{2,3}(k)=\pm \frac{2 \pi}{3}$ and hence the corresponding peak velocities are $0.$
    If $\theta$ equals to $-\pi$ or $\pi,$ accordingly the coin operator is the Grover matrix, the right and left peak velocities are $\frac{1}{\sqrt3}$ and $-\frac{1}{\sqrt3}$ respectively \cite{vstefavnak2012continuous}.

 From now onward we consider coins from the family $\X_{\theta}$ except the Grover and permutation matrices i.e. $(1+2 \cos{\theta})\neq 0,3,-1.$ 
Now, to find the peak velocities we determine for which values of $k$ the second derivatives of $\omega_{2,3}(k)$ vanish.
Thus,
\beano \frac{d^2\omega_{2}(k)}{dk^2} &=&\frac{\cos{k} (1+2 \cos{\theta})}{{\left(9-{\left(-1+\cos{k}+\cos{\theta}(1+2\cos{k})\right)}^2\right)}^{\frac{1}{2}}}- \\ && \frac{{\sin^2{k}}(1+2\cos{k})^2(-1+\cos{k}+\cos{\theta}(1+2 \cos{k}))}{{\left(9-{\left(-1+\cos{k}+\cos{\theta}(1+2\cos{k})\right)}^2\right)}^{\frac{3}{2}}}\eeano
vanishes for
$$ k_0^1=\pi-\arccos\left(\frac{5\cos{\theta}+7-3\sqrt{\cos^2{\theta}+6\cos{\theta}+5}}{2(2\cos{\theta}+1)}\right)$$ and $$k_0^2=\pi-\arccos\left(\frac{5\cos{\theta}+7+3\sqrt{\cos^2{\theta}+6\cos{\theta}+5}}{2(2\cos{\theta}+1)}\right).$$

Then we determine the group velocities given by  \beano \frac{d\omega_{2}(k)}{dk} &=& -\frac{d\omega_{3}(k)}{dk} \\ &=& \frac{\frac{1}{3} \sin{k}(1+2 \cos{\theta})}{\sqrt{1-\left(-\frac{1}{3}(1- \cos{k})+\frac{1}{3} \cos{\theta} (1+2 \cos{k})\right)^2}}\eeano at the stationary points $k_1^0$ and $k_2^0.$
Hence the velocities of the right and left going probability peaks $v_{R}$ and $v_{L}$ are given by

\beano 
v_{L}^{(1)}(\theta) &=&\left.\frac{d\omega_{2}(k)}{dk}\right\vert_{k=k_1^0}\\
&=&\left\{
  \begin{array}{l}
    \dfrac{1}{3}\sqrt{\dfrac{9\sqrt{\cos^2{\theta}+6\cos{\theta}+5}-(15\cos{\theta}+21)}{ \sqrt{\cos^2{\theta}+6\cos{\theta}+5}-(\cos{\theta}+3)}},\\  \hfill{\mbox{if}\,\,-\dfrac{2\pi}{3} < \theta < -\dfrac{2\pi}{3};} \\
    -\dfrac{1}{3}\sqrt{\dfrac{9\sqrt{\cos^2{\theta}+6\cos{\theta}+5}-(15\cos{\theta}+21)}{ \sqrt{\cos^2{\theta}+6\cos{\theta}+5}-(\cos{\theta}+3)}}\\ 
    \hfill{\mbox{if}\,\,-\pi<\theta<-\dfrac{2\pi}{3},\,\,\dfrac{2\pi}{3}<\theta<\pi}, \\
\end{array}
\right.\\
v_{R}^{(1)}(\theta) &=& \left.\frac{d\omega_{3}(k)}{dk}\right\vert_{k=k_1^0} = -v_{L}^{(1)}(\theta), \\
v_{L}^{(2)}(\theta) &=& \left.\frac{d\omega_{2}(k)}{dk}\right\vert_{k=k_2^0} \\
&=& \left\{
  \begin{array}{l}
  \dfrac{1}{3}\sqrt{\dfrac{9\sqrt{\cos^2{\theta}+6\cos{\theta}+5}+(15\cos{\theta}+21)}{ \sqrt{\cos^2{\theta}+6\cos{\theta}+5}+(\cos{\theta}+3)}}\\\hfill{\mbox{if}\,\,-\dfrac{2\pi}{3}<\theta<-\dfrac{2\pi}{3}};\\
   -\dfrac{1}{3}\sqrt{\dfrac{9\sqrt{\cos^2{\theta}+6\cos{\theta}+5}+(15\cos{\theta}+21)}{ \sqrt{\cos^2{\theta}+6\cos{\theta}+5}+(\cos{\theta}+3)}}\\\hfill{\mbox{if}\,\,-\pi<\theta<-\frac{2\pi}{3},\,\,\dfrac{2\pi}{3}<\theta<\pi},\\
    \end{array}
\right. \\
v_{R}^{(2)}(\theta) &=& \left.\frac{d\omega_{3}(k)}{dk}\right\vert_{k=k_2^0} =-v_{L}^{(2)}(\theta),
\eeano
where $(1+2 \cos{\theta})\neq 0,3,-1.$ It is worthy to mention that $v_{S}^{(1)}(\theta)=v_{S}^{(2)}(\theta)=0.$

Here we mention that $v_{R}(\theta)$ and $v_{L}(\theta)$ do not keep same sign for all the coin parametric values $\theta$ through the interval $[\pi,\pi],$ in contrast to the study in \cite{vstefavnak2012continuous}.
Henceforth, in order to get rid of the conflict with the names `velocity of the right going peaks'
and `velocity of the the left going peaks' which are generally used in case of positive and negative quantities; we emphasize that we represent $v_{R}(\theta)$ and $v_{L}(\theta),$ respectively, by the expressions $\left.\frac{d\omega_{2}(k)}{dk}\right\vert_{k=k^0}$ and $\left.\frac{d\omega_{3}(k)}{dk}\right\vert_{k=k^0}$ only, not in the sense of sign. 
    
\begin{figure}[H]
    \centering
    {\includegraphics[height=4 cm,width=5.5 cm]{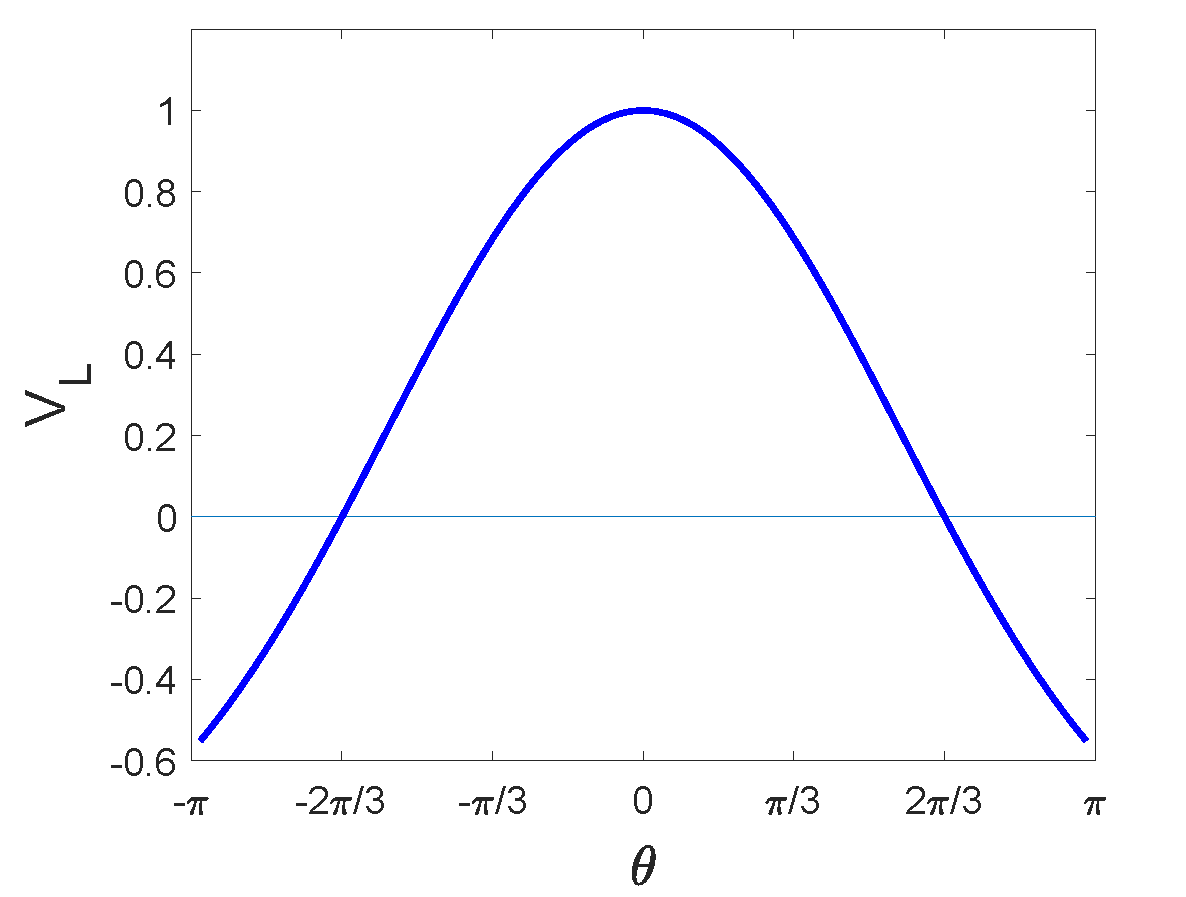}}
    \caption{The velocity $v_L^{(1)}{(\theta)}$ for the one-parameter family of quantum walks defined by the coin class $\mathcal{X}_{\theta}.$ The velocity of the left travelling probability peak takes the maximal value $1$ at $\theta=0$ and when $|\theta|$ increases it decreases and becomes $0$ at $\theta=\pm2 \pi/3.$ $v_L^{(1)}{(\theta)}$ takes negative value for $|\theta|>2\pi/3.$ The curve for peak velocity is symmetric about the vertical line through $\theta=0.$ For the Grover walk,  $v_L^{(1)}(\theta)$ attains the value $-\frac{1}{\sqrt3}$ i.e. $v_L^{(1)}(\pi)=v_L^{(1)}(-\pi)=-\frac{1}{\sqrt3}$. } \label{continuous deformation: root1}
\end{figure}

\begin{figure}[H]
    \centering
    {\includegraphics[height=4 cm,width=5.5 cm]{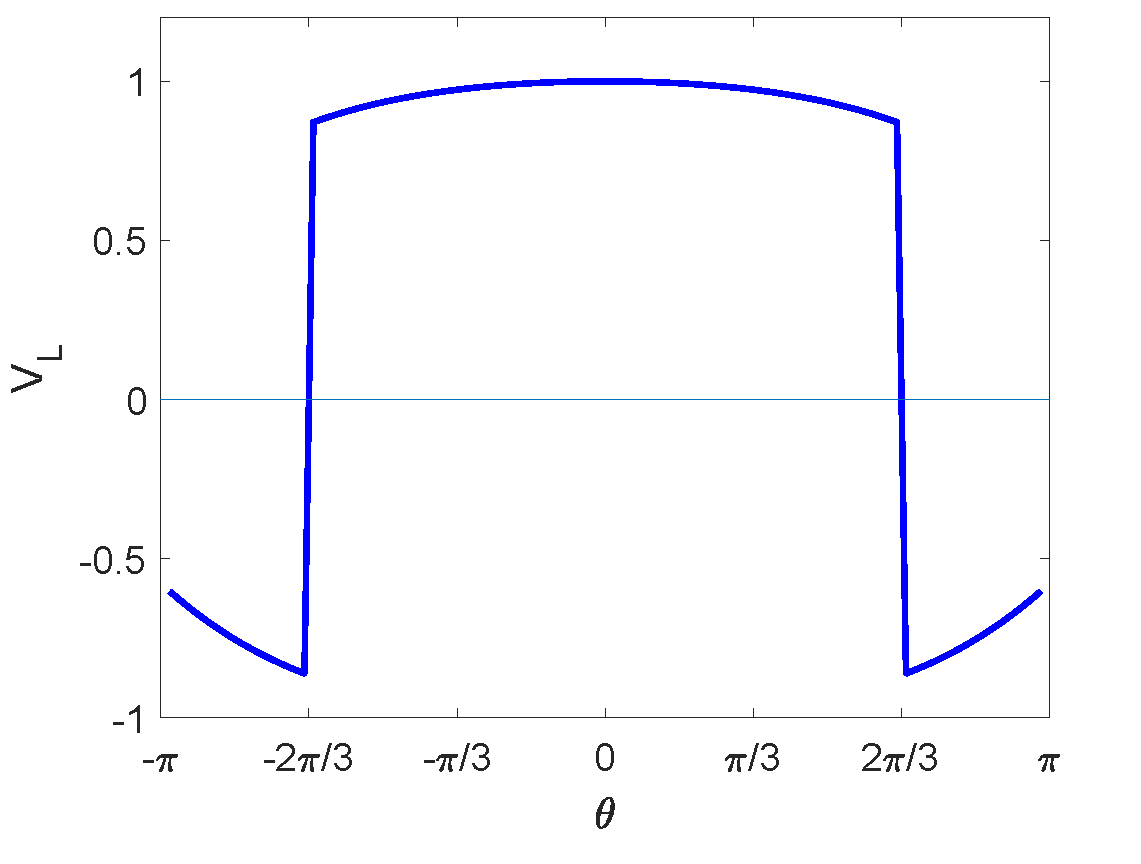}}
    \caption{The velocity $v_L^{(2)}{(\theta)}$ for the one-parameter family of quantum walks defined by the coin class $\mathcal{X}_{\theta}.$ We see $v_L^{(2)}(\theta)$ attains positive values for $-2\pi/3<\theta<2\pi/3$ and take maximum value $1$ at $\theta=0.$  For the Grover walk $v_L^{(2)}(\theta)=-\frac{1}{\sqrt3},\theta= \pi,-\pi.$ Moreover, the peak velocity curve is symmetric about the vertical line through $\theta=0.$ There is a certain jump discontinuity in the peak velocity curve at $\theta=\pm 2\pi/3.$ However the peak velocity $v_L^{(2)}{(\theta)}$ is not feasible to carry out. }\label{continuous deformation: root2}
\end{figure}

We plot the function $v_L^{(1)}{(\theta)}$ and $v_L^{(2)}{(\theta)}$ with $\theta$ in Figure \ref{continuous deformation: root1} and Figure \ref{continuous deformation: root2} respectively, whereas the underlying coin operator is $\X_{\theta},-\pi\leq\theta\leq\pi$.   
    
\begin{figure}[H]
    \centering
    \subfigure[$t=50$]{\includegraphics[height=3.7 cm,width=4.3 cm ]{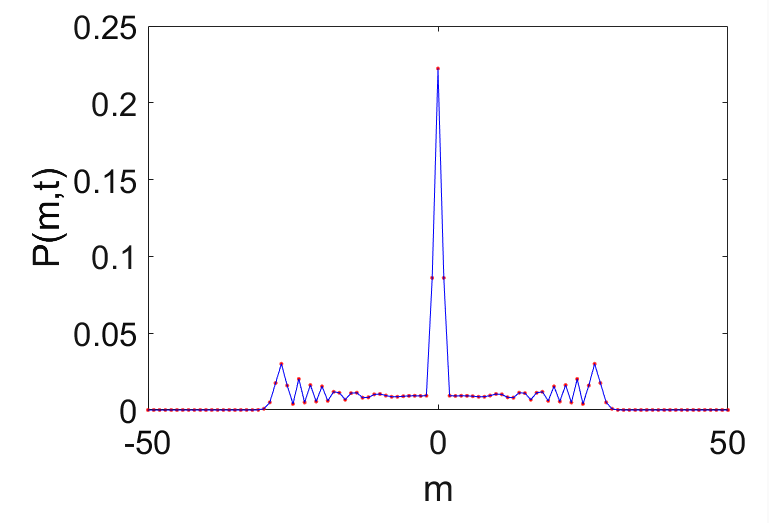}}
     \hspace{-0.3 cm}
     \subfigure[$t=100$]{\includegraphics[height=3.7 cm,width=4.3 cm]{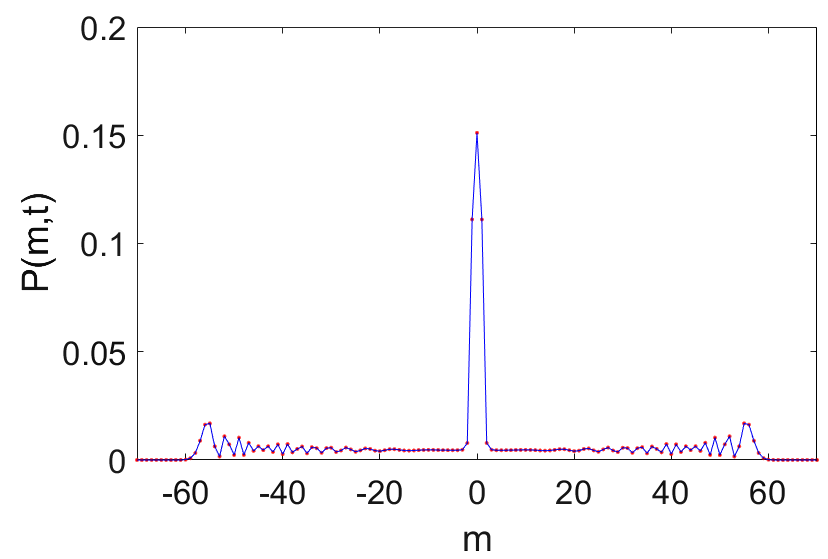}}
   % \subfigure[Probability distribution for the Grover coin with an initial state $(\frac{1}{\sqrt{2}},0,\frac{1}{\sqrt{2}}).$ ]{\includegraphics[height=4 cm,width=5 cm]{prb dis grov initial state (1,1,1).pdf}}
   % \caption{The probability distribution of the three state quantum walk using the Grover matrix as a coin operator after $T=50$ steps starting with two different initial states. In both the cases clearly the probability distribution contains three dominant peaks whose positions $-27,27$ and $0$ are determined by the velocities $v_L=-\frac{1}{\sqrt 3},v_R=\frac{1}{\sqrt 3}$ and $v_S=0$ respectively. The peak at the origin do not propagates, which emphasizes the localization of the Grover walk.}\label{fig:probability distri pi}
    \hspace{-0.2 cm}
     \subfigure[$t=500$]{\includegraphics[height=3.7 cm,width=4.3 cm]{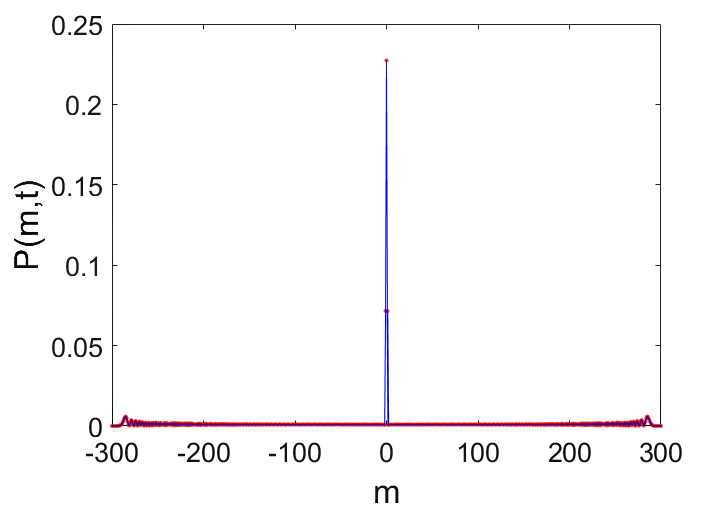}}
   \caption{The probability distribution of the Grover walk after time steps $t=50, t=100$ and $t=500$ with  initial state $(\frac{1}{\sqrt{2}},0,\frac{1}{\sqrt{2}}).$ Clearly the probability distribution contains three dominant peaks whose positions are determined by the velocities $v_L^{(1)}(\pi)=-\frac{1}{\sqrt 3}, v_S^{(1)}(\pi)=0, v_R^{(1)}(\pi)=\frac{1}{\sqrt 3}.$ In $(a)$ the peak positions correspond to $-27,0,27$ and in $(b)$ the peak positions appear at $-56,56,0.$ As time increases we see that the probability values at the peaks diminish gradually. The peak at the origin does not propagate with time, which emphasizes the localization of the Grover walk.}\label{fig:probability distri pi}
\end{figure}

   In Figure \ref{fig:probability distri pi}, we plot the probability distributions of the Grover walk for different positions of the walker after three certain time steps. 
   
\begin{figure}[H]
    \centering
    \subfigure[$\mathcal{X}_{\theta},\theta=\frac{\pi}{2}.$ ]{\includegraphics[height=4 cm,width=4 cm]{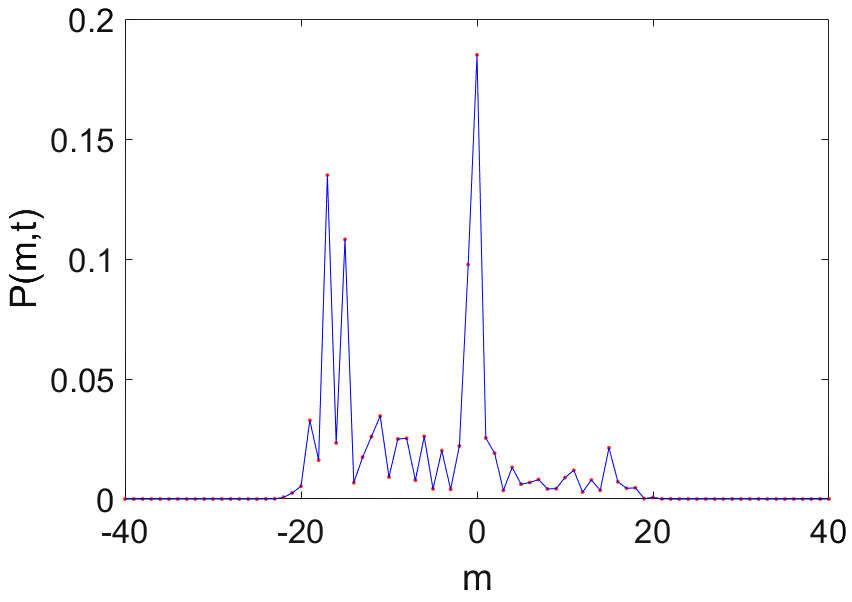}}
    \subfigure[$\mathcal{X}_{\theta},\theta=-\frac{\pi}{2}.$ ]{\includegraphics[height=4 cm,width=4 cm]{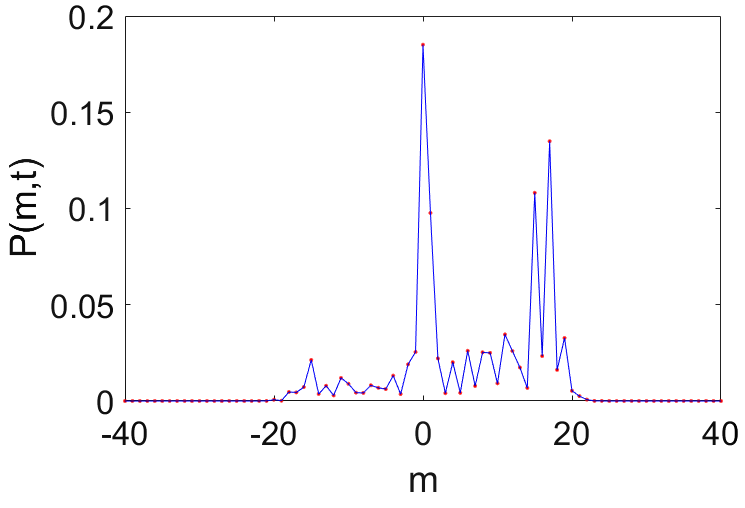}}
     \vspace{0.2 cm}
   % \subfigure[The probability distributions after combining $(a)$ and $(b)$ together for the coin $\mathcal{X}_{\theta}$ for $\theta=\frac{\pi}{2}$ and $\theta=-\frac{\pi}{2}.$ While the red and green colored nodes come from positive and negative $\theta$ respectively. ]{\includegraphics[height=6.5 cm,width=10 cm]{prb dis grov at half.pdf}}
    \caption{The probability distribution of the walk using coins from $\mathcal{X}_{\theta}$ for $\theta=\frac{\pi}{2}$ and $\theta=-\frac{\pi}{2}$ after $t=50$ time steps with the initial state $(\frac{1}{\sqrt{2}},0,\frac{1}{\sqrt{2}}).$ The peaks on the left and  the right side move with velocities $v^{(1)}_{L,R}(\frac{\pi}{2})=v^{(1)}_{L,R}(-\frac{\pi}{2}) \approx \mp 0.3568$ and appear at positions $Tv^{(1)}_{L,R}(\frac{\pi}{2})=Tv^{(1)}_{L,R}(-\frac{\pi}{2})\approx \mp 17.$ Here the probability distribution curves spread much slower in compared with  the Grover walk shown in Figure \ref{fig:probability distri pi}.  } \label{fig:probability distri pi/2}
\end{figure}

      Figure \ref{fig:probability distri pi/2} shows the probability distributions of the walks after time $t=50$  at different position of the walker with some coins from $\X_{\theta}.$
      
      From Figure \ref{continuous deformation: root1} we see $| v^{(1)}_{L}(\frac{\pi}{2})|<|v^{(1)}_{L}({\pi})|.$ Accordingly in Figure \ref{fig:probability distri pi/2} the right and left peaks travel slower than the Grover walk in Figure \ref{fig:probability distri pi} through the line, whereas  the initial state is $(\frac{1}{\sqrt{2}},0,\frac{1}{\sqrt{2}})$ and $t=50$ for both the cases. Indeed, it is to be noted that unlike the Grover coin the coins $\mathcal{X}_{\theta},\theta=\frac{\pi}{2},-\frac{\pi}{2}$ are not symmetric, which cause the asymmetric probability distributions with respect to the position $m=0.$  
      
 %$\theta=\pi/2,\pi/6$ 

\begin{figure}[H]
    \centering
   %\subfigure[with initial state $(\frac{i}{\sqrt2},0,\frac{1}{\sqrt2}).$ ]{\includegraphics[height=3 cm,width=6 cm]{prob dis all with time.pdf}}
    % \subfigure[with initial state $(\frac{1}{\sqrt3},\frac{1}{\sqrt3},\frac{1}{\sqrt3})$]{\includegraphics[height=3.2 cm,width=2.9 cm]{prob dis all with time initial111.pdf}}
    \centering
    \includegraphics[height=5 cm,width=6 cm]{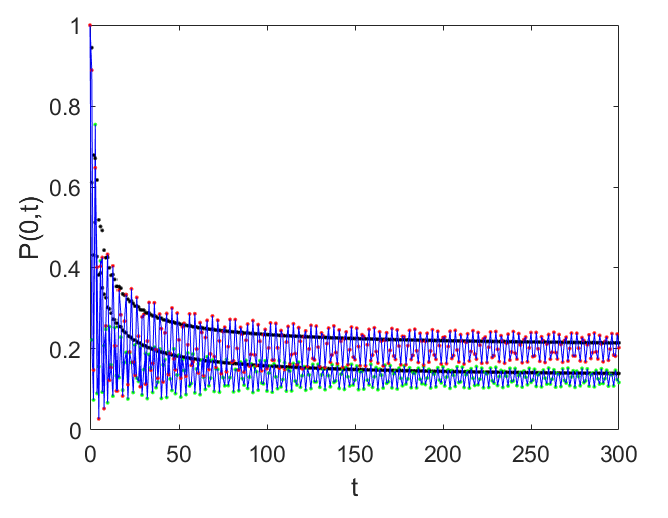}
     \caption{The probability distribution of the walker with time at the origin $m=0,$ starting with the initial state $(\frac{1}{\sqrt2},0,\frac{1}{\sqrt2}),$ while the coins are $\X_{\theta}$ for $\theta=\pi,\frac{\pi}{2}.$ The red and green colored nodes correspond to $\theta=\pi$ and $\theta=\frac{\pi}{2}$ respectively.  Here the black dotted curves passing through the probability distributions are the averaged probabilities of the corresponding walks. It looks apparently that as time propagates, the averaged probabilities of the walks at $m=0$ with the Grover coin $\X_{\pi}$ and with coin $\X_{\frac{\pi}{2}}$ converge to $0.2$ and $0.13$ }\label{probability distri 1: time}
\end{figure}

Now we see in Figure \ref{probability distri 1: time} the probability distribution of the walk with coin from $\X_{\theta}$ with time changes, at a certain position of the walker on the line. 

In Figure \ref{probability distri 1: time} we plot the probability of the walker to be at position $m=0$, over time. Starting from $m=0$, the quantum walk with coins from $\X_{\theta}$ propagates to other vertices of the line with time. However, the probability to localize (i.e.\ to remain at $m=0$) converges to a constant as $t$ increases.

\subsection{With coin operators $C\in \Y_{\theta}$}
Now we consider the walk with coins from $\Y_{\theta},$ while for $x=\frac{ 2\cos\theta-1}{3},$ by Theorem \ref{thm:eig 1d line y} the eigenvalues of $\tilde{U}_{\mathcal Y}(k)$ are $\lambda_j=e^{i\omega_j(k)},j=1,2,3$ where
 $\omega_1(k)=\pi,\omega_2(k)=-\omega_3({ k})=\omega({k})$ with 
 $\cos{\omega(k)}=-\frac{1}{3}(1-\cos{k})-\frac{\cos{\theta}}{3}(1-2 \cos{k}).$ Here also the eigenvalue $\lambda_1=-1$ is independent of wave number $k$ and hence the walk shows localization.
 Similarly as the case $\X_{\theta}$ we determine the peak velocities using the coin operators from $\Y_{\theta}.$ 
 %We get negative times the permutation matrices in the collection $\Y_{\theta}$ for $\theta=\pi,\frac{\pi}{3}, -\frac{\pi}{3},\pi.$ 
 For ${\theta=\pi},-\pi,$ $\Y_{\theta}$ becomes $-I_3,$ so that $\omega_{2,3}(k)=\pm k$ and
$v_R=1,v_L=-1.$  Similarly, if $\theta=\frac{ \pi}{3}$ or $\theta=-\frac{\pi}{3}$ we get permutation matrices $P_{(132)}\in\Y_{\theta}$ or $-P_{(123)}\in\Y_{\theta}$ respectively. For both the cases peak velocities are $0.$
    If $\theta=0$ the coin operator is the negative times Grover matrix, whose right and left peak velocities are $-\frac{1}{\sqrt3}$ and $\frac{1}{\sqrt3}$ respectively.

Now We get
$\frac{d^2\omega_{2}(k)}{dk^2}$
vanishes for
$$k_0^1=\pi-\arccos\left(\frac{5\cos{\theta}-7+3\sqrt{\cos^2{\theta}-6\cos{\theta}+5}}{2(2\cos{\theta}-1)}\right)$$ and $$k_0^2=\pi-\arccos\left(\frac{5\cos{\theta}-7-3\sqrt{\cos^2{\theta}-6\cos{\theta}+5}}{2(2\cos{\theta}-1)}\right),$$
for $(2 \cos{\theta}-1)\neq 0,-3,1$ i.e. $\theta\neq \frac{\pi}{3}, -\frac{\pi}{3},\pi,-\pi,0.$  

Then the group velocities \beano \frac{d\omega_{2}(k)}{dk} &=& -\frac{d\omega_{3}(k)}{dk} \\ &=& \frac{\frac{1}{3} \sin{k}(1-2 \cos{\theta})}{\sqrt{1-\left(\frac{1}{3}(1- \cos{k})+\frac{1}{3} \cos{\theta} (1+2 \cos{k})\right)^2}}\eeano at the stationary points $k_1^0$ and $k_2^0$ are as follows.

\beano 
v_{L}^{(1)}(\theta) &=& \left.\frac{d\omega_{2}(k)}{dk}\right\vert_{k=k_1^0} \\
&=&  \left\{
  \begin{array}{l}
-\dfrac{1}{3}\sqrt{\dfrac{9\sqrt{\cos^2{\theta}-6\cos{\theta}+5}+(15\cos{\theta}-21)}{ \sqrt{\cos^2{\theta}-6\cos{\theta}+5}+(\cos{\theta}-3)}}\\
\hfill{\mbox{if}\,\,-\dfrac{\pi}{3}<\theta<\dfrac{\pi}{3}},\\
\dfrac{1}{3}\sqrt{\dfrac{9\sqrt{\cos^2{\theta}-6\cos{\theta}+5}+(15\cos{\theta}-21)}{ \sqrt{\cos^2{\theta}-6\cos{\theta}+5}+(\cos{\theta}-3)}}\\
\hfill{\mbox{if}\,\,-\pi<\theta<-\dfrac{\pi}{3},\,\,\dfrac{\pi}{3}<\theta<\pi},\\
    \end{array}
\right. \\
v_{R}^{(1)}(\theta) &=& -v_{L}^{(1)}(\theta),\\
v_{L}^{(2)}(\theta) &=& \left.\frac{d\omega_{2}(k)}{dk}\right\vert_{k=k_2^0}\\
&=& \left\{
  \begin{array}{l}
-\dfrac{1}{3}\sqrt{\dfrac{9\sqrt{\cos^2{\theta}-6\cos{\theta}+5}-(15\cos{\theta}-21)}{ \sqrt{\cos^2{\theta}-6\cos{\theta}+5}-(\cos{\theta}-3)}}\\
\hfill{\mbox{if}\,\,-\dfrac{\pi}{3}<\theta<\dfrac{\pi}{3}},\\
\dfrac{1}{3}\sqrt{\dfrac{9\sqrt{\cos^2{\theta}-6\cos{\theta}+5}-(15\cos{\theta}-21)}{ \sqrt{\cos^2{\theta}-6\cos{\theta}+5}-(\cos{\theta}-3)}}\\
\hfill{\mbox{if}\,\,-\pi<\theta<-\dfrac{\pi}{3},\,\,\dfrac{\pi}{3}<\theta<\pi},\\
    \end{array}
\right. \\
v_{R}^{(2)}(\theta) &=& -v_{L}^{(2)}(\theta),
\eeano
whenever $(2 \cos{\theta}-1)\neq 0,-3,1.$ 

\begin{figure}[H]
    \centering
    \subfigure[Group velocity $v_L^{(1)}{(\theta)}$]{\includegraphics[height=4 cm,width=4 cm]{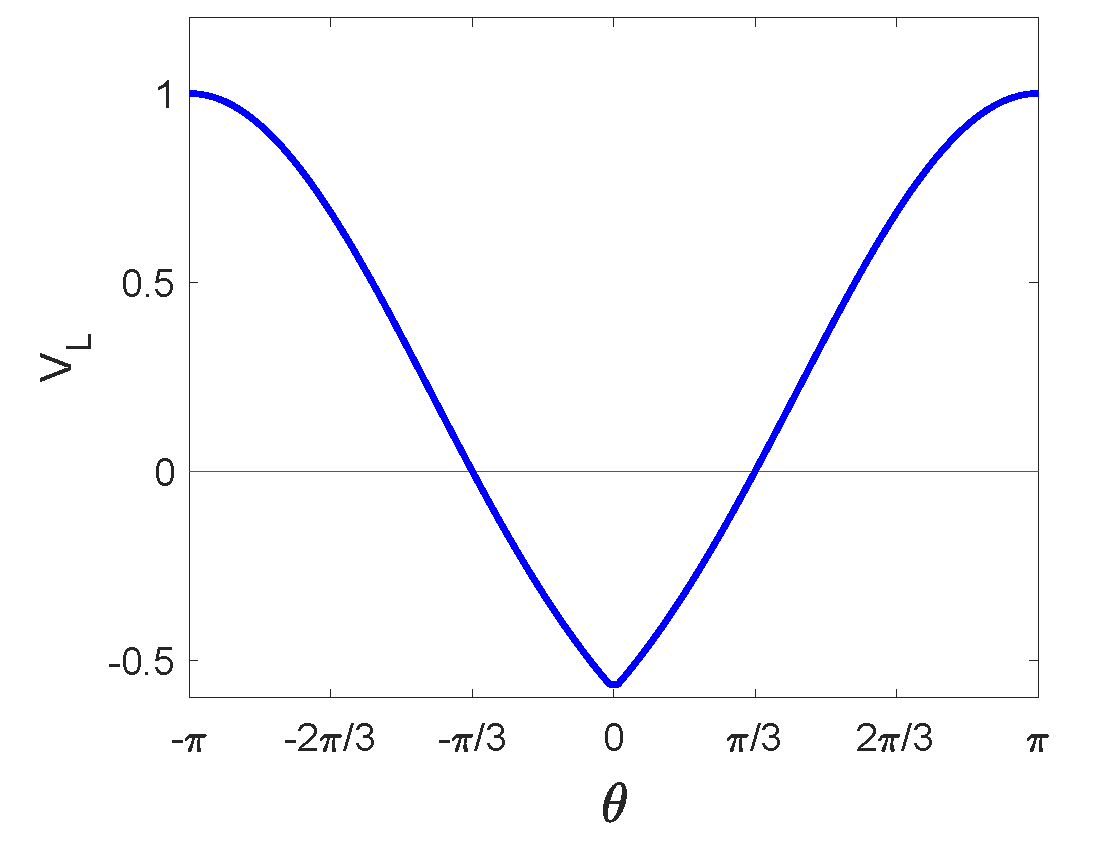}}
    \subfigure[Group velocity $v_L^{(2)}{(\theta)}$ ]{\includegraphics[height=4 cm,width=4 cm]{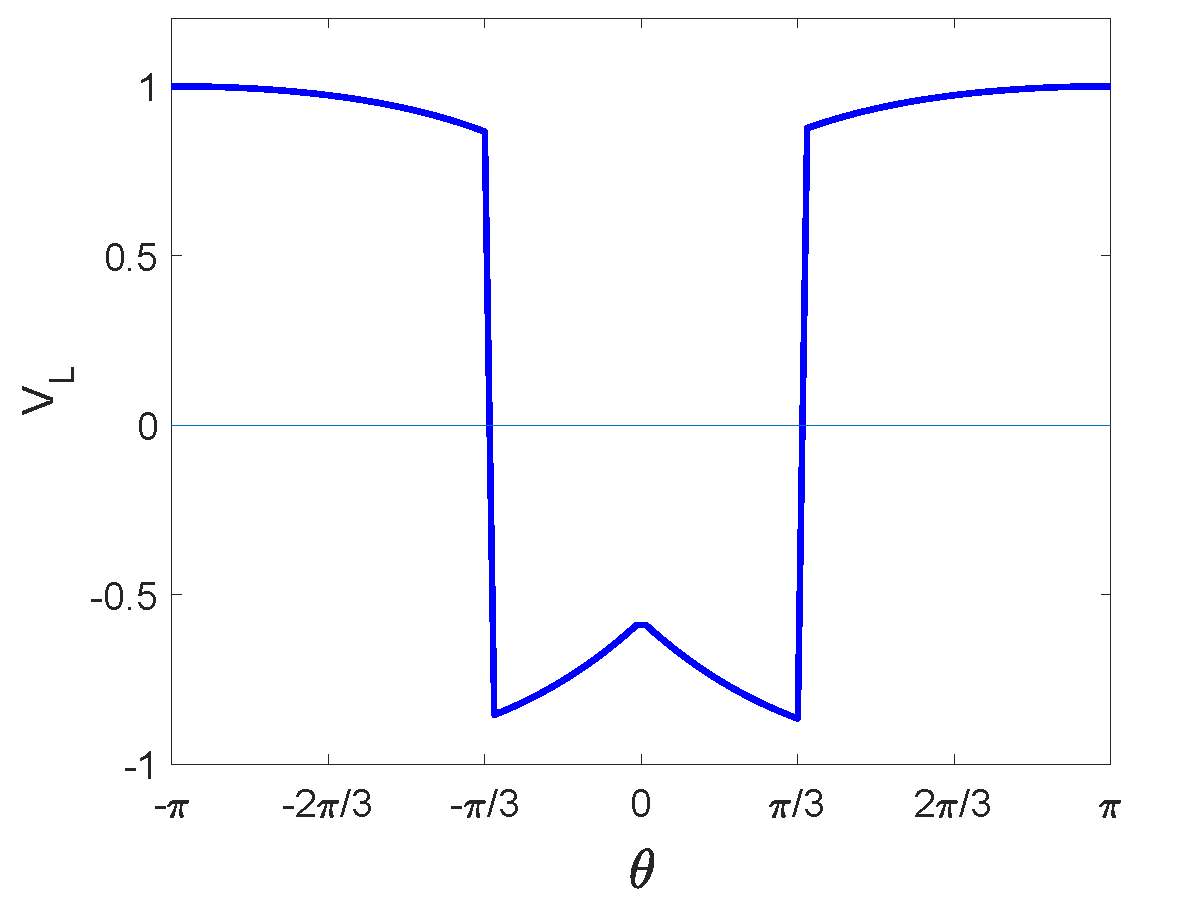}}
     \vspace{0.2 cm}
    \caption{$v_L^{(1)}{(\theta)}$ and $v_L^{(2)}{(\theta)}$ for the one-parameter family of quantum walks defined by the coin class $\Y_{\theta}.$ At $\theta=0$ $v_L^{(1)}(0)$ attains the value $-\frac{1}{\sqrt3}.$ $v_L^{(1)}{(\theta)}$ takes the maximal value $1$ at $\theta=-\pi,\pi$ and is  negative for $|\theta|<2\pi/3.$ In $(a)$ velocity increases as $|\theta|$ increases and  becomes $0$ at $\theta=\pm\pi/3.$ Whereas in $(b)$ velocity decreases as $|\theta|$ increases and there is a discontinuity of the left peak velocity curve at $\theta=\pm\pi/3,$ in contrast to $(a).$ } \label{fig:gp vel y}
\end{figure}

We show the changes in the functional values of $v_L^{(1)}{(\theta)}$ and $v_L^{(2)}{(\theta)}$ with $\theta$ in Figure \ref{fig:gp vel y}.

\begin{figure}[H]
    \centering
    \subfigure[$t=50$]{\includegraphics[height=4 cm,width=4 cm ]{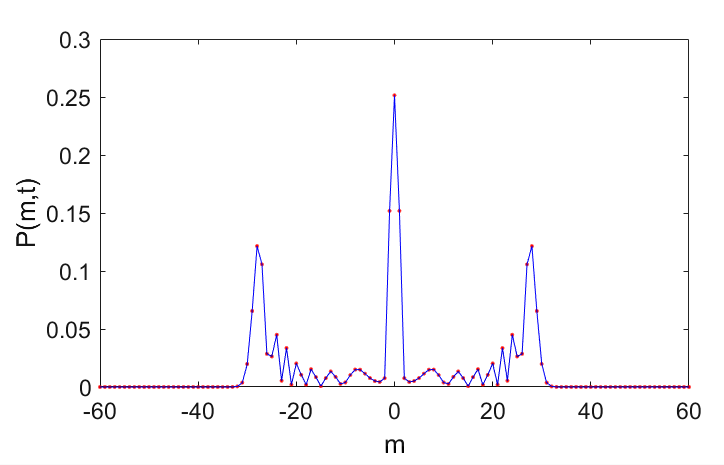}}
     \hspace{-0.3 cm}
     \subfigure[$t=100$]{\includegraphics[height=4 cm,width=4 cm]{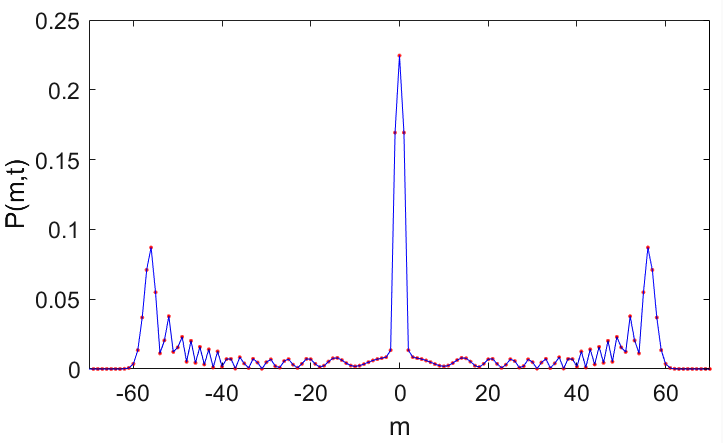}}
     \hspace{-0.3 cm}
     \subfigure[$t=500$]{\includegraphics[height=4 cm,width=4 cm]{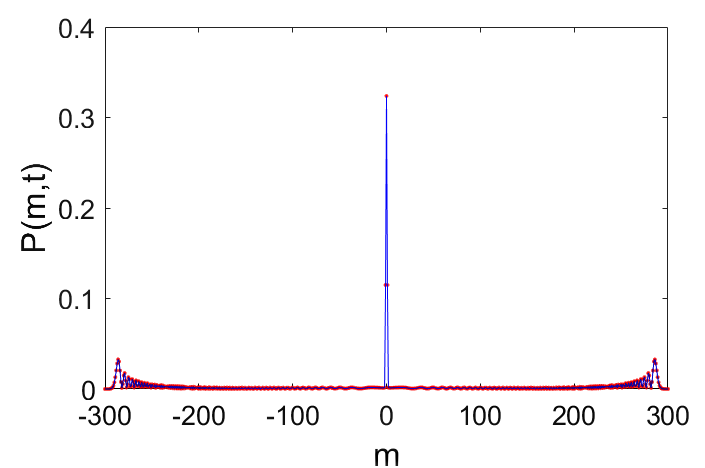}}
   \caption{The probability distribution corresponding to negative of the Grover coin after time steps $t=50,100,500$ with  initial state $(\frac{1}{\sqrt{3}},\frac{1}{\sqrt{3}},\frac{1}{\sqrt{3}}).$ Clearly the probability distribution contains three dominant peaks whose positions are determined by the velocities $v_L^{(1)}(0)=\frac{1}{\sqrt 3}, v_S^{(1)}(0)=0, v_R^{(1)}(0)=-\frac{1}{\sqrt 3}.$ In $(a)$ the peak positions correspond to $-28,0,28$ and in $(b)$ the peak positions appear at $-56,0,56.$ In contrast to $(a)$ and $(b)$ for the probability distribution in $(c)$ corresponds to  large time step $t=500,$ the left and right peaks take small probability values. In all the figures the central peak at the origin does not propagate with time, which emphasizes the localization of the walk with negative Grover coin.}\label{fig:probability distri y pi}
\end{figure}

Next Figure \ref{fig:probability distri y pi} shows the probability distributions of the walk with negative times the Grover coin i.e. $\theta=0$, at different position of the walker after three certain time steps.

\begin{figure}[H]
    \centering
    \subfigure[$\mathcal{Y}_{\theta},\theta=\frac{\pi}{6}.$ ]{\includegraphics[height=4 cm,width=4 cm]{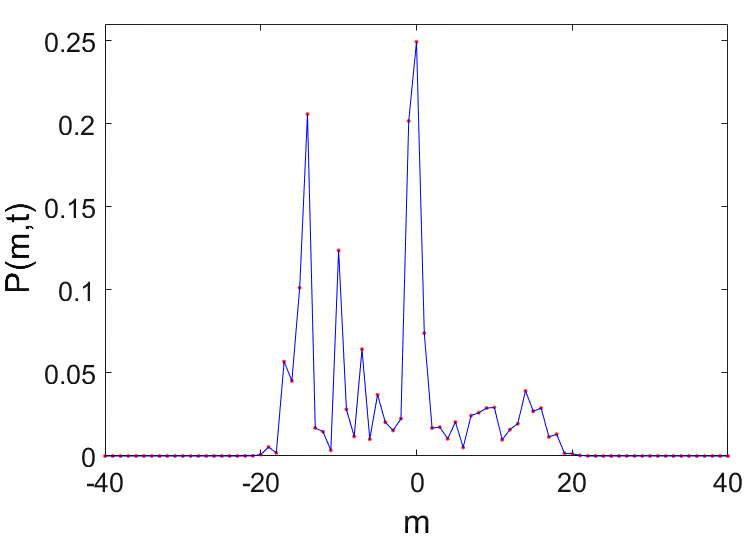}}
    \subfigure[$\mathcal{Y}_{\theta},\theta=-\frac{\pi}{6}.$ ]{\includegraphics[height=4 cm,width=4 cm]{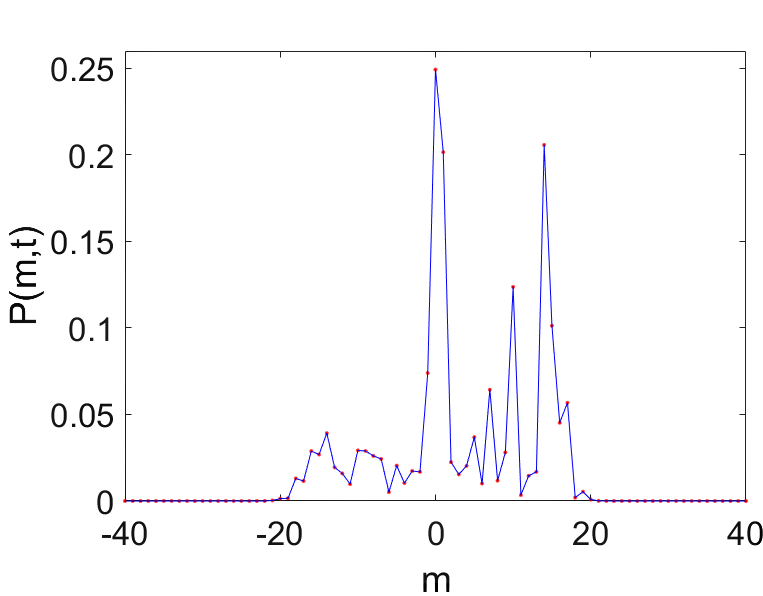}}
     \vspace{0.2 cm}
    \caption{The probability distribution of the walk using coins from $\mathcal{Y}_{\theta}$ for $\theta=\frac{\pi}{6}$ and $\theta=-\frac{\pi}{6}$ after $t=50$ time steps with the initial state $(\frac{1}{\sqrt{3}},\frac{1}{\sqrt{3}},\frac{1}{\sqrt{3}}).$ The peaks on the left and  the right side move with velocities $v^{(1)}_{L,R}(\frac{\pi}{6})=v^{(1)}_{L,R}(-\frac{\pi}{6}) $ and appear at positions $Tv^{(1)}_{L,R}(\frac{\pi}{6})=Tv^{(1)}_{L,R}(-\frac{\pi}{6})\approx \mp 14.$ Clearly the probability distribution curves spread much slower in compared with the walk shown in Figure \ref{fig:probability distri y pi}.  } \label{fig:probability distri y pi/6}
\end{figure}

 Figure \ref{fig:probability distri y pi/6} shows the probability distributions of the walks after time $t=50$  at different position of the walker with some coins from $\Y_{\theta}.$ %$\theta=\pi/2,\pi/6$ 

From Figure \ref{fig:gp vel y} we see $|v^{(1)}_{L}(\frac{\pi}{6})|<|v^{(1)}_{L}({0})|.$ Hence in Figure \ref{fig:probability distri y pi/6} the right and left peaks travel slower than the walk in Figure \ref{fig:probability distri y pi} through the line, whereas  the initial state is $(\frac{1}{\sqrt{3}},\frac{1}{\sqrt{3}},\frac{1}{\sqrt{3}})$ and $t=50$ for both the cases. In contrast to the walk with negative Grover coin the walks with coins $\mathcal{Y}_{\theta}, \theta=\frac{\pi}{6},-\frac{\pi}{6}$ have  asymmetric probability distributions with respect to the position $m=0.$ \\

\section{Conclusion} 
We have proved two limit theorems for three-state quantum walks on one dimensional lattices when the coin operators are considered as generalized Grover operators which are recently introduced in literature by characterizing orthogonal matrices of dimension $3\times 3$ that can be written as linear sum of permutation matrices. Furthermore, we have analyzed the localization properties of the walks using the obtained formula of probability measure of finding the walker at any position with a given generic initial state. Indeed, we have shown that the walks show localization phenomena generically, and we also have classified the initial states for which the walks do not show localization phenomena when the initial position is assumed to be the origin. In support of these results we have analyzed peak velocities associated with the limiting distributions of the walks. All our analytical results have been thoroughly corroborated with numerical examples.

Our results can be used to determine how fast the quantum walks described here converge to their time averaged limiting distribution. As mentioned earlier, this phenomena, known as mixing, has been crucial in analyzing the running times of several quantum algorithms. In fact, it has been demonstrated that discrete-time quantum walks on a one-dimensional lattice with a Hadamard coin mixes quadratically faster than its classical counterpart. It would be interesting to investigate whether this speedup is retained for quantum walks with the parametrized coins introduced here. Other topologies, such as higher dimensional lattices, can also be explored in this regard. Another future plan of our research includes developing quantum circuit models for the proposed quantum walks.

\begin{acknowledgments} AM thanks Council for Scientific and Industrial Research
(CSIR), India for financial support in the form of a junior/senior research fellowship. RSS acknowledges support through Prime Minister Research Fellowship (PMRF),
Government of India. SC acknowledges IIT Kharagpur for providing necessary support for his visit to IIT Kharagpur.  \end{acknowledgments}

\appendix

\section{Derivation related to limit theorems}\label{appdx:A}

Below we derive a compact expression for the quantum state $\ket{\psi(m,t)},$ the probability amplitude vector as defined in eqn. (\ref{eqn:prob am}) when the coin operator $C\in \X.$ Similar expression can be obtained for coins in $\Y.$ 

\begin{widetext}
\begin{align*}
&\sum_{j=2}^{3}\frac{1}{2\pi}\int_{-\pi}^{\pi} e^{i (- k m+ w_j( k)t)}\bra{v_j( k)} \Psi(k,0) \ket{v_j( k)} dk \\
&=\frac{1}{2\pi}\int_{-\pi}^{\pi}\frac{1}{\|f_2(\omega(k))\|^2}\left(\left[\frac{y+(1-x-y)e^{i\omega(k)}}{(1-x-y)+ye^{i(\omega(k) -k)}}, 1 , \frac{1-x-y+ye^{i\omega(k)}}{y+(1-x-y)e^{i(\omega(k)+k)}}\right]^T \right. \\
& \left.\left[\alpha\frac{y+(1-x-y)e^{-i\omega(k)}}{(1-x-y)+ye^{-i(\omega(k) -k)}}+\beta+\gamma\frac{1-x-y+ye^{-i\omega(k)}}{y+(1-x-y)e^{-i(\omega(k)+k)}} \right]e^{i(\omega(k) t-km)}\right) dk\\ &+\frac{1}{2\pi}\int_{-\pi}^{\pi}\frac{1}{\|f_3(-\omega(k))\|^2}\left(\left[\frac{y+(1-x-y)e^{-i\omega(k)}}{(1-x-y)+ye^{-i(\omega(k)+k)}}, 1 , \frac{1-x-y+ye^{-i\omega(k)}}{y+(1-x-y)e^{-i(\omega(k)-k)}}\right]^T \right. \\ 
& \left. \left[\alpha \frac{y+(1-x-y)e^{i\omega(k)}}{(1-x-y)+ye^{i(\omega(k)+k)}}+ \beta +\gamma \frac{1-x-y+ye^{i\omega(k)}}{y+(1-x-y)e^{i(\omega(k)-k)}}\right]e^{i(-\omega(k) t -km)} \right)dk \\
&=[a_{11}\alpha + a_{12}\beta + a_{13}\gamma,a_{21}\alpha + a_{22}\beta + a_{23}\gamma,a_{31}\alpha + a_{32}\beta + a_{33}\gamma]^T %= A[\alpha,\beta,\gamma]^T.
\end{align*} 
\end{widetext}

Then comparing both sides for the coefficients of $\alpha,\beta,\gamma$ we get $a_{ij},i,j\in\{1,2,3\}.$ Here we derive $a_{11}$ and the other $a_{ij}$ can be obtained  similarly.

By Theorem \ref{thm:eig 1d line} $\|f_2(\omega(k))\|^2=\|f_3(-\omega(k))\|^2$ and hence  
\begin{widetext}
\beano a_{11}=\frac{1}{2\pi}\int_{-\pi}^{\pi} && \frac{1}{\|f_2(\omega(k))\|^2}\left(\left|\frac{y+(1-x-y)e^{i\omega(k)}}{(1-x-y)+ye^{i(\omega(k) -k)}}\right|^2 e^{i(\omega(k) t-km)} + \right.\\  && \left. \left|\frac{y+(1-x-y)e^{-i\omega(k)}}{(1-x-y)+ye^{-i(\omega(k)+k)}}\right|^2e^{i-(\omega(k) t+km)}\right) dk.\eeano
\end{widetext}

Simplifying the above expression it becomes 
\begin{widetext}
\beano a_{11}=\frac{1}{2\pi}\int_{-\pi}^{\pi} && e^{-ikm} \left(\frac{1+x-2x\cos{\omega(k)}}{1+x-2x\cos{(\omega(k) -k)}}e^{i\omega(k) t}+ \right.\\ && \left. \frac{1+x-2x\cos{\omega(k)}}{1+x-2x\cos{(\omega(k) +k)}}e^{-i\omega(k) t}\right) dk.\eeano
\end{widetext}

Clearly the above expression simplifies $a_{11}$ to the form $\frac{1}{2\pi}\int_{-\pi}^{\pi}G(\omega(k),k)\cos{(\omega(k) t)}\cos{(km)} dk,$ where $G(\omega(k),k)$ is an quotient function 
%of two polynomials in $x,$ 
which is a continuous functions of $k.$ Hence for all coins in $\mathcal{X}$ which are not permutation matrices i.e. $x\neq 0,1,$ we get $G(\omega(k),k)$ is Riemann integrable as well as Lebesgue integrable. %Thus $G(\omega(k),k)$ is an $L^1$ function.

Thus using Riemann-Lebesgue Lemma we say $a_{11}\sim 0$ as $t \rightarrow \infty $.
Similarly it can be done for others $a_{ij} ,i,j\in\{1,2,3\}.$

\section{Derivation related to localization condition using the definition given in \cite{machida2014limit} }\label{appdx:B}

Recall that the localization condition for a DTQW as proposed in \cite{machida2014limit} is $\sum_{m\in \mathbb{Z}}\lim_{t \to \infty}P(m,t)$ to be positive. 

Then for coins in $\X,$ the parameter $\nu$ as mentioned in Theorem \ref{thm: limit law 1d}, is such that $-5+2\sqrt{6}\leq \nu< 1$ for $x\neq 0,1,$ so that $1-\nu^2\neq 0.$ Suppose $x\neq 0,1$ and $T_1=\alpha(1-x-y)+\beta(1+x)+\gamma y,T_2=\alpha y -\beta x,T_3=-\beta x+\gamma(1-x-y).$ Then considering the approximation of $\lim_{t \to \infty}P(m,t)$  in Theorem \ref{thm: limit law 1d}, taking sum over all the positions on the line we obtain 
\begin{widetext}
\beano  \sum_{m\in \mathbb{Z}}\lim_{t \to \infty}P(m,t)= 
 \frac{1}{3(1-x)(x+3)} &&  \left[2\left\{\left(|A|^2+|B|^2\right)\sum_{m\in \mathbb{Z}}\nu^{2|m|}+  2 \mbox{Re}(A\overline{B})\sum_{m\in \mathbb{Z}}\nu^{|m|+|m+1|}\right\}+\right.\\
&&\left.\left\{\left(|T_1|^2+|T_2|^2+|T_3|^2\right)\sum_{m\in \mathbb{Z}}\nu^{2|m|}+ \right.\right.\\ && \left.\left. 2\left(\mbox{Re}(T_1\overline{T_2})+\mbox{Re}(T_1\overline{T_3})\right)\sum_{m\in \mathbb{Z}}\nu^{|m|+|m+1|}\right\}+ \left\{2\mbox{Re}(T_2\overline T_3)\sum_{m\in \mathbb{Z}}\nu^{|-m+1|+|m+1|}\right\}\right] \\ 
=\frac{1}{3(1-x)(x+3)} && \left[\left\{\left(2\left(|A|^2+|B|^2\right) +\left(|T_1|^2+|T_2|^2+|T_3|^2\right)\right)\sum_{m\in \mathbb{Z}}\nu^{2|m|}\right\}+\right.\\ && \left\{2\left(\mbox{Re}(A\overline{B})+\mbox{Re}(T_1\overline{T_2})+\mbox{Re}(T_1\overline{T_3})\right)\sum_{m\in \mathbb{Z}}\nu^{|m|+|m+1|}\right\}+ \\ && \left.\left\{2\mbox{Re}(T_2\overline T_3)\sum_{m\in \mathbb{Z}}\nu^{|-m+1|+|m+1|}\right\}\right] \\
=\frac{1}{3(1-x)(x+3)}  && \left[\frac{1+\nu^2}{1-\nu^2}\left\{2|A|^2+2|B|^2+|T_1|^2+|T_2|^2+|T_3|^2+\right.\right.\\ && \left.\left.2\nu^{-1}\left(\mbox{Re}(A\overline{B})+\mbox{Re}(T_1\overline{T_2})+\mbox{Re}(T_1\overline{T_3})\right)+2\mbox{Re}\left(T_2\overline T_3\right)\right\}-\right.\\
&&\left.2\nu^{-1}\left(\mbox{Re}(A\overline{B})+\mbox{Re}(T_1\overline{T_2})+\mbox{Re}(T_1\overline{T_3})\right) -2(1-\nu^2)\mbox{Re}(T_2\overline T_3)\right] \\ =\frac{1}{3(1-x)(x+3)} &&  \left[\frac{1+\nu^2}{1-\nu^2}\left(2|A|^2+2|B|^2+|T_1|^2+|T_2|^2+|T_3|^2\right)+\right.\\ 
&&\left.\frac{4 \nu}{1-\nu^2}\left(\mbox{Re}(A\overline{B})+\mbox{Re}(T_1\overline{T_2})+\mbox{Re}(T_1\overline{T_3})\right)+  2\nu^2\frac{3-\nu^2}{1-\nu^2}\mbox{Re}\left(T_2\overline T_3\right)\right].
%\geq & \frac{2}{3(1-x)(x+3)}\left[\left(|A|^2+|B|^2\right)\sum_{m\in \mathbb{Z}}\nu^{2|m|}+2 \mbox{Re}(A\overline{B})\sum_{m\in \mathbb{Z}}\nu^{|m|+|m+1|}\right]\\ 
%\geq&\frac{2}{3(1-x)(x+3)(1-\nu^2)} \left[|A|^2+|B|^2+2 \nu^{-1}\mbox{Re}(A\overline{B})\right].
\eeano 
\end{widetext}

Further, for coins in $\Y,$ in order to derive a criterion for localization and to analyze the dependency of it on the initial state, we obtain  from Theorem \ref{thm:limit law 1d y} for $x\neq 0,-1,$ 
\begin{widetext}
\beano \sum_{m\in \mathbb{Z}}\lim_{t \to \infty}P(m,t)
%=&\frac{1}{3(1+x)(3-x)}\left[\frac{1+\mu^2}{1-\mu^2}\left\{2|D|^2+2|E|^2+|S_1|^2+|S_2|^2+|S_3|^2+\right.\right.\\&\left.\left.2\mu^{-1}\left(\mbox{Re}(D\overline{E})+\mbox{Re}(S_1\overline{S_2})+\mbox{Re}(S_1\overline{S_3})\right)+2\mbox{Re}\left(S_2\overline S_3\right)\right\}\right.\\&\left.-2\mu^{-1}\left(\mbox{Re}(D\overline{E})+\mbox{Re}(S_1\overline{S_2})+\mbox{Re}(S_1\overline{S_3})\right)-2(1-\mu^2)\mbox{Re}(S_2\overline S_3)\right].\\
=\frac{1}{3(1+x)(3-x)} && \left[\frac{1+\mu^2}{1-\mu^2}\left(2|D|^2+2|E|^2+|S_1|^2+|S_2|^2+|S_3|^2\right)+\right.\\ && \left.\frac{4 \mu}{1-\mu^2}\left(\mbox{Re}(D\overline{E})+\mbox{Re}(S_1\overline{S_2})+\mbox{Re}(S_1\overline{S_3})\right)  +2\mu^2\frac{3-\mu^2}{1-\mu^2}\mbox{Re}\left(S_2\overline S_3\right)\right].
%&\geq \frac{2}{3(1+x)(3-x)}\left[\left(|D|^2+|E|^2\right)\sum_{m\in \mathbb{Z}}\mu^{2|m|}+2 \mbox{Re}(D\overline{E})\sum_{m\in \mathbb{Z}}\mu^{|m|+|m+1|}\right]\\
%&=\frac{2}{3(1+x)(3-x)(1-\mu^2)} \left[|D|^2+|E|^2+2 \mu\mbox{Re}(D\overline{E})\right],
\eeano
where $S_1=\alpha(1+x+y)+\beta(1-x)-\gamma y,
S_2=-\alpha y +\beta x, S_3=\beta x+\gamma(1+x+y)$ and $-5+2\sqrt{6}\leq \mu< 1,\mu^2<1.$
\end{widetext}

% The \nocite command causes all entries in a bibliography to be printed out
% whether or not they are actually referenced in the text. This is appropriate
% for the sample file to show the different styles of references, but authors
% most likely will not want to use it.
%\nocite{*}

\bibliography{Bibliography}% Produces the bibliography via BibTeX.

%apsrev4-2.bst 2019-01-14 (MD) hand-edited version of apsrev4-1.bst
%Control: key (0)
%Control: author (8) initials jnrlst
%Control: editor formatted (1) identically to author
%Control: production of article title (0) allowed
%Control: page (0) single
%Control: year (1) truncated
%Control: production of eprint (0) enabled
\providecommand{\noopsort}[1]{}\providecommand{\singleletter}[1]{#1}%
\begin{thebibliography}{51}%
\makeatletter
\providecommand \@ifxundefined [1]{%
 \@ifx{#1\undefined}
}%
\providecommand \@ifnum [1]{%
 \ifnum #1\expandafter \@firstoftwo
 \else \expandafter \@secondoftwo
 \fi
}%
\providecommand \@ifx [1]{%
 \ifx #1\expandafter \@firstoftwo
 \else \expandafter \@secondoftwo
 \fi
}%
\providecommand \natexlab [1]{#1}%
\providecommand \enquote  [1]{``#1''}%
\providecommand \bibnamefont  [1]{#1}%
\providecommand \bibfnamefont [1]{#1}%
\providecommand \citenamefont [1]{#1}%
\providecommand \href@noop [0]{\@secondoftwo}%
\providecommand \href [0]{\begingroup \@sanitize@url \@href}%
\providecommand \@href[1]{\@@startlink{#1}\@@href}%
\providecommand \@@href[1]{\endgroup#1\@@endlink}%
\providecommand \@sanitize@url [0]{\catcode `\\12\catcode `\$12\catcode
  `\&12\catcode `\#12\catcode `\^12\catcode `\_12\catcode `\%12\relax}%
\providecommand \@@startlink[1]{}%
\providecommand \@@endlink[0]{}%
\providecommand \url  [0]{\begingroup\@sanitize@url \@url }%
\providecommand \@url [1]{\endgroup\@href {#1}{\urlprefix }}%
\providecommand \urlprefix  [0]{URL }%
\providecommand \Eprint [0]{\href }%
\providecommand \doibase [0]{https://doi.org/}%
\providecommand \selectlanguage [0]{\@gobble}%
\providecommand \bibinfo  [0]{\@secondoftwo}%
\providecommand \bibfield  [0]{\@secondoftwo}%
\providecommand \translation [1]{[#1]}%
\providecommand \BibitemOpen [0]{}%
\providecommand \bibitemStop [0]{}%
\providecommand \bibitemNoStop [0]{.\EOS\space}%
\providecommand \EOS [0]{\spacefactor3000\relax}%
\providecommand \BibitemShut  [1]{\csname bibitem#1\endcsname}%
\let\auto@bib@innerbib\@empty
%</preamble>
\bibitem [{\citenamefont {Aharonov}\ \emph {et~al.}(2001)\citenamefont
  {Aharonov}, \citenamefont {Ambainis}, \citenamefont {Kempe},\ and\
  \citenamefont {Vazirani}}]{aharonov2001quantum}%
  \BibitemOpen
  \bibfield  {author} {\bibinfo {author} {\bibfnamefont {D.}~\bibnamefont
  {Aharonov}}, \bibinfo {author} {\bibfnamefont {A.}~\bibnamefont {Ambainis}},
  \bibinfo {author} {\bibfnamefont {J.}~\bibnamefont {Kempe}},\ and\ \bibinfo
  {author} {\bibfnamefont {U.}~\bibnamefont {Vazirani}},\ }\bibfield  {title}
  {\bibinfo {title} {Quantum walks on graphs},\ }in\ \href@noop {} {\emph
  {\bibinfo {booktitle} {Proceedings of the thirty-third annual ACM symposium
  on Theory of computing}}}\ (\bibinfo {year} {2001})\ pp.\ \bibinfo {pages}
  {50--59}\BibitemShut {NoStop}%
\bibitem [{\citenamefont {Childs}(2009)}]{childs2009universal}%
  \BibitemOpen
  \bibfield  {author} {\bibinfo {author} {\bibfnamefont {A.~M.}\ \bibnamefont
  {Childs}},\ }\bibfield  {title} {\bibinfo {title} {Universal computation by
  quantum walk},\ }\href@noop {} {\bibfield  {journal} {\bibinfo  {journal}
  {Physical review letters}\ }\textbf {\bibinfo {volume} {102}},\ \bibinfo
  {pages} {180501} (\bibinfo {year} {2009})}\BibitemShut {NoStop}%
\bibitem [{\citenamefont {Lovett}\ \emph {et~al.}(2010)\citenamefont {Lovett},
  \citenamefont {Cooper}, \citenamefont {Everitt}, \citenamefont {Trevers},\
  and\ \citenamefont {Kendon}}]{lovett2010universal}%
  \BibitemOpen
  \bibfield  {author} {\bibinfo {author} {\bibfnamefont {N.~B.}\ \bibnamefont
  {Lovett}}, \bibinfo {author} {\bibfnamefont {S.}~\bibnamefont {Cooper}},
  \bibinfo {author} {\bibfnamefont {M.}~\bibnamefont {Everitt}}, \bibinfo
  {author} {\bibfnamefont {M.}~\bibnamefont {Trevers}},\ and\ \bibinfo {author}
  {\bibfnamefont {V.}~\bibnamefont {Kendon}},\ }\bibfield  {title} {\bibinfo
  {title} {Universal quantum computation using the discrete-time quantum
  walk},\ }\href@noop {} {\bibfield  {journal} {\bibinfo  {journal} {Physical
  Review A}\ }\textbf {\bibinfo {volume} {81}},\ \bibinfo {pages} {042330}
  (\bibinfo {year} {2010})}\BibitemShut {NoStop}%
\bibitem [{\citenamefont {Childs}\ \emph {et~al.}(2003)\citenamefont {Childs},
  \citenamefont {Cleve}, \citenamefont {Deotto}, \citenamefont {Farhi},
  \citenamefont {Gutmann},\ and\ \citenamefont
  {Spielman}}]{childs2003exponential}%
  \BibitemOpen
  \bibfield  {author} {\bibinfo {author} {\bibfnamefont {A.~M.}\ \bibnamefont
  {Childs}}, \bibinfo {author} {\bibfnamefont {R.}~\bibnamefont {Cleve}},
  \bibinfo {author} {\bibfnamefont {E.}~\bibnamefont {Deotto}}, \bibinfo
  {author} {\bibfnamefont {E.}~\bibnamefont {Farhi}}, \bibinfo {author}
  {\bibfnamefont {S.}~\bibnamefont {Gutmann}},\ and\ \bibinfo {author}
  {\bibfnamefont {D.~A.}\ \bibnamefont {Spielman}},\ }\bibfield  {title}
  {\bibinfo {title} {Exponential algorithmic speedup by a quantum walk},\ }in\
  \href@noop {} {\emph {\bibinfo {booktitle} {Proceedings of the thirty-fifth
  annual ACM symposium on Theory of computing}}}\ (\bibinfo {year} {2003})\
  pp.\ \bibinfo {pages} {59--68}\BibitemShut {NoStop}%
\bibitem [{\citenamefont {Ambainis}(2007)}]{ambainis2007quantum}%
  \BibitemOpen
  \bibfield  {author} {\bibinfo {author} {\bibfnamefont {A.}~\bibnamefont
  {Ambainis}},\ }\bibfield  {title} {\bibinfo {title} {Quantum walk algorithm
  for element distinctness},\ }\href@noop {} {\bibfield  {journal} {\bibinfo
  {journal} {SIAM Journal on Computing}\ }\textbf {\bibinfo {volume} {37}},\
  \bibinfo {pages} {210} (\bibinfo {year} {2007})}\BibitemShut {NoStop}%
\bibitem [{\citenamefont {Magniez}\ \emph {et~al.}(2011)\citenamefont
  {Magniez}, \citenamefont {Nayak}, \citenamefont {Roland},\ and\ \citenamefont
  {Santha}}]{magniez2011search}%
  \BibitemOpen
  \bibfield  {author} {\bibinfo {author} {\bibfnamefont {F.}~\bibnamefont
  {Magniez}}, \bibinfo {author} {\bibfnamefont {A.}~\bibnamefont {Nayak}},
  \bibinfo {author} {\bibfnamefont {J.}~\bibnamefont {Roland}},\ and\ \bibinfo
  {author} {\bibfnamefont {M.}~\bibnamefont {Santha}},\ }\bibfield  {title}
  {\bibinfo {title} {Search via quantum walk},\ }\href@noop {} {\bibfield
  {journal} {\bibinfo  {journal} {SIAM journal on computing}\ }\textbf
  {\bibinfo {volume} {40}},\ \bibinfo {pages} {142} (\bibinfo {year}
  {2011})}\BibitemShut {NoStop}%
\bibitem [{\citenamefont {Apers}\ \emph {et~al.}(2021)\citenamefont {Apers},
  \citenamefont {Chakraborty}, \citenamefont {Novo},\ and\ \citenamefont
  {Roland}}]{apers2021quadratic}%
  \BibitemOpen
  \bibfield  {author} {\bibinfo {author} {\bibfnamefont {S.}~\bibnamefont
  {Apers}}, \bibinfo {author} {\bibfnamefont {S.}~\bibnamefont {Chakraborty}},
  \bibinfo {author} {\bibfnamefont {L.}~\bibnamefont {Novo}},\ and\ \bibinfo
  {author} {\bibfnamefont {J.}~\bibnamefont {Roland}},\ }\bibfield  {title}
  {\bibinfo {title} {Quadratic speedup for spatial search by continuous-time
  quantum walk},\ }\href@noop {} {\bibfield  {journal} {\bibinfo  {journal}
  {arXiv preprint arXiv:2112.12746}\ } (\bibinfo {year} {2021})}\BibitemShut
  {NoStop}%
\bibitem [{\citenamefont {Mohseni}\ \emph {et~al.}(2008)\citenamefont
  {Mohseni}, \citenamefont {Rebentrost}, \citenamefont {Lloyd},\ and\
  \citenamefont {Aspuru-Guzik}}]{mohseni2008environment}%
  \BibitemOpen
  \bibfield  {author} {\bibinfo {author} {\bibfnamefont {M.}~\bibnamefont
  {Mohseni}}, \bibinfo {author} {\bibfnamefont {P.}~\bibnamefont {Rebentrost}},
  \bibinfo {author} {\bibfnamefont {S.}~\bibnamefont {Lloyd}},\ and\ \bibinfo
  {author} {\bibfnamefont {A.}~\bibnamefont {Aspuru-Guzik}},\ }\bibfield
  {title} {\bibinfo {title} {Environment-assisted quantum walks in
  photosynthetic energy transfer},\ }\href@noop {} {\bibfield  {journal}
  {\bibinfo  {journal} {The Journal of chemical physics}\ }\textbf {\bibinfo
  {volume} {129}},\ \bibinfo {pages} {11B603} (\bibinfo {year}
  {2008})}\BibitemShut {NoStop}%
\bibitem [{\citenamefont {Rebentrost}\ \emph {et~al.}(2009)\citenamefont
  {Rebentrost}, \citenamefont {Mohseni}, \citenamefont {Kassal}, \citenamefont
  {Lloyd},\ and\ \citenamefont {Aspuru-Guzik}}]{rebentrost2009environment}%
  \BibitemOpen
  \bibfield  {author} {\bibinfo {author} {\bibfnamefont {P.}~\bibnamefont
  {Rebentrost}}, \bibinfo {author} {\bibfnamefont {M.}~\bibnamefont {Mohseni}},
  \bibinfo {author} {\bibfnamefont {I.}~\bibnamefont {Kassal}}, \bibinfo
  {author} {\bibfnamefont {S.}~\bibnamefont {Lloyd}},\ and\ \bibinfo {author}
  {\bibfnamefont {A.}~\bibnamefont {Aspuru-Guzik}},\ }\bibfield  {title}
  {\bibinfo {title} {Environment-assisted quantum transport},\ }\href@noop {}
  {\bibfield  {journal} {\bibinfo  {journal} {New Journal of Physics}\ }\textbf
  {\bibinfo {volume} {11}},\ \bibinfo {pages} {033003} (\bibinfo {year}
  {2009})}\BibitemShut {NoStop}%
\bibitem [{\citenamefont {Chin}\ \emph {et~al.}(2010)\citenamefont {Chin},
  \citenamefont {Datta}, \citenamefont {Caruso}, \citenamefont {Huelga},\ and\
  \citenamefont {Plenio}}]{chin2010noise}%
  \BibitemOpen
  \bibfield  {author} {\bibinfo {author} {\bibfnamefont {A.~W.}\ \bibnamefont
  {Chin}}, \bibinfo {author} {\bibfnamefont {A.}~\bibnamefont {Datta}},
  \bibinfo {author} {\bibfnamefont {F.}~\bibnamefont {Caruso}}, \bibinfo
  {author} {\bibfnamefont {S.~F.}\ \bibnamefont {Huelga}},\ and\ \bibinfo
  {author} {\bibfnamefont {M.~B.}\ \bibnamefont {Plenio}},\ }\bibfield  {title}
  {\bibinfo {title} {Noise-assisted energy transfer in quantum networks and
  light-harvesting complexes},\ }\href@noop {} {\bibfield  {journal} {\bibinfo
  {journal} {New Journal of Physics}\ }\textbf {\bibinfo {volume} {12}},\
  \bibinfo {pages} {065002} (\bibinfo {year} {2010})}\BibitemShut {NoStop}%
\bibitem [{\citenamefont {Meyer}(1996)}]{meyer1996kt}%
  \BibitemOpen
  \bibfield  {author} {\bibinfo {author} {\bibfnamefont {D.~A.}\ \bibnamefont
  {Meyer}},\ }\bibfield  {title} {\bibinfo {title} {From quantum cellular
  automata to quantum lattice gases},\ }\href@noop {} {\bibfield  {journal}
  {\bibinfo  {journal} {Journal of Statistical Physics}\ }\textbf {\bibinfo
  {volume} {85}},\ \bibinfo {pages} {551} (\bibinfo {year} {1996})}\BibitemShut
  {NoStop}%
\bibitem [{\citenamefont {Farhi}\ and\ \citenamefont
  {Gutmann}(1998)}]{farhi1998quantum}%
  \BibitemOpen
  \bibfield  {author} {\bibinfo {author} {\bibfnamefont {E.}~\bibnamefont
  {Farhi}}\ and\ \bibinfo {author} {\bibfnamefont {S.}~\bibnamefont
  {Gutmann}},\ }\bibfield  {title} {\bibinfo {title} {Quantum computation and
  decision trees},\ }\href@noop {} {\bibfield  {journal} {\bibinfo  {journal}
  {Physical Review A}\ }\textbf {\bibinfo {volume} {58}},\ \bibinfo {pages}
  {915} (\bibinfo {year} {1998})}\BibitemShut {NoStop}%
\bibitem [{\citenamefont {Venegas-Andraca}(2012)}]{venegas2012quantum}%
  \BibitemOpen
  \bibfield  {author} {\bibinfo {author} {\bibfnamefont {S.~E.}\ \bibnamefont
  {Venegas-Andraca}},\ }\bibfield  {title} {\bibinfo {title} {Quantum walks: a
  comprehensive review},\ }\href@noop {} {\bibfield  {journal} {\bibinfo
  {journal} {Quantum Information Processing}\ }\textbf {\bibinfo {volume}
  {11}},\ \bibinfo {pages} {1015} (\bibinfo {year} {2012})}\BibitemShut
  {NoStop}%
\bibitem [{\citenamefont {Chakraborty}\ \emph
  {et~al.}(2020{\natexlab{a}})\citenamefont {Chakraborty}, \citenamefont
  {Novo},\ and\ \citenamefont {Roland}}]{chakraborty2020finding}%
  \BibitemOpen
  \bibfield  {author} {\bibinfo {author} {\bibfnamefont {S.}~\bibnamefont
  {Chakraborty}}, \bibinfo {author} {\bibfnamefont {L.}~\bibnamefont {Novo}},\
  and\ \bibinfo {author} {\bibfnamefont {J.}~\bibnamefont {Roland}},\
  }\bibfield  {title} {\bibinfo {title} {Finding a marked node on any graph via
  continuous-time quantum walks},\ }\href@noop {} {\bibfield  {journal}
  {\bibinfo  {journal} {Physical Review A}\ }\textbf {\bibinfo {volume}
  {102}},\ \bibinfo {pages} {022227} (\bibinfo {year}
  {2020}{\natexlab{a}})}\BibitemShut {NoStop}%
\bibitem [{\citenamefont {Atia}\ and\ \citenamefont
  {Chakraborty}(2021)}]{atia2021improved}%
  \BibitemOpen
  \bibfield  {author} {\bibinfo {author} {\bibfnamefont {Y.}~\bibnamefont
  {Atia}}\ and\ \bibinfo {author} {\bibfnamefont {S.}~\bibnamefont
  {Chakraborty}},\ }\bibfield  {title} {\bibinfo {title} {Improved upper bounds
  for the hitting times of quantum walks},\ }\href@noop {} {\bibfield
  {journal} {\bibinfo  {journal} {Physical Review A}\ }\textbf {\bibinfo
  {volume} {104}},\ \bibinfo {pages} {032215} (\bibinfo {year}
  {2021})}\BibitemShut {NoStop}%
\bibitem [{\citenamefont {Aharonov}\ \emph {et~al.}(2008)\citenamefont
  {Aharonov}, \citenamefont {Van~Dam}, \citenamefont {Kempe}, \citenamefont
  {Landau}, \citenamefont {Lloyd},\ and\ \citenamefont
  {Regev}}]{aharonov2008adiabatic}%
  \BibitemOpen
  \bibfield  {author} {\bibinfo {author} {\bibfnamefont {D.}~\bibnamefont
  {Aharonov}}, \bibinfo {author} {\bibfnamefont {W.}~\bibnamefont {Van~Dam}},
  \bibinfo {author} {\bibfnamefont {J.}~\bibnamefont {Kempe}}, \bibinfo
  {author} {\bibfnamefont {Z.}~\bibnamefont {Landau}}, \bibinfo {author}
  {\bibfnamefont {S.}~\bibnamefont {Lloyd}},\ and\ \bibinfo {author}
  {\bibfnamefont {O.}~\bibnamefont {Regev}},\ }\bibfield  {title} {\bibinfo
  {title} {Adiabatic quantum computation is equivalent to standard quantum
  computation},\ }\href@noop {} {\bibfield  {journal} {\bibinfo  {journal}
  {SIAM review}\ }\textbf {\bibinfo {volume} {50}},\ \bibinfo {pages} {755}
  (\bibinfo {year} {2008})}\BibitemShut {NoStop}%
\bibitem [{\citenamefont {Caha}\ \emph {et~al.}(2018)\citenamefont {Caha},
  \citenamefont {Landau},\ and\ \citenamefont {Nagaj}}]{caha2018clocks}%
  \BibitemOpen
  \bibfield  {author} {\bibinfo {author} {\bibfnamefont {L.}~\bibnamefont
  {Caha}}, \bibinfo {author} {\bibfnamefont {Z.}~\bibnamefont {Landau}},\ and\
  \bibinfo {author} {\bibfnamefont {D.}~\bibnamefont {Nagaj}},\ }\bibfield
  {title} {\bibinfo {title} {Clocks in feynman's computer and kitaev's local
  hamiltonian: Bias, gaps, idling, and pulse tuning},\ }\href@noop {}
  {\bibfield  {journal} {\bibinfo  {journal} {Physical Review A}\ }\textbf
  {\bibinfo {volume} {97}},\ \bibinfo {pages} {062306} (\bibinfo {year}
  {2018})}\BibitemShut {NoStop}%
\bibitem [{\citenamefont {Richter}(2007)}]{richter2007quantum}%
  \BibitemOpen
  \bibfield  {author} {\bibinfo {author} {\bibfnamefont {P.~C.}\ \bibnamefont
  {Richter}},\ }\bibfield  {title} {\bibinfo {title} {Quantum speedup of
  classical mixing processes},\ }\href@noop {} {\bibfield  {journal} {\bibinfo
  {journal} {Physical Review A}\ }\textbf {\bibinfo {volume} {76}},\ \bibinfo
  {pages} {042306} (\bibinfo {year} {2007})}\BibitemShut {NoStop}%
\bibitem [{\citenamefont {Chakraborty}\ \emph
  {et~al.}(2020{\natexlab{b}})\citenamefont {Chakraborty}, \citenamefont
  {Luh},\ and\ \citenamefont {Roland}}]{chakraborty2020fast}%
  \BibitemOpen
  \bibfield  {author} {\bibinfo {author} {\bibfnamefont {S.}~\bibnamefont
  {Chakraborty}}, \bibinfo {author} {\bibfnamefont {K.}~\bibnamefont {Luh}},\
  and\ \bibinfo {author} {\bibfnamefont {J.}~\bibnamefont {Roland}},\
  }\bibfield  {title} {\bibinfo {title} {How fast do quantum walks mix?},\
  }\href@noop {} {\bibfield  {journal} {\bibinfo  {journal} {Physical review
  letters}\ }\textbf {\bibinfo {volume} {124}},\ \bibinfo {pages} {050501}
  (\bibinfo {year} {2020}{\natexlab{b}})}\BibitemShut {NoStop}%
\bibitem [{\citenamefont {Chakraborty}\ \emph
  {et~al.}(2020{\natexlab{c}})\citenamefont {Chakraborty}, \citenamefont
  {Luh},\ and\ \citenamefont {Roland}}]{chakraborty2020analog}%
  \BibitemOpen
  \bibfield  {author} {\bibinfo {author} {\bibfnamefont {S.}~\bibnamefont
  {Chakraborty}}, \bibinfo {author} {\bibfnamefont {K.}~\bibnamefont {Luh}},\
  and\ \bibinfo {author} {\bibfnamefont {J.}~\bibnamefont {Roland}},\
  }\bibfield  {title} {\bibinfo {title} {Analog quantum algorithms for the
  mixing of markov chains},\ }\href@noop {} {\bibfield  {journal} {\bibinfo
  {journal} {Physical Review A}\ }\textbf {\bibinfo {volume} {102}},\ \bibinfo
  {pages} {022423} (\bibinfo {year} {2020}{\natexlab{c}})}\BibitemShut
  {NoStop}%
\bibitem [{\citenamefont {Inui}\ \emph {et~al.}(2004)\citenamefont {Inui},
  \citenamefont {Konishi},\ and\ \citenamefont {Konno}}]{inui2004localization}%
  \BibitemOpen
  \bibfield  {author} {\bibinfo {author} {\bibfnamefont {N.}~\bibnamefont
  {Inui}}, \bibinfo {author} {\bibfnamefont {Y.}~\bibnamefont {Konishi}},\ and\
  \bibinfo {author} {\bibfnamefont {N.}~\bibnamefont {Konno}},\ }\bibfield
  {title} {\bibinfo {title} {Localization of two-dimensional quantum walks},\
  }\href@noop {} {\bibfield  {journal} {\bibinfo  {journal} {Physical Review
  A}\ }\textbf {\bibinfo {volume} {69}},\ \bibinfo {pages} {052323} (\bibinfo
  {year} {2004})}\BibitemShut {NoStop}%
\bibitem [{\citenamefont {Inui}\ and\ \citenamefont
  {Konno}(2005)}]{inui2005localization}%
  \BibitemOpen
  \bibfield  {author} {\bibinfo {author} {\bibfnamefont {N.}~\bibnamefont
  {Inui}}\ and\ \bibinfo {author} {\bibfnamefont {N.}~\bibnamefont {Konno}},\
  }\bibfield  {title} {\bibinfo {title} {Localization of multi-state quantum
  walk in one dimension},\ }\href@noop {} {\bibfield  {journal} {\bibinfo
  {journal} {Physica A: Statistical Mechanics and its Applications}\ }\textbf
  {\bibinfo {volume} {353}},\ \bibinfo {pages} {133} (\bibinfo {year}
  {2005})}\BibitemShut {NoStop}%
\bibitem [{\citenamefont {Inui}\ \emph {et~al.}(2005)\citenamefont {Inui},
  \citenamefont {Konno},\ and\ \citenamefont {Segawa}}]{inui2005one}%
  \BibitemOpen
  \bibfield  {author} {\bibinfo {author} {\bibfnamefont {N.}~\bibnamefont
  {Inui}}, \bibinfo {author} {\bibfnamefont {N.}~\bibnamefont {Konno}},\ and\
  \bibinfo {author} {\bibfnamefont {E.}~\bibnamefont {Segawa}},\ }\bibfield
  {title} {\bibinfo {title} {One-dimensional three-state quantum walk},\
  }\href@noop {} {\bibfield  {journal} {\bibinfo  {journal} {Physical Review
  E}\ }\textbf {\bibinfo {volume} {72}},\ \bibinfo {pages} {056112} (\bibinfo
  {year} {2005})}\BibitemShut {NoStop}%
\bibitem [{\citenamefont {Segawa}(2013)}]{segawa2013localization}%
  \BibitemOpen
  \bibfield  {author} {\bibinfo {author} {\bibfnamefont {E.}~\bibnamefont
  {Segawa}},\ }\bibfield  {title} {\bibinfo {title} {Localization of quantum
  walks induced by recurrence properties of random walks},\ }\href@noop {}
  {\bibfield  {journal} {\bibinfo  {journal} {Journal of Computational and
  Theoretical Nanoscience}\ }\textbf {\bibinfo {volume} {10}},\ \bibinfo
  {pages} {1583} (\bibinfo {year} {2013})}\BibitemShut {NoStop}%
\bibitem [{\citenamefont {Koll{\'a}r}\ \emph {et~al.}(2015)\citenamefont
  {Koll{\'a}r}, \citenamefont {Kiss},\ and\ \citenamefont
  {Jex}}]{kollar2015strongly}%
  \BibitemOpen
  \bibfield  {author} {\bibinfo {author} {\bibfnamefont {B.}~\bibnamefont
  {Koll{\'a}r}}, \bibinfo {author} {\bibfnamefont {T.}~\bibnamefont {Kiss}},\
  and\ \bibinfo {author} {\bibfnamefont {I.}~\bibnamefont {Jex}},\ }\bibfield
  {title} {\bibinfo {title} {Strongly trapped two-dimensional quantum walks},\
  }\href@noop {} {\bibfield  {journal} {\bibinfo  {journal} {Physical Review
  A}\ }\textbf {\bibinfo {volume} {91}},\ \bibinfo {pages} {022308} (\bibinfo
  {year} {2015})}\BibitemShut {NoStop}%
\bibitem [{\citenamefont {Tate}(2019)}]{tate2019eigenvalues}%
  \BibitemOpen
  \bibfield  {author} {\bibinfo {author} {\bibfnamefont {T.}~\bibnamefont
  {Tate}},\ }\bibfield  {title} {\bibinfo {title} {Eigenvalues, absolute
  continuity and localizations for periodic unitary transition operators},\
  }\href@noop {} {\bibfield  {journal} {\bibinfo  {journal} {Infinite
  Dimensional Analysis, Quantum Probability and Related Topics}\ }\textbf
  {\bibinfo {volume} {22}},\ \bibinfo {pages} {1950011} (\bibinfo {year}
  {2019})}\BibitemShut {NoStop}%
\bibitem [{\citenamefont {Konno}(2010)}]{konno2009localization}%
  \BibitemOpen
  \bibfield  {author} {\bibinfo {author} {\bibfnamefont {N.}~\bibnamefont
  {Konno}},\ }\bibfield  {title} {\bibinfo {title} {Localization of an
  inhomogeneous discrete-time quantum walk on the line},\ }\href@noop {}
  {\bibfield  {journal} {\bibinfo  {journal} {Quantum Information Processing}\
  }\textbf {\bibinfo {volume} {9}},\ \bibinfo {pages} {405} (\bibinfo {year}
  {2010})}\BibitemShut {NoStop}%
\bibitem [{\citenamefont {Sarkar}\ \emph {et~al.}(2020)\citenamefont {Sarkar},
  \citenamefont {Mandal},\ and\ \citenamefont
  {Adhikari}}]{sarkar2020periodicity}%
  \BibitemOpen
  \bibfield  {author} {\bibinfo {author} {\bibfnamefont {R.~S.}\ \bibnamefont
  {Sarkar}}, \bibinfo {author} {\bibfnamefont {A.}~\bibnamefont {Mandal}},\
  and\ \bibinfo {author} {\bibfnamefont {B.}~\bibnamefont {Adhikari}},\
  }\bibfield  {title} {\bibinfo {title} {Periodicity of lively quantum walks on
  cycles with generalized grover coin},\ }\href@noop {} {\bibfield  {journal}
  {\bibinfo  {journal} {Linear Algebra and its Applications}\ }\textbf
  {\bibinfo {volume} {604}},\ \bibinfo {pages} {399} (\bibinfo {year}
  {2020})}\BibitemShut {NoStop}%
\bibitem [{\citenamefont {Koll{\'a}r}\ \emph {et~al.}(2010)\citenamefont
  {Koll{\'a}r}, \citenamefont {{\v{S}}tefa{\v{n}}{\'a}k}, \citenamefont
  {Kiss},\ and\ \citenamefont {Jex}}]{kollar2010recurrences}%
  \BibitemOpen
  \bibfield  {author} {\bibinfo {author} {\bibfnamefont {B.}~\bibnamefont
  {Koll{\'a}r}}, \bibinfo {author} {\bibfnamefont {M.}~\bibnamefont
  {{\v{S}}tefa{\v{n}}{\'a}k}}, \bibinfo {author} {\bibfnamefont
  {T.}~\bibnamefont {Kiss}},\ and\ \bibinfo {author} {\bibfnamefont
  {I.}~\bibnamefont {Jex}},\ }\bibfield  {title} {\bibinfo {title} {Recurrences
  in three-state quantum walks on a plane},\ }\href@noop {} {\bibfield
  {journal} {\bibinfo  {journal} {Physical Review A}\ }\textbf {\bibinfo
  {volume} {82}},\ \bibinfo {pages} {012303} (\bibinfo {year}
  {2010})}\BibitemShut {NoStop}%
\bibitem [{\citenamefont {Koll{\'a}r}\ \emph {et~al.}(2020)\citenamefont
  {Koll{\'a}r}, \citenamefont {Gily{\'e}n}, \citenamefont
  {Tk{\'a}{\v{c}}ov{\'a}}, \citenamefont {Kiss}, \citenamefont {Jex},\ and\
  \citenamefont {{\v{S}}tefa{\v{n}}{\'a}k}}]{kollar2020complete}%
  \BibitemOpen
  \bibfield  {author} {\bibinfo {author} {\bibfnamefont {B.}~\bibnamefont
  {Koll{\'a}r}}, \bibinfo {author} {\bibfnamefont {A.}~\bibnamefont
  {Gily{\'e}n}}, \bibinfo {author} {\bibfnamefont {I.}~\bibnamefont
  {Tk{\'a}{\v{c}}ov{\'a}}}, \bibinfo {author} {\bibfnamefont {T.}~\bibnamefont
  {Kiss}}, \bibinfo {author} {\bibfnamefont {I.}~\bibnamefont {Jex}},\ and\
  \bibinfo {author} {\bibfnamefont {M.}~\bibnamefont
  {{\v{S}}tefa{\v{n}}{\'a}k}},\ }\bibfield  {title} {\bibinfo {title} {Complete
  classification of trapping coins for quantum walks on the two-dimensional
  square lattice},\ }\href@noop {} {\bibfield  {journal} {\bibinfo  {journal}
  {Physical Review A}\ }\textbf {\bibinfo {volume} {102}},\ \bibinfo {pages}
  {012207} (\bibinfo {year} {2020})}\BibitemShut {NoStop}%
\bibitem [{\citenamefont {{\v{S}}tefa{\v{n}}{\'a}k}\ \emph
  {et~al.}(2012)\citenamefont {{\v{S}}tefa{\v{n}}{\'a}k}, \citenamefont
  {Bezd{\v{e}}kov{\'a}},\ and\ \citenamefont {Jex}}]{vstefavnak2012continuous}%
  \BibitemOpen
  \bibfield  {author} {\bibinfo {author} {\bibfnamefont {M.}~\bibnamefont
  {{\v{S}}tefa{\v{n}}{\'a}k}}, \bibinfo {author} {\bibfnamefont
  {I.}~\bibnamefont {Bezd{\v{e}}kov{\'a}}},\ and\ \bibinfo {author}
  {\bibfnamefont {I.}~\bibnamefont {Jex}},\ }\bibfield  {title} {\bibinfo
  {title} {Continuous deformations of the grover walk preserving
  localization},\ }\href@noop {} {\bibfield  {journal} {\bibinfo  {journal}
  {The European Physical Journal D}\ }\textbf {\bibinfo {volume} {66}},\
  \bibinfo {pages} {1} (\bibinfo {year} {2012})}\BibitemShut {NoStop}%
\bibitem [{\citenamefont {Mandal}\ and\ \citenamefont
  {Adhikari}(2022)}]{MANDAL2022}%
  \BibitemOpen
  \bibfield  {author} {\bibinfo {author} {\bibfnamefont {A.}~\bibnamefont
  {Mandal}}\ and\ \bibinfo {author} {\bibfnamefont {B.}~\bibnamefont
  {Adhikari}},\ }\bibfield  {title} {\bibinfo {title} {A characterization of
  orthogonal permutative matrices of order 4},\ }\href@noop {} {\bibfield
  {journal} {\bibinfo  {journal} {Linear Algebra and its Applications}\
  }\textbf {\bibinfo {volume} {654}},\ \bibinfo {pages} {102} (\bibinfo {year}
  {2022})}\BibitemShut {NoStop}%
\bibitem [{\citenamefont {Watabe}\ \emph {et~al.}(2008)\citenamefont {Watabe},
  \citenamefont {Kobayashi}, \citenamefont {Katori},\ and\ \citenamefont
  {Konno}}]{watabe2008limit}%
  \BibitemOpen
  \bibfield  {author} {\bibinfo {author} {\bibfnamefont {K.}~\bibnamefont
  {Watabe}}, \bibinfo {author} {\bibfnamefont {N.}~\bibnamefont {Kobayashi}},
  \bibinfo {author} {\bibfnamefont {M.}~\bibnamefont {Katori}},\ and\ \bibinfo
  {author} {\bibfnamefont {N.}~\bibnamefont {Konno}},\ }\bibfield  {title}
  {\bibinfo {title} {Limit distributions of two-dimensional quantum walks},\
  }\href@noop {} {\bibfield  {journal} {\bibinfo  {journal} {Physical Review
  A}\ }\textbf {\bibinfo {volume} {77}},\ \bibinfo {pages} {062331} (\bibinfo
  {year} {2008})}\BibitemShut {NoStop}%
\bibitem [{\citenamefont {Mandal}\ \emph {et~al.}(2021)\citenamefont {Mandal},
  \citenamefont {Sarkar},\ and\ \citenamefont
  {Adhikari}}]{mandal2021localization}%
  \BibitemOpen
  \bibfield  {author} {\bibinfo {author} {\bibfnamefont {A.}~\bibnamefont
  {Mandal}}, \bibinfo {author} {\bibfnamefont {R.~S.}\ \bibnamefont {Sarkar}},\
  and\ \bibinfo {author} {\bibfnamefont {B.}~\bibnamefont {Adhikari}},\
  }\bibfield  {title} {\bibinfo {title} {Localization of two-dimensional
  quantum walks defined by generalized grover coins},\ }\href@noop {}
  {\bibfield  {journal} {\bibinfo  {journal} {arXiv preprint arXiv:2103.00515}\
  } (\bibinfo {year} {2021})}\BibitemShut {NoStop}%
\bibitem [{\citenamefont {Di~Franco}\ \emph {et~al.}(2011)\citenamefont
  {Di~Franco}, \citenamefont {Mc~Gettrick}, \citenamefont {Machida},\ and\
  \citenamefont {Busch}}]{di2011alternate}%
  \BibitemOpen
  \bibfield  {author} {\bibinfo {author} {\bibfnamefont {C.}~\bibnamefont
  {Di~Franco}}, \bibinfo {author} {\bibfnamefont {M.}~\bibnamefont
  {Mc~Gettrick}}, \bibinfo {author} {\bibfnamefont {T.}~\bibnamefont
  {Machida}},\ and\ \bibinfo {author} {\bibfnamefont {T.}~\bibnamefont
  {Busch}},\ }\bibfield  {title} {\bibinfo {title} {Alternate two-dimensional
  quantum walk with a single-qubit coin},\ }\href@noop {} {\bibfield  {journal}
  {\bibinfo  {journal} {Physical Review A}\ }\textbf {\bibinfo {volume} {84}},\
  \bibinfo {pages} {042337} (\bibinfo {year} {2011})}\BibitemShut {NoStop}%
\bibitem [{\citenamefont {Ide}\ \emph {et~al.}(2011)\citenamefont {Ide},
  \citenamefont {Konno}, \citenamefont {Machida},\ and\ \citenamefont
  {Segawa}}]{ide2011return}%
  \BibitemOpen
  \bibfield  {author} {\bibinfo {author} {\bibfnamefont {Y.}~\bibnamefont
  {Ide}}, \bibinfo {author} {\bibfnamefont {N.}~\bibnamefont {Konno}}, \bibinfo
  {author} {\bibfnamefont {T.}~\bibnamefont {Machida}},\ and\ \bibinfo {author}
  {\bibfnamefont {E.}~\bibnamefont {Segawa}},\ }\bibfield  {title} {\bibinfo
  {title} {Return probability of one-dimensional discrete-time quantum walks
  with final-time dependence},\ }\href@noop {} {\bibfield  {journal} {\bibinfo
  {journal} {Quantum Information and Computation}\ }\textbf {\bibinfo {volume}
  {11}},\ \bibinfo {pages} {761} (\bibinfo {year} {2011})}\BibitemShut
  {NoStop}%
\bibitem [{\citenamefont {{\v{S}}tefa{\v{n}}{\'a}k}\ \emph
  {et~al.}(2014)\citenamefont {{\v{S}}tefa{\v{n}}{\'a}k}, \citenamefont
  {Bezd{\v{e}}kov{\'a}},\ and\ \citenamefont {Jex}}]{vstefavnak2014limit}%
  \BibitemOpen
  \bibfield  {author} {\bibinfo {author} {\bibfnamefont {M.}~\bibnamefont
  {{\v{S}}tefa{\v{n}}{\'a}k}}, \bibinfo {author} {\bibfnamefont
  {I.}~\bibnamefont {Bezd{\v{e}}kov{\'a}}},\ and\ \bibinfo {author}
  {\bibfnamefont {I.}~\bibnamefont {Jex}},\ }\bibfield  {title} {\bibinfo
  {title} {Limit distributions of three-state quantum walks: the role of coin
  eigenstates},\ }\href@noop {} {\bibfield  {journal} {\bibinfo  {journal}
  {Physical Review A}\ }\textbf {\bibinfo {volume} {90}},\ \bibinfo {pages}
  {012342} (\bibinfo {year} {2014})}\BibitemShut {NoStop}%
\bibitem [{\citenamefont {Chen}\ and\ \citenamefont
  {Zhang}(2016)}]{chen2016defect}%
  \BibitemOpen
  \bibfield  {author} {\bibinfo {author} {\bibfnamefont {T.}~\bibnamefont
  {Chen}}\ and\ \bibinfo {author} {\bibfnamefont {X.}~\bibnamefont {Zhang}},\
  }\bibfield  {title} {\bibinfo {title} {The defect-induced localization in
  many positions of the quantum random walk},\ }\href@noop {} {\bibfield
  {journal} {\bibinfo  {journal} {Scientific Reports}\ }\textbf {\bibinfo
  {volume} {6}},\ \bibinfo {pages} {1} (\bibinfo {year} {2016})}\BibitemShut
  {NoStop}%
\bibitem [{\citenamefont {Bataille}(2022)}]{bataille2022quantum}%
  \BibitemOpen
  \bibfield  {author} {\bibinfo {author} {\bibfnamefont {M.}~\bibnamefont
  {Bataille}},\ }\bibfield  {title} {\bibinfo {title} {Quantum circuits of
  \texttt{CNOT} gates: optimization and entanglement},\ }\href@noop {}
  {\bibfield  {journal} {\bibinfo  {journal} {Quantum Information Processing}\
  }\textbf {\bibinfo {volume} {21}},\ \bibinfo {pages} {1} (\bibinfo {year}
  {2022})}\BibitemShut {NoStop}%
\bibitem [{\citenamefont {Knight}\ \emph {et~al.}(2003)\citenamefont {Knight},
  \citenamefont {Rold{\'a}n},\ and\ \citenamefont {Sipe}}]{knight2003quantum}%
  \BibitemOpen
  \bibfield  {author} {\bibinfo {author} {\bibfnamefont {P.~L.}\ \bibnamefont
  {Knight}}, \bibinfo {author} {\bibfnamefont {E.}~\bibnamefont {Rold{\'a}n}},\
  and\ \bibinfo {author} {\bibfnamefont {J.~E.}\ \bibnamefont {Sipe}},\
  }\bibfield  {title} {\bibinfo {title} {Quantum walk on the line as an
  interference phenomenon},\ }\href@noop {} {\bibfield  {journal} {\bibinfo
  {journal} {Physical Review A}\ }\textbf {\bibinfo {volume} {68}},\ \bibinfo
  {pages} {020301} (\bibinfo {year} {2003})}\BibitemShut {NoStop}%
\bibitem [{\citenamefont {Kempf}\ and\ \citenamefont
  {Portugal}(2009)}]{kempf2009group}%
  \BibitemOpen
  \bibfield  {author} {\bibinfo {author} {\bibfnamefont {A.}~\bibnamefont
  {Kempf}}\ and\ \bibinfo {author} {\bibfnamefont {R.}~\bibnamefont
  {Portugal}},\ }\bibfield  {title} {\bibinfo {title} {Group velocity of
  discrete-time quantum walks},\ }\href@noop {} {\bibfield  {journal} {\bibinfo
   {journal} {Physical Review A}\ }\textbf {\bibinfo {volume} {79}},\ \bibinfo
  {pages} {052317} (\bibinfo {year} {2009})}\BibitemShut {NoStop}%
\bibitem [{\citenamefont {Orthey~Jr}\ and\ \citenamefont
  {Amorim}(2019)}]{orthey2019connecting}%
  \BibitemOpen
  \bibfield  {author} {\bibinfo {author} {\bibfnamefont {A.~C.}\ \bibnamefont
  {Orthey~Jr}}\ and\ \bibinfo {author} {\bibfnamefont {E.~P.}\ \bibnamefont
  {Amorim}},\ }\bibfield  {title} {\bibinfo {title} {Connecting velocity and
  entanglement in quantum walks},\ }\href@noop {} {\bibfield  {journal}
  {\bibinfo  {journal} {Physical Review A}\ }\textbf {\bibinfo {volume} {99}},\
  \bibinfo {pages} {032320} (\bibinfo {year} {2019})}\BibitemShut {NoStop}%
\bibitem [{\citenamefont {{\v{S}}tefa{\v{n}}{\'a}k}\ \emph
  {et~al.}(2008{\natexlab{a}})\citenamefont {{\v{S}}tefa{\v{n}}{\'a}k},
  \citenamefont {Kiss},\ and\ \citenamefont {Jex}}]{vstefavnak2008recurrence}%
  \BibitemOpen
  \bibfield  {author} {\bibinfo {author} {\bibfnamefont {M.}~\bibnamefont
  {{\v{S}}tefa{\v{n}}{\'a}k}}, \bibinfo {author} {\bibfnamefont
  {T.}~\bibnamefont {Kiss}},\ and\ \bibinfo {author} {\bibfnamefont
  {I.}~\bibnamefont {Jex}},\ }\bibfield  {title} {\bibinfo {title} {Recurrence
  properties of unbiased coined quantum walks on infinite d-dimensional
  lattices},\ }\href@noop {} {\bibfield  {journal} {\bibinfo  {journal}
  {Physical Review A}\ }\textbf {\bibinfo {volume} {78}},\ \bibinfo {pages}
  {032306} (\bibinfo {year} {2008}{\natexlab{a}})}\BibitemShut {NoStop}%
\bibitem [{\citenamefont {Ko}\ \emph {et~al.}(2016)\citenamefont {Ko},
  \citenamefont {Segawa},\ and\ \citenamefont {Yoo}}]{ko2016one}%
  \BibitemOpen
  \bibfield  {author} {\bibinfo {author} {\bibfnamefont {C.~K.}\ \bibnamefont
  {Ko}}, \bibinfo {author} {\bibfnamefont {E.}~\bibnamefont {Segawa}},\ and\
  \bibinfo {author} {\bibfnamefont {H.~J.}\ \bibnamefont {Yoo}},\ }\bibfield
  {title} {\bibinfo {title} {One-dimensional three-state quantum walks: Weak
  limits and localization},\ }\href@noop {} {\bibfield  {journal} {\bibinfo
  {journal} {Infinite Dimensional Analysis, Quantum Probability and Related
  Topics}\ }\textbf {\bibinfo {volume} {19}},\ \bibinfo {pages} {1650025}
  (\bibinfo {year} {2016})}\BibitemShut {NoStop}%
\bibitem [{\citenamefont {{\v{S}}tefa{\v{n}}{\'a}k}\ \emph
  {et~al.}(2008{\natexlab{b}})\citenamefont {{\v{S}}tefa{\v{n}}{\'a}k},
  \citenamefont {Jex},\ and\ \citenamefont {Kiss}}]{vstefavnak2008polya}%
  \BibitemOpen
  \bibfield  {author} {\bibinfo {author} {\bibfnamefont {M.}~\bibnamefont
  {{\v{S}}tefa{\v{n}}{\'a}k}}, \bibinfo {author} {\bibfnamefont
  {I.}~\bibnamefont {Jex}},\ and\ \bibinfo {author} {\bibfnamefont
  {T.}~\bibnamefont {Kiss}},\ }\bibfield  {title} {\bibinfo {title} {Recurrence
  and p{\'o}lya number of quantum walks},\ }\href@noop {} {\bibfield  {journal}
  {\bibinfo  {journal} {Physical review letters}\ }\textbf {\bibinfo {volume}
  {100}},\ \bibinfo {pages} {020501} (\bibinfo {year}
  {2008}{\natexlab{b}})}\BibitemShut {NoStop}%
\bibitem [{\citenamefont {Machida}(2014)}]{machida2014limit}%
  \BibitemOpen
  \bibfield  {author} {\bibinfo {author} {\bibfnamefont {T.}~\bibnamefont
  {Machida}},\ }\bibfield  {title} {\bibinfo {title} {Limit theorems of a
  3-state quantum walk and its application for discrete uniform measures},\
  }\href@noop {} {\bibfield  {journal} {\bibinfo  {journal} {arXiv preprint
  arXiv:1404.1522}\ } (\bibinfo {year} {2014})}\BibitemShut {NoStop}%
\bibitem [{\citenamefont {Segawa}\ and\ \citenamefont
  {Suzuki}(2016)}]{segawa2016generator}%
  \BibitemOpen
  \bibfield  {author} {\bibinfo {author} {\bibfnamefont {E.}~\bibnamefont
  {Segawa}}\ and\ \bibinfo {author} {\bibfnamefont {A.}~\bibnamefont
  {Suzuki}},\ }\bibfield  {title} {\bibinfo {title} {Generator of an abstract
  quantum walk},\ }\href@noop {} {\bibfield  {journal} {\bibinfo  {journal}
  {Quantum Studies: Mathematics and Foundations}\ }\textbf {\bibinfo {volume}
  {3}},\ \bibinfo {pages} {11} (\bibinfo {year} {2016})}\BibitemShut {NoStop}%
\bibitem [{\citenamefont {Klappenecker}\ and\ \citenamefont
  {Roetteler}(2003)}]{klappenecker2003quantum}%
  \BibitemOpen
  \bibfield  {author} {\bibinfo {author} {\bibfnamefont {A.}~\bibnamefont
  {Klappenecker}}\ and\ \bibinfo {author} {\bibfnamefont {M.}~\bibnamefont
  {Roetteler}},\ }\bibfield  {title} {\bibinfo {title} {Quantum software
  reusability},\ }\href@noop {} {\bibfield  {journal} {\bibinfo  {journal}
  {International Journal of Foundations of Computer Science}\ }\textbf
  {\bibinfo {volume} {14}},\ \bibinfo {pages} {777} (\bibinfo {year}
  {2003})}\BibitemShut {NoStop}%
\bibitem [{\citenamefont {Machida}\ and\ \citenamefont
  {Chandrashekar}(2015)}]{machida2015localization}%
  \BibitemOpen
  \bibfield  {author} {\bibinfo {author} {\bibfnamefont {T.}~\bibnamefont
  {Machida}}\ and\ \bibinfo {author} {\bibfnamefont {C.}~\bibnamefont
  {Chandrashekar}},\ }\bibfield  {title} {\bibinfo {title} {Localization and
  limit laws of a three-state alternate quantum walk on a two-dimensional
  lattice},\ }\href@noop {} {\bibfield  {journal} {\bibinfo  {journal}
  {Physical Review A}\ }\textbf {\bibinfo {volume} {92}},\ \bibinfo {pages}
  {062307} (\bibinfo {year} {2015})}\BibitemShut {NoStop}%
\bibitem [{\citenamefont {Stein}\ and\ \citenamefont
  {Shakarchi}(2011)}]{stein2011fourier}%
  \BibitemOpen
  \bibfield  {author} {\bibinfo {author} {\bibfnamefont {E.~M.}\ \bibnamefont
  {Stein}}\ and\ \bibinfo {author} {\bibfnamefont {R.}~\bibnamefont
  {Shakarchi}},\ }\href@noop {} {\emph {\bibinfo {title} {Fourier analysis: an
  introduction}}},\ Vol.~\bibinfo {volume} {1}\ (\bibinfo  {publisher}
  {Princeton University Press},\ \bibinfo {year} {2011})\BibitemShut {NoStop}%
\bibitem [{\citenamefont {Wong}(2001)}]{wong2001asymptotic}%
  \BibitemOpen
  \bibfield  {author} {\bibinfo {author} {\bibfnamefont {R.}~\bibnamefont
  {Wong}},\ }\href@noop {} {\emph {\bibinfo {title} {Asymptotic approximations
  of integrals}}}\ (\bibinfo  {publisher} {SIAM},\ \bibinfo {year}
  {2001})\BibitemShut {NoStop}%
\end{thebibliography}%

\end{document}